\documentclass[preprint,aps,nofootinbib,tightenlines]{revtex4}

\usepackage{epsfig}
\usepackage{amsmath,amssymb}
\usepackage{graphicx}
\usepackage{bm}
\usepackage{comment}

\newcommand{\Choose}[2]{{\begin{pmatrix} {#1} \\ {#2} \end{pmatrix}}}
\newcommand{\lsim}{\raisebox{-0.7ex}{$\stackrel{\textstyle <}{\sim}$ }}
\newcommand{\gsim}{\raisebox{-0.7ex}{$\stackrel{\textstyle >}{\sim}$ }}
\def\sheep{$\overline{\overline{\eta}}_3^L$}
\def\apipi{a_{\pi^+\pi^+}}
\def\si{^1 \hskip -0.02in S _0}
\def\siii{^3 \hskip -0.02in S _1}
\def\diii{^3 \hskip -0.04in D _1}

\begin{document}

\begin{figure}[!t]
\vskip -1.8cm
\leftline{\includegraphics[width=0.22\textwidth]{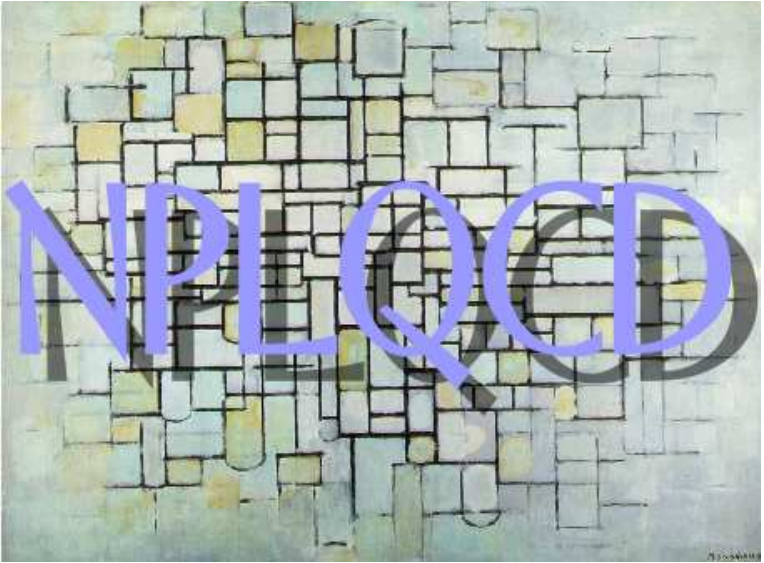}}
\end{figure}

\preprint{\vbox{
\hbox{UNH-08-02}
\hbox{JLAB-THY-08-830}
\hbox{NT@UW-08-11}
}}

\vskip 1.0cm

\title{{\Large\bf 
  HADRONIC INTERACTIONS FROM \\ LATTICE QCD}~\footnote{Review prepared for the
  {\it  International Journal of Modern Physics C}.
} \\
}

\author{\bf Silas R. Beane}
\affiliation{Department of Physics, University of New Hampshire,
Durham, NH 03824-3568, USA.\\
}
\author{\bf Kostas Orginos}
\affiliation{Department of Physics, College of William and Mary, Williamsburg,
  VA 23187-8795, USA \\
and Jefferson Laboratory, 12000 Jefferson Avenue, 
Newport News, VA 23606, USA.\\
}
\author{\bf Martin J.~Savage}
\affiliation{Department of Physics, University of Washington, 
Seattle, WA 98195-1560, USA.\\
}

\begin{abstract}
\noindent
We present an overview of recent efforts to calculate the interactions
among hadrons using lattice QCD.  After outlining the techniques
that are used to extract scattering parameters, we detail the latest
calculations of meson-meson scattering, baryon-baryon scattering and
multi-meson systems obtained with domain-wall valence quarks on the
staggered MILC lattices by the NPLQCD collaboration.  Estimates of the
computational resources required to achieve precision results in
the baryon sector are presented.  
\end{abstract}

\maketitle
\vfill\eject
\tableofcontents
\vfill\eject

\section{Introduction}
\label{sec:Intro}
\noindent
Lattice Quantum Chromodynamics (QCD) is emerging from ---what should be
considered to be--- a thirty year research and development (R and D)
phase, into the production phase, where it will produce precise and
accurate calculations of non-perturbative strong interaction
quantities.  The last five years has seen the field evolve from an era
of quenched calculations in small spatial volumes at large lattice
spacings, to the present day, where fully-dynamical calculations with
chiral symmetry at finite lattice-spacing are standard, the lattice
volumes have a spatial extent of $L \sim 2.5 - 3.5~{\rm fm}$ and
lattice spacings are $b\lsim 0.13~{\rm fm}$.  Unfortunately, it is
still the case that the light-quark masses, $m_q$, are larger than
those of nature, with typical pion masses $m_\pi\sim 300~{\rm MeV}$.
However, the next five years will see calculations at the physical
light-quark masses, $m_\pi\sim 140~{\rm MeV}$, in large volumes,
$L\gsim 6~{\rm fm}$, and at small lattice spacings, $b\lsim 0.06~{\rm
fm}$ become standard.  This impressive evolution will continue with
the increasing computer power dedicated to Lattice QCD calculations,
and development of new and faster algorithms with which to generate
lattice configurations, and to compute quark propagators and hadronic
observables.

The impact on nuclear physics of being able to reliably compute strong
interaction quantities directly from QCD, the underlying theory of the strong
interactions, 
cannot be overstated.  While
it is important to recover what is known experimentally to high
precision, that is obviously not the main objective, and certainly not
a good reason to pursue such an effort.  The reason for investing
(both money and careers) in this area is to be able to calculate
quantities of importance that cannot be accessed experimentally, or
which can be measured with only limited precision.  Nuclear physics is
an incredibly rich field, whose phenomenology has been explored for
decades.  However, there is still little understanding of the
connection to QCD and the basic building blocks of nature, quarks and
gluons.  Such a connection will be firmly established with lattice
QCD, and as a consequence, the calculation of quantities that are not
experimentally accessible will finally become possible.  On a more
academic level, Lattice QCD will allow for an exploration of how
nuclei and nuclear interactions depend upon the fundamental parameters
of nature, and an understanding of the fine-tunings that permeate
nuclear physics will finally be translated into (the possible)
fine-tunings of the light-quark masses.

Two important examples of how Lattice QCD calculations can impact
nuclear physics are in the evolution of a supernova and in the structure
of nuclei.  In the evolution of supernova, one of the key ingredients
that determines whether a supernova evolves into a black-hole or a neutron
star is the nuclear equation of state (NEOS).  The NEOS is determined
by the dynamical degrees of freedom and their interactions. The
degrees of freedom at a given density are determined by the mass of the
hadrons in the medium, which depends upon their interactions.  For
instance, the mass of the $\Sigma^-$ is expected to be significantly
less than in free-space due to its interactions with neutrons,
however, the precise mass-shift is uncertain due to the model
dependence of existing calculations~\cite{Page:2006ud}. As the
$\Sigma^-$ carries the same charge as the electron, it can become
energetically favorable to have the nuclear material composed of
neutrons, protons and $\Sigma^-$'s, as opposed to just neutrons and protons, 
due to the location of the fermi-levels.
Experimentally, little is known about the $\Sigma^-$-neutron
interaction, and a precise Lattice QCD calculation of this interaction, and others,
will greatly reduce the theoretical uncertainty in this particular
potential contribution to the NEOS in hadronic systems with densities
of a few times that of nuclear matter.  As regards the structure of nuclei, we
now have refined many-body techniques, such as Greens function
Monte-Carlo (GFMC)~\cite{Pieper:2007ax} with which to calculate the
ground states and excited states of light nuclei, with atomic number
$A\lsim 14$.  Using only the modern nucleon-nucleon (NN) potentials
that reproduce all scattering data below inelastic thresholds with
$\chi^2/dof\sim 1$, such as ${\rm AV}_{18}$~\cite{Wiringa:1994wb}, ones fails quite
dramatically to recover the structure of light nuclei.  The inclusion
of a three-nucleon interaction greatly improves the predicted
structure of nuclei~\cite{Pieper:2007ax}. Lattice QCD will be able to
calculate the interactions of multiple nucleons, 
bound or unbound in the same way it can be
used to determine the two-body scattering parameters.  For instance, a
calculation of the three-neutron interaction will be possible.

\subsection{Aspects of QCD and Nuclear Physics}
\label{sec:QCDNP}
\noindent
The structure and interactions of all nuclei are determined completely
by QCD and by the
electroweak interactions, primarily electromagnetism.  The Lagrange
density of QCD is written in terms of the quark and gluon fields, and
is manifestly invariant under local $SU(3)_c$ transformations.  At
short-distances the coupling between the quarks and gluons, and
between the gluons and themselves, $\alpha_s (Q)$, is small and
allows for processes to be computed as an asymptotic series in
$\alpha_s (Q)$--{\it asymptotic freedom}.  However, as the typical
length scale of the process grows, $\alpha_s (Q)$ becomes large, and
perturbative calculations fail for $Q^2\lsim 1~{\rm GeV}^2$.  At
long-distances, and hence low energies, the appropriate degrees of
freedom are not the quarks and gluons that the QCD Lagrange density is
written in terms of, but the hadrons, such as pions and nucleons.  It
is the properties of collections of nucleons, hyperons and mesons that
defines the field of nuclear physics, a field that, simply put, is the
exploration of the non-perturbative regime of QCD.  In this low-energy
regime, the approximate chiral symmetry of the QCD Lagrange density is
spontaneously broken by the vacuum, giving rise to pseudo-Goldstone
bosons and the approximate isospin symmetry.  With such rich dynamics
present in what naively looks like a simple theory it is not
surprising that relatively little progress has been made in
determining the properties and interactions of nuclei, even the
simplest nuclei, directly from QCD.  Understanding and calculating the
properties and interactions of nuclei is further complicated by the
fact that nuclear physics, and hence QCD, exhibits one or more
fine-tunings.  The values of the light quark masses, the scale of
chiral symmetry breaking, $\Lambda_\chi$, and the electromagnetic
coupling constant take values such that the scattering lengths in the
two-nucleon systems are both unnaturally large compared with the range
of the nuclear interaction.  Further, the location of energy-levels in
carbon and oxygen are such that the triple-$\alpha$ process can proceed to
produce enough carbon for us to be writing this review.  Of course,
one looks toward an anthropic explanation (for a lengthy discussion of
the anthropic principle see Ref.~\cite{BarrowTipler}) of these
fine-tunings, but at present we have no understanding of
how the fine-tunings exhibited in nuclear systems translate into
fine-tunings in the fundamental parameters of nature (if in fact they are fine-tuned).

\subsection{Aspects of Lattice QCD}
\label{sec:LQCD}
\noindent
The only known way to solve QCD in the low-energy regime 
is numerically using Lattice QCD.
Lattice QCD is a technique in which Euclidean space correlation
functions are calculated by a Monte-Carlo evaluation of the Euclidean
space path integral.  The calculations are performed in Euclidean
space so that contributions to any given correlation function that
have a large action are exponentially suppressed.  This is in contrast
with Minkowski space in which large action contributions result in a
complex phase which will average to an exponentially small
contribution with nearby paths.  Space-time is discretized with the
quarks residing on the lattice sites, and the gluon fields residing on
the links between lattice sites.  The lattice spacing, $b$, (the
distance between adjacent lattice sites) is required to be much
smaller than the scale of chiral symmetry breaking (or any physical
length scale in the system) for its effects to be small.  The effects
of a finite lattice spacing can be systematically removed by combining
lattice calculations of the same correlation functions at several
lattice spacing with the low-energy effective field theory (EFT)
constructed to explicitly include the lattice spacing.  The EFTs
describing such calculations are somewhat more complicated than their
continuum partners as they must reproduce that matrix elements of the
Symanzik action constructed with higher dimension operators induced by
the lattice spacing\cite{Symanzik:1983gh}.  While the action lacks
Lorentz invariance and rotational symmetry, it does possess hypercubic
symmetry.  As computers have finite memory and performance, the
lattice volumes are finite in all four space-time directions.
Generally, periodic boundary conditions (BC's) are imposed on the
fields in the space-directions (a three torus), while anti-periodic
BC's are imposed on the fields in the time-direction, which in many
cases is much larger than the space-directions~\footnote{A linear
combination of propagators generated with periodic and anti-periodic
boundary conditions can be used to effectively double the length of
the time-direction}.  For the calculations we will be discussing in
this review, the lattice volumes are large compared with the Compton
wavelength of the pion, and deviations in single particle properties
from their infinite volume values are exponentially small, generically
$\sim e^{-m_\pi L}$.  Finally, the cost of performing lattice
calculations increases dramatically with decreasing quark mass, and
presently, calculations at the physical quark masses, $m_\pi\sim
140~{\rm MeV}$ are not possible.  Currently, lattice calculations are
performed at unphysical values of the quark masses, and the light
quark mass dependence of the observable of interest, which can be
determined perturbatively in the low-energy EFTs, is used to
extrapolate to the physical light quark masses.  
Therefore, the
practical situation with current  Lattice QCD calculations is that they are performed at
finite lattice spacing, within finite volumes and at unphysical quark
masses.
The appropriate EFT (e.g. $\chi$-PT, heavy-baryon-$\chi$-PT) 
is then used to extrapolate to the infinite volume, continuum
limit of QCD where the parameters of nature reside.

\subsection{Diversion into Sociology}
\label{sec:nplqcdformation}
\begin{figure}[!ht]
\vspace*{4pt}
\centerline{\psfig{file=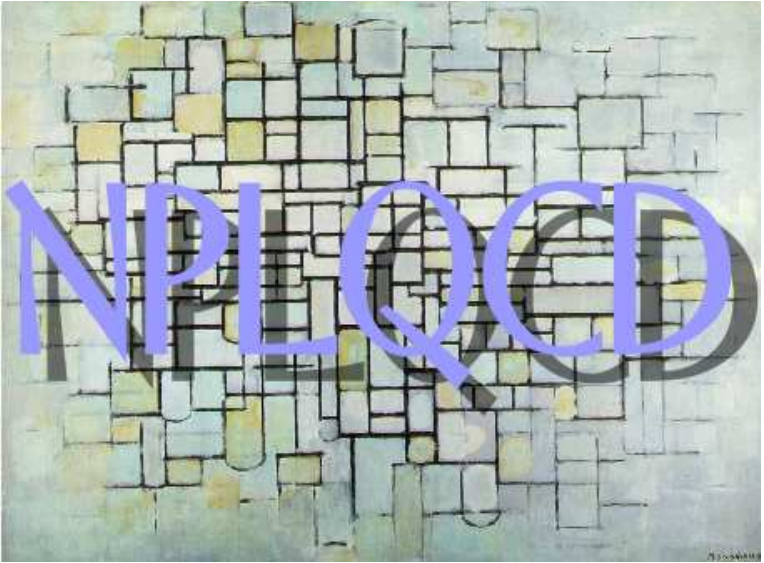,width=5.0cm}}
\vspace*{8pt}
\caption{
The NPLQCD logo.
}
\label{fig:NPLQCDlogo}
\end{figure}

\noindent
Many of the members of the NPLQCD collaboration 
(of which the authors are a part of and with the logo displayed in Fig.~\ref{fig:NPLQCDlogo})
spent years developing
EFTs for nuclear physics, and more generally performing calculations of strong
interaction processes in the low-energy regime.
Around 2005, it started to become apparent to us that in order to make progress in
low-energy QCD, the approximate chiral symmetries of QCD were, in many
cases,  no longer sufficient
to make predictions for quantities at the precision required to impact experiments, or to shed
light on fundamental questions in strong interaction physics.
Basically, the number of counterterms that were commonly appearing in
calculations beyond leading order (LO) or next-to-leading order (NLO) were
large, and in many instances exceeded the number of experimental measurements
that were available or would become available in the near future.
We decided that the only path forward in this area was to use Lattice QCD to
calculate these counterterms.
The focus of the NPLQCD collaboration is to calculate nuclear reactions from Lattice QCD, and our initial
efforts have focused on calculating the elastic scattering of the lightest
hadrons.  It is these calculations that are the focus of this review.

We wish to stress that we would not have been able to perform the
calculations in the time frame that we have if it were not for the
Lattice QCD infrastructure that existed in 2005 and continues to exist
and grow in the United States, namely USQCD and the efforts of the
lattice group at Jefferson Laboratory.  The USQCD collaborative
effort~\footnote{\rm http://www.usqcd.org/} has allowed for the large
scale production, and more importantly for us, sharing of lattice
resources.  The MILC collaboration has produced, and continues to
produce, multiple staggered (Kogut-Susskind) lattice ensembles at different quark
masses, lattice spacing and lattice volumes.  Further, NPLQCD has
shared resources with the LHP collaboration~\footnote{\rm
http://www.jlab.org/~dgr/lhpc/}.  Another significant contribution to
our program is the software package {\it
Chroma}~\cite{Edwards:2004sx,McClendon}, developed by Robert Edwards
and his team at Jefferson Laboratory, which we have used extensively
to perform our calculations.

\section{Formal Aspects of Lattice Calculations for Nuclear Physics}
\label{sec:formal}
\noindent
In this section we highlight the formal aspects of Lattice QCD
that have direct bearing on the extraction of scattering and reaction processes
in multi-hadron systems.  
Intricate details of Lattice QCD calculations are beyond the scope of this
review.  In later sections,
however, we discuss specific aspects of Lattice QCD calculations that are 
relevant to the discussions at hand.  For even finer details,
we refer the reader to a number of excellent texts on the
subject~\cite{CreutzBOOK,RotheBOOK,DeGrandDeTarBOOK}.
Throughout, we will be discussing lattices with cubic symmetry in the spatial
directions, each of length $L$ in lattice units.  The time direction is assumed to be much larger than the
spatial directions so that all calculations are assumed to be at 
zero temperature, $T=0$.
Further, while the lattice spacing is conventionally  denoted by ``a'', we
will use ``b'' in order to avoid conflict with the notation
for scattering lengths which are universally denoted by ``a''.

\subsection{Physics From Euclidean Space Correlation Functions}
\noindent
In the continuum limit, the Euclidean space correlations functions
(suitably Fourier transformed) are the sums of exponential functions,
whose arguments are the product of Euclidean time with the eigenvalues
of the Hamiltonian in the finite-volume associated with eigenstates
that couple to the hadronic sources and sinks.  At large times, the
correlation function becomes a single exponential dictated by the
ground state energy and the overlap of the source and sink with the
ground state.  As an example, consider the pion two-point function,
$C_{\pi^+}(t)$ generated by a source (and sink) of the form
$\pi^+({\bf x},t)=\overline{u}({\bf x},t)\gamma_5 d({\bf x},t)$,
\begin{eqnarray}
C_{\pi^+}(t) & = & 
\sum_{\bf x}\ \langle 0|\ \pi^- ({\bf x},t)\ \pi^+ ({\bf 0},0)\ |0\rangle
\ =\ 
\sum_{\bf x}\ \langle\pi^- ({\bf x},t)\ \pi^+ ({\bf 0},0)\rangle
\ \ \ .
\label{eq:singlepioncorrelator}
\end{eqnarray}
The sum over all lattice sites at each time-slice, $t$, projects onto the
spatial momentum ${\bf p}={\bf 0}$ states.
The source $\pi^+({\bf x},t)$ not only produces single pion states, but also
all possible states with the same quantum numbers as a single pion.
More generally, the source and sink are smeared over lattice sites in the
vicinity of $({\bf x},t)$ to increase the overlap onto the ground state and
lowest-lying excited states.
Translating the sink operator in time via $\pi^+({\bf x},t)=e^{\hat H t}
\pi^+({\bf x},0)e^{-\hat H t}$
gives~\footnote{We assume the absence of external electroweak fields that
  exert forces on hadrons in the lattice volume.}
\begin{eqnarray}
C_{\pi^+}(t) & = & 
\sum_n\ {e^{-E_n t}\over 2 E_n}\  \sum_{\bf x}\ \langle 0|\ \pi^- ({\bf
  x},0) |n\rangle 
\langle n|\pi^+ ({\bf 0},0) |0\rangle
\ \rightarrow\ A_0\ {e^{-m_\pi t}\over 2 m_\pi}
\ \ \ .
\label{eq:singlepioncorrelatorASYMP}
\end{eqnarray}
At finite lattice spacing, the correlation functions for Wilson fermions remain
the sum over exponential functions, but for particular choices of parameters
used in domain-wall fermions, the correlation functions exhibit additional sinusoidally
modulated exponential behavior at short-times (with a period set by the lattice
spacing).

It is straightforward to show that the lowest energy extracted from the
correlation function in
Eqs.~(\ref{eq:singlepioncorrelator}) and (\ref{eq:singlepioncorrelatorASYMP})
correspond to the mass of the $\pi^+$ (and more generally the mass of the lightest
hadronic state  that couples to the source and sink) in the finite volume. 
Finite volume effects are exponentially suppressed~\cite{Luscher:1985dn},
therefore, while every lattice calculation is performed in a finite volume, as
long as the volume is large compared to the pion mass, the mass of stable
single particle states can be extracted with high accuracy.

Once any correlation function, e.g. $C_{\pi^+}(t)$, is calculated, the most
common objective is to extract the argument of the exponential function that
persists at large times. One way to do this is to simply fit the function
over a finite number of time-slices to a single exponential function.  A second
method that is perhaps a bit more ``pleasing to the eye'', is to form the effective
mass function, e.g.
\begin{eqnarray}
M_{\rm eff.}(t) & = & 
\log\left({ C_{\pi^+}(t)\over  C_{\pi^+}(t+1)}\right)\ \rightarrow\ m_{\pi}
\ \ \ ,
\label{eq:effectivemassfunction}
\end{eqnarray}
where both $t$ and $M_{\rm eff.}(t)$ are in lattice units.
At large times, $M_{\rm eff.}(t)$ becomes  a constant equal to the mass of the
lightest state contributing to the correlation function~\footnote{This is
  obviously the most simplistic approach one can take to this problem.  To
  extract the ground state and excited states one can implement the method of
  L\"uscher and Wolff~\cite{Luscher:1990ck} in which the correlation functions resulting from different
sources and sinks are calculated.   The resulting matrix of correlation
functions is diagonalized, and the effective mass function for each resulting
eigenvalue can be used to extract the spectrum.}.

\subsection{Hadronic Interactions, the Maiani-Testa Theorem and L\"uscher's
  Method}
\noindent
Extracting hadronic interactions from Lattice QCD calculations is far
more complicated than the determination of the spectrum of stable particles.
This is encapsulated in the
Maiani-Testa theorem~\cite{Maiani:1990ca}, which states that S-matrix elements
cannot be extracted from infinite-volume Euclidean-space Green functions except
at kinematic thresholds~\footnote{An infinite number of infinitely precise
  calculations would allow one to circumvent this theorem.}.
This is clearly problematic from the nuclear physics perspective, as a main
motivation for pursuing Lattice QCD is to be able to compute nuclear reactions
involving multiple nucleons.
Of course, it is clear from the statement of this theorem how it can be evaded,
one computes Euclidean-space correlation functions at finite volume to extract
S-matrix elements, the formulation of which was known for decades in the
context of non-relativistic quantum mechanics~\cite{Huang:1957im} and extended to quantum field theory
by L\"uscher~\cite{Luscher:1986pf,Luscher:1990ux}.
L\"uscher showed that the energy of two particles in a  finite volume depends
in a calculable way upon their elastic scattering amplitude and their masses
for energies below the inelastic threshold. 
As a concrete example consider $\pi^+\pi^+$ scattering.
A $\pi^+\pi^+$ correlation function 
in the $A_1$ representation of the cubic group~\cite{Mandula:ut} 
(that projects onto the s-wave state in the continuum limit) is
\begin{eqnarray}
C_{\pi^+\pi^+}(p, t) & = & 
\sum_{|{\bf p}|=p}\ 
\sum_{\bf x , y}
e^{i{\bf p}\cdot({\bf x}-{\bf y})} 
\langle \pi^-(t,{\bf x})\ \pi^-(t, {\bf y})\ \pi^+(0, {\bf 0})\ \pi^+(0, {\bf 0})
\rangle
\ \ \ . 
\label{pipi_correlator} 
\end{eqnarray}
In relatively large lattice volumes the energy
difference between the interacting and non-interacting two-meson states
is a small fraction of the total energy, which is dominated by the
masses of the mesons.  In order to extract this energy difference 
the ratio of correlation functions, $G_{\pi^+ \pi^+}(p, t)$, can be formed, 
where
\begin{eqnarray}
G_{\pi^+ \pi^+}(p, t) & \equiv &
\frac{C_{\pi^+\pi^+}(p, t)}{C_{\pi^+}(t) C_{\pi^+}(t)} 
\ \rightarrow \ \sum_{n=0}^\infty\ {\cal A}_n\ e^{-\Delta E_n\ t} 
\  \ ,
\label{ratio_correlator} 
\end{eqnarray}
and the arrow denotes the large-time behavior of $G_{\pi^+ \pi^+}$.  
The energy eigenvalue,  $E_n$, and its deviation from the sum of the rest
masses of the particle, $\Delta E_n$, are related to the
center-of-mass momentum $p_n$ by
\begin{eqnarray}
\Delta E_n \ & \equiv & 
E_n\ -\  2m_\pi\ =\ 
\ 2\sqrt{\ p_n^2\ +\ m_\pi^2\ } 
\ -\ 2m_\pi \ .
\label{eq:energieshift}
\end{eqnarray}
To obtain $p\cot\delta(p)$, where $\delta(p)$ is the phase shift, the
square of the center-of-mass momentum, $p$, is extracted from this
energy shift and inserted
into~\cite{Huang:1957im,Luscher:1986pf,Luscher:1990ux,Hamber:1983vu}
\begin{eqnarray}
p\cot\delta(p) \ =\ {1\over \pi L}\ {\bf
  S}\left(\,\left(\frac{p L}{2\pi}\right)^2\,\right)
\ \ ,
\label{eq:energies}
\end{eqnarray}
which is valid below the inelastic threshold. The regulated three-dimensional sum is~\cite{Beane:2003da}
\begin{eqnarray}
{\bf S}\left(\, x \, \right)\ \equiv \ \sum_{\bf j}^{ |{\bf j}|<\Lambda}
{1\over |{\bf j}|^2-x}\ -\  {4 \pi \Lambda}
\ \ \  ,
\label{eq:Sdefined}
\end{eqnarray}
where the summation is over all triplets of integers ${\bf j}$ such that $|{\bf j}| < \Lambda$ and the
limit $\Lambda\rightarrow\infty$ is implicit.
\begin{figure}[!ht]
\vskip 0.15in
\vspace*{4pt}
\centerline{\psfig{file=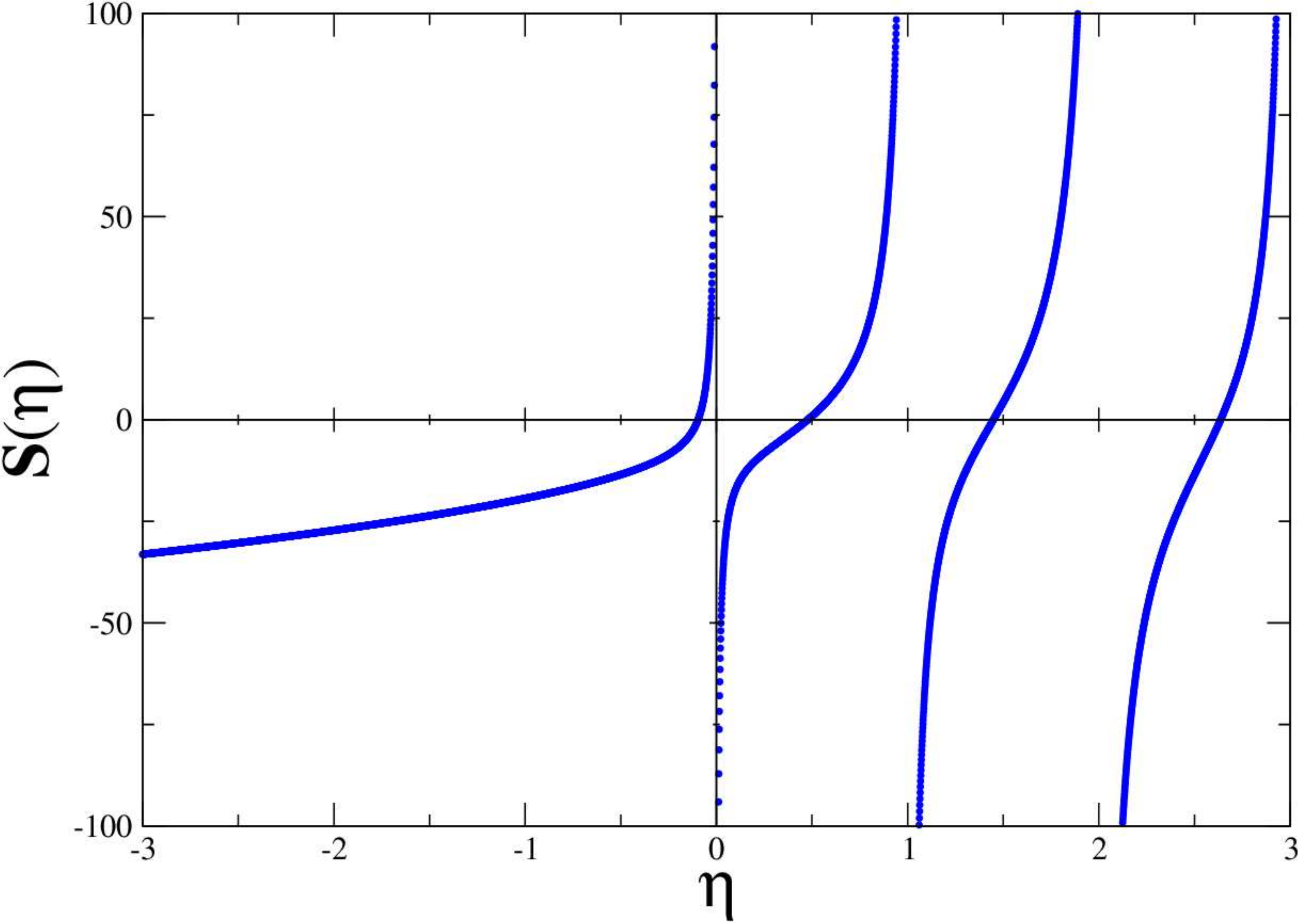,width=8.0cm}}
\vspace*{8pt}
\caption{The function ${\bf S}({\eta})$ vs.~${\eta}$, defined in
  Eq.~(\protect\ref{eq:Sdefined}), has poles only for ${\eta}\geq 0$.
}
\label{fig:Sfunction}
\end{figure}
Therefore, by measuring the energy-shift, $\Delta E_n$,  of the two particles in the finite lattice
volume, the scattering phase-shift is determined at $\Delta E_n$.
In the absence of interactions between the particles,
$|p\cot\delta|=\infty$, and the energy eigenstates in the finite volume  
occur at momenta ${\bf p} =2\pi{\bf j}/L$.  
It is important to re-emphasize that this relation 
in Eq.~(\ref{eq:energies})
is valid relativistically~\cite{Luscher:1986pf,Luscher:1990ux}.
Perhaps most important for nuclear physics is that this expression is valid
for large and  even infinite scattering lengths~\cite{Beane:2003da}.  
The only restriction is that the lattice volume be much larger than the
range of the interaction between the hadrons, which for two nucleons, is set by
the mass of the pion.

For the scattering of two nucleons, the scattering length is known to
be unnaturally large at the physical pion mass, and therefore, the
relation in Eq.~(\ref{eq:energies}) will need to be used to extract
the scattering parameters.  For systems that are not finely-tuned,
such as the $\pi^+\pi^+$ system, an expansion in the volume can be
used.  In the large volume limit ($L\gg |a|$) the energy of the two
lowest-lying continuum states in the $A_1$ representation of the cubic
group~\cite{Mandula:ut} are~\cite{Luscher:1986pf,Luscher:1990ux}
\begin{eqnarray}
\Delta E_0 & = & + {4\pi a\over M L^3}\left[\ 1\ -\ c_1 {a\over L}\ 
+\ c_2 \left({a\over L}\right)^2\ +\ ...\right]
\ +\ {\cal O}(L^{-6})
\nonumber\\
\Delta E_1 & = & {4\pi^2\over M L^2} - 
{12\tan\delta_0\over M L^2}\left[\ 
1 + c_1^\prime\tan\delta_0 + c_2^\prime \tan^2\delta_0\ +\ ...\ \right]
\ +\ {\cal O}(L^{-6})
\ \ \ ,
\label{eq:e0}
\end{eqnarray}
where $\delta_0$ is the s-wave phase-shift evaluated at $p=2\pi/L$.
The coefficients, which result from sums over the allowed
momenta~\cite{Luscher:1986pf,Luscher:1990ux} in the finite cubic volume,
are $c_1=-2.837297$, $c_2=+6.375183$, $c_1^\prime=-0.061367 $,  and $c_2^\prime=-0.354156$.
(Note that we use the nuclear physics sign convention for the scattering length.)
In addition, 
for $a>0$ with an attractive interaction~\footnote{
The extension of the 
Chowla-Selberg formula to higher dimensions~\cite{Elizalde:1997jv} gives
\begin{eqnarray}
{\bf S}\left(-x^2 \right)\
&&\ \ 
\rightarrow\ \ 
- 2 \pi^2 x \ +\  6\pi e^{-2\pi x} + ...
\ \ \ ,
\end{eqnarray}
for large $x$,
where the ellipses denote terms exponentially suppressed by factors of
$e^{-2\sqrt{2}\pi x}$, or more.}
a bound state exists with energy~\cite{Beane:2003da} (in the large volume limit)
\begin{eqnarray}
\Delta E_{-1} & = & -{\gamma^2\over M}\left[\ 
1\ +\ {12\over \gamma L}\  {1\over 1-2\gamma (p\cot\delta)^\prime}\ 
e^{-\gamma L}\ +\ ...
\right]
\ \ \ ,
\label{eq:eb}
\end{eqnarray}
where $(p\cot\delta)^\prime={d\over dp^2}\ p\cot\delta$ evaluated at
$p^2=-\gamma^2$. The quantity $\gamma$ is the solution of
\begin{eqnarray}
\gamma\  +\  p\cot\delta |_{p^2=-\gamma^2} \ & = & 0
\ \ \ ,
\label{eq:pctdg}
\end{eqnarray}
which yields the bound-state binding energy in the infinite-volume limit.
As expected, the finite volume corrections are exponentially suppressed by the
binding momentum.
This is consistent with the corrections to a single particle state where the
lightest hadronic excitation is the zero-momentum two-particle continuum state,
as opposed to a state containing an additional pion for, say, the finite volume
corrections to the single nucleon mass~\footnote{The finite volume dependence
  of bound states has been explored numerically in Ref.~\cite{Sasaki:2006jn}.}.

In the limit where $L\ll |(p\cot\delta)^{-1}|$, which is a useful limit to consider
when systems have unnaturally-large scattering lengths, 
the solution of Eq.~(\ref{eq:energies}) gives the energy of the lowest-lying state to be
\begin{eqnarray}
\Delta\tilde E_0 & = & 
{4\pi^2\over M L^2}\left[\ d_1 \ +\   d_2\  L p\cot\delta_0\  + ...\
  \right]
\nonumber\\
\Delta\tilde E_1 & = & 
{4\pi^2\over M L^2}\left[\  d_1^\prime \ +\  d_2^\prime \  L
  p\cot\delta_0
\  + ...\  \right]
\ \ \ ,
\label{eq:usE0}
\end{eqnarray}
where the coefficients are $d_1 = -0.095901$, $d_2 = +0.0253716$,
$d_1^\prime = +0.472895$, $d_2^\prime = +0.0790234$
and where $p\cot\delta_0$ in $\Delta\tilde E_0$ is evaluated at an energy
$\Delta E={4\pi^2\over M L^2}\ d_1$, while 
$p\cot\delta_0$ in $\Delta\tilde E_1$ is evaluated at an energy
$\Delta E={4\pi^2\over M L^2}\ d_1^\prime$.
The values of the $d_i^{(\prime)}$ are determined by zeros of the 
three-dimensional zeta-functions, and 
the expressions for $\Delta E_i$ and $\Delta\tilde E_i$, excluding $\Delta E_{-1}$,
are valid for both positive and negative scattering lengths.

Recently, we have performed Lattice QCD calculations of the energy of multiple
$\pi^+$'s, from which both the two-body scattering
parameters and three-body interaction were determined~\cite{Beane:2007es,Detmold:2008fn}.  In order to make use of
the lattice calculations, the ground-state energy of $n$ identical bosons in a finite volume
was required. 
The calculation of the energy-shift of $n$ identical bosons at 
${\cal O}(L^{-7})$ is straightforward but tedious.
The energy-shift of the ground state is~\cite{Beane:2007qr,Tan:2007bg,Detmold:2008gh}
\begin{eqnarray}
\label{eq:Lm7}
\Delta  E_0(n,L) &=&
  \frac{4\pi\, a}{M\,L^3}\Choose{n}{2}\Bigg\{1
-\left(\frac{a}{\pi\,L}\right){\cal I}
+\left(\frac{a}{\pi\,L}\right)^2\left[{\cal I}^2+(2n-5){\cal J}\right]
\nonumber 
\\&&\hspace*{1cm}
-
\left(\frac{a}{\pi\,L}\right)^3\Big[{\cal I}^3 + (2 n-7)
  {\cal I}{\cal J} + \left(5 n^2-41 n+63\right){\cal K}\Big]
\nonumber
\\&&\hspace*{1cm}
+
\left(\frac{a}{\pi\,L}\right)^4\Big[
{\cal I}^4 - 6 {\cal I}^2 {\cal J} + (4 + n - n^2){\cal J}^2 
+ 4 (27-15 n + n^2) {\cal I} \ {\cal K}
\nonumber\\
&&\hspace*{3cm}
+(14 n^3-227 n^2+919 n-1043) {\cal L}\ 
\Big]
\Bigg\}
\nonumber
\\&&\hspace*{0cm}
+\Choose{n}{2} \frac{8\pi^2 a^3 r }{M\, L^6}\ 
\Big[\  1\ +\ \left(\frac{a}{\pi\,L}\right) 3(n-3) {\cal I}\ 
\Big]
+\Choose{n}{3} {\overline{\eta}_3^L\over L^6}\ 
\left[ 1\ -\ 6 \ \left({a\over \pi L}\right) \ {\cal I} \ \right]
\nonumber\\
&&
+\Choose{n}{3}\left[\ 
{192 \ a^5\over M\pi^3 L^7} \left( {\cal T}_0\ +\ {\cal T}_1\ n \right)
\ +\ 
{6\pi a^3\over M^3 L^7}\ (n+3)\ {\cal I}\ 
\right]
+ {\cal O}\left(L^{-8}\right)
\ .
\label{eq:energyshift}
\end{eqnarray}
where the geometric constants that enter are
\begin{align} 
&{\cal I}\ =\  -8.9136329
&{\cal J}\ =\  16.532316
\qquad\qquad {\cal K}\  = \  8.4019240
\nonumber\\
&{\cal L}\  = \  6.9458079
&{\cal T}_0\  = -4116.2338
\qquad\qquad {\cal T}_1\  = \ 450.6392
\nonumber\\
&{\cal S}_{\rm MS}\ = \ -185.12506
&
\label{eq:sums}
\end{align}
and ${\tiny \Choose{n}{k}}$=$n!/(n-k)!/k!$.  
The last term in the last bracket of Eq.~(\ref{eq:energyshift}) is the
leading relativistic contribution to the energy-shift. 
Deviations from the energy-shift of $n$-bosons computed
with non-relativistic quantum mechanics arise only for  three or more
particles as the two-particle energy-shift has the same form when computed in 
non-relativistic quantum mechanics and in quantum field
theory~\cite{Luscher:1986pf,Luscher:1990ux}. 
The renormalization-scale independent, but volume dependent, three-body quantity
\begin{align} 
\overline{\eta}_3^L & =  
\eta_3(\mu)\ +\ {64\pi a^4\over m}\left(3\sqrt{3}-4\pi\right)\ \log\left(\mu
  L\right)\ -\ 
{96 a^4\over\pi^2 m} {\cal S}
\label{eq:etathreebar}
\end{align}
was determined in recent Lattice QCD calculations~\cite{Beane:2007es,Detmold:2008fn},
where $\eta_3(\mu)$ is the coefficient of the three-body interaction in the Hamiltonian.
In Eq.~(\ref{eq:sums}), ${\cal S}_{\rm MS}$ is the value of
the scheme-dependent quantity
${\cal S}$ in the Minimal Subtraction (MS) scheme used to renormalize the theory.
The naive dimensional analysis (NDA) estimate of $\overline{\eta}_3^L$ is 
$\overline{\eta}_3^L\sim 1/(m_\pi f_\pi^4)$.
The leading terms in Eq.~(\ref{eq:Lm7}) evaluated at $n=2$ reproduce those
terms shown in   Eq.~(\ref{eq:e0}).

\subsubsection{Is it a Bound-State or a Scattering-State ?}
\label{sec:BSorSS}
\noindent
Obviously it is important to understand what infinite-volume physics
can be extracted from finite-volume calculations.  One important
question that arises is: if a negative energy shifted state is
calculated on the lattice at finite-volume, does it correspond to a
bound state or to a scattering state?  Clearly, calculations of the
same correlation function in multiple volumes will allow for the
exponential volume dependence of a bound state~\footnote{
Corrections to the bound-state
pole-condition in terms of $i\cot\delta$
that depend exponentially upon the volume~\cite{Sasaki:2006jn} 
are equivalent to 
the corrections
to the bound state mass~\cite{Beane:2003da} that have the same form.
}
to be distinguished
from the power-law volume dependence of a scattering state.  However,
one can make an educated guess about the nature of the state by the
magnitude of the energy shift.  Consider a simple system whose
scattering amplitude is dominated at low-energies by the scattering
length and effective range, as is the case for the scattering of two
nucleons.  The location of the states in the lattice volume are
determined by the solution of Eq.~(\ref{eq:energies}).
\begin{figure}[!ht]
\vskip 0.15in
\vspace*{4pt}
\centerline{\psfig{file=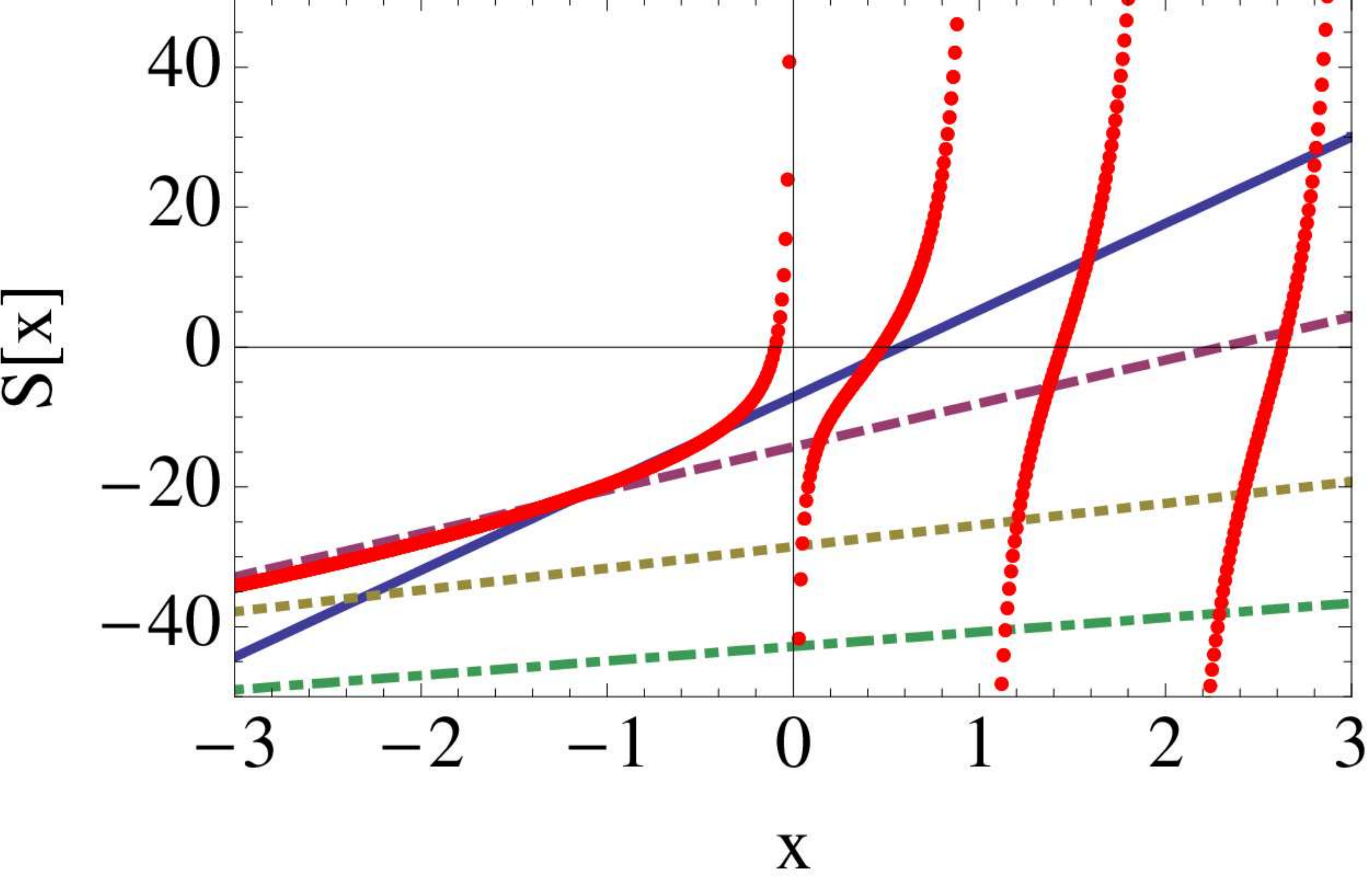,width=7.5cm}\qquad
\psfig{file=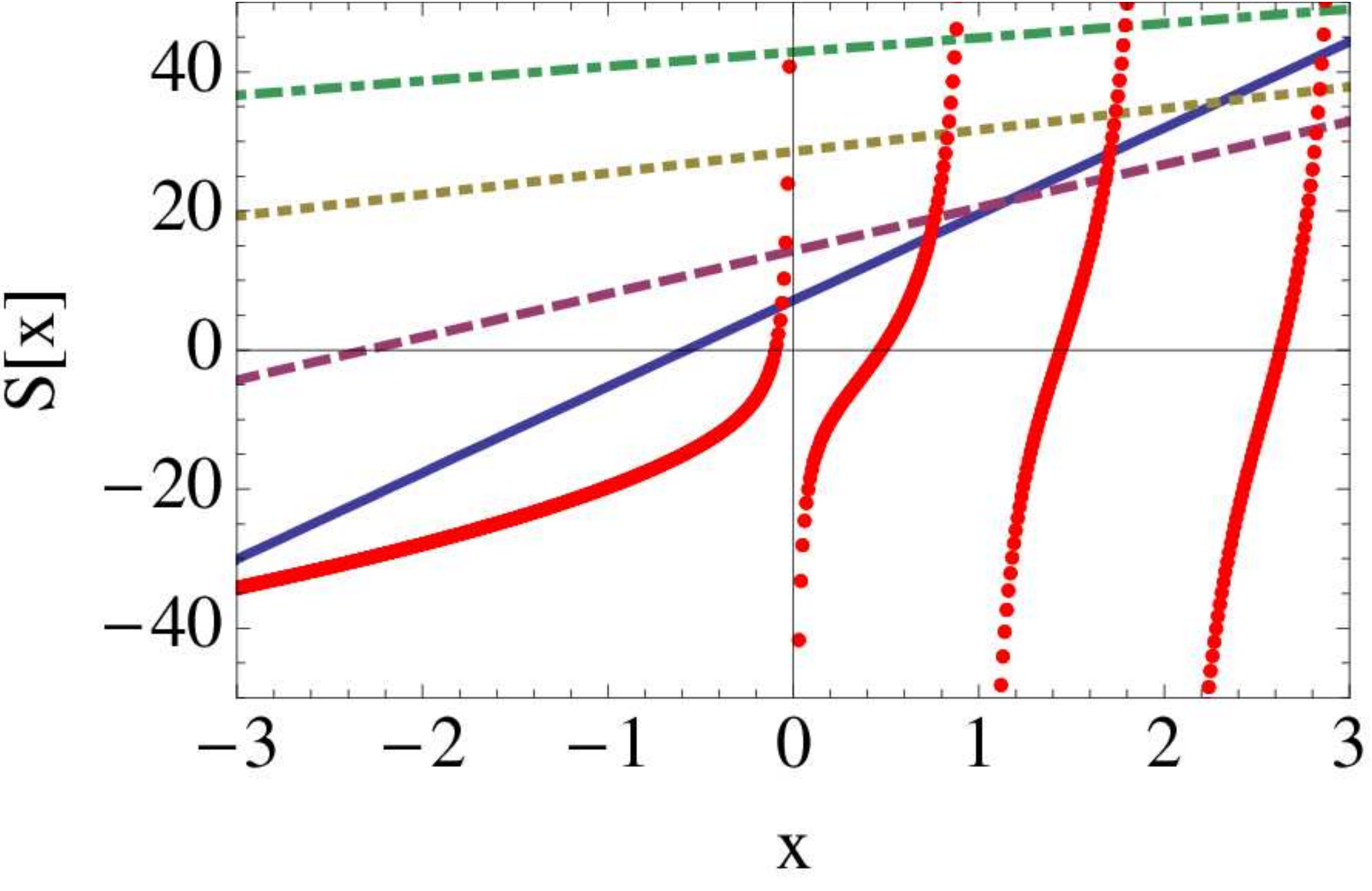,width=7.5cm}}
\vspace*{8pt}
\caption{The functions ${\bf S}(x)$ and $p\cot\delta$ vs.~$x$, where 
$p\cot\delta$ is evaluated at $p^2 = {4\pi^2\over L^2} x$.
The left panel is evaluated for scattering parameters $a=+1.1~{\rm fm}$ and
$r=0.5~{\rm fm}$ and all other scattering parameters vanish.  The volume
parameters are those of coarse MILC lattices with $L=20$ (solid, blue), $L=40$
(dashed, purple), $L=80$ (dashed, tan), and $L=120$ (dot-dashed, green)
and $b=0.125~{\rm
  fm}$.
The right panel is evaluated at the same parameters as the left except
$a=-1.1~{\rm fm}$.
The intercepts of the two curves corresponds to the location of the
energy-eigenstates in the finite volume.
}
\label{fig:Sfunpcots}
\end{figure}
In Fig.~\ref{fig:Sfunpcots} we show the graphical solution to
Eq.~(\ref{eq:energies}) for two systems, one with 
$a=+1.1~{\rm fm}$ and $r=0.5~{\rm fm}$ (left panel) and the other with
$a=-1.1~{\rm fm}$ and $r=0.5~{\rm fm}$ (right panel).
One finds that for energies giving a
value of $x=\left({p L\over2\pi}\right)^2$ less than $x_0=-0.0959006$ ($p\cot\delta < 0$)
the state will likely become a bound
state in the infinite-volume limit.
In contrast, states with energies giving a
value of $x\gsim x_0$ ($p\cot\delta > 0$) will 
likely become a continuum
state in the infinite-volume limit.  These statements are at best 
rules of thumb, and one can construct obvious exceptions.
Further, one can imagine scattering parameters in $p\cot\delta$ that modify
these rules.

Numerical explorations of the spectrum of compact scalar QED 
containing  a loose bound state have been performed in multiple
volumes~\cite{Sasaki:2006jn}.
The results of that work
confirm the large-volume behavior of the spectrum expected from
Eq.~(\ref{eq:energies})~\cite{Beane:2003da}
 and shown in Fig.~\ref{fig:Sfunpcots}.

\subsection{Finite-Lattice Spacing Artifacts and Mixed-Action Effective Field Theories}
\noindent
As space-time has been discretized in the process of performing the
Lattice QCD calculation of any given observable, there is a systematic
deviation between the calculation and the actual physical value of the
quantity.  One anticipates that in the limit that the lattice spacing
is much smaller than any physical length scale associated with the
strong interactions, the finite-lattice spacing effects can be
neglected for practical purposes.  For a fixed lattice volume, this
requires a lattice with a large number of lattice sites, and at
present time typical lattice spacing are $b\sim 0.125~{\rm fm}$ on the
coarse MILC lattices, $b\sim 0.09~{\rm fm}$ on the fine MILC lattices,
and $b\sim 0.06~{\rm fm}$ on the super-fine MILC lattices.  It is clear
that the lattice spacing of the coarse MILC lattices is not that much
smaller than the scale of chiral symmetry breaking, $\Lambda_\chi\sim
1~{\rm GeV}\sim 0.2~{\rm fm}$.  Naively, such lattice spacing would
clearly be problematic if the leading lattice-spacing artifacts
entered at order ${\cal O}(b)$ which is the case for naive Wilson
fermions, but most calculations today are performed with
improved-fermions, such as Clover, Domain-Wall, Staggered or Overlap
fermions which all have lattice spacing artifacts entering at ${\cal
O}(b^2)$~\footnote{There are chiral symmetry breaking corrections
entering at ${\cal O}(b)$, but they are exponentially suppressed, a
measure of which is the value of the residual mass.}.  The work we will review is performed
with a mixed-action.  With the exception of the potentials between
B-mesons, the staggered lattice discretization of the quark action is
used to describe the sea-quarks, while the Domain-Wall lattice
discretization is used for the valence quarks.  The two
discretizations have their advantages and disadvantages, and the
mixed-action scheme has proved to be an effective way to perform the
calculations that we are interested in.

Once the lattice calculation has been performed and the quantities of
interest, such as scattering lengths, have been extracted using the
finite-volume techniques described above, the finite-lattice spacing
effects must be removed.  Ideally, one would calculate the same
quantity at multiple lattice spacings, and then extrapolate to the
continuum using the appropriate finite-lattice spacing
expression~\cite{Bar:2002nr,Bar:2003mh,Bar:2005tu} computed in the
low-energy EFT, such as mixed-action chiral perturbation theory
(MA$\chi$-PT)~\footnote{MA$\chi$-PT is the low energy effective field
theory describing the dynamics of the pseudo-Goldstone bosons in the
situation where the sea-quarks are governed by one lattice action,
while the valence quarks are governed by a different one, in our case
domain-wall valence quarks and staggered sea quarks.  This is a
partially-quenched theory that allows for the lattice spacing to be
systematically included as a small expansion parameter, starting at
${\cal O}(b^2)$.  }.  At present, we do not have sufficient computer
resources to perform calculations at multiple lattice spacings.  We
have just completed our work on the available coarse MILC lattices and
we are currently calculating with the fine MILC lattices. However, one
of the very exciting results to emerge from MA$\chi$-PT is that when
certain mesonic observables are written as an expansion in
$m_\pi/f_\pi$ (for two-flavors), and more generally $M_M/f_M$ for
three-flavors, as measured in the lattice calculation~\footnote{ We
denote quantities that are computed directly from the correlation
functions, such as $m_\pi$, as lattice-physical quantities. These are
not extrapolated to the continuum, to infinite-volume or to the
physical point. }, lattice spacing artifacts can be eliminated at
low-orders in the expansion by a field redefinition, due to the good
chiral symmetry properties of the domain-wall (more generally
Ginsparg-Wilson) fermions used in the valence sector.  The
mixed-action corrections for the $\pi^+\pi^+$ scattering length have
been determined in Ref.~\cite{Chen:2005ab}.  It was demonstrated that
when the extrapolation formulae for this system are expressed in terms
of the lattice-physical parameters there are no
lattice-spacing-dependent counterterms at $\mathcal{O}(b^2)$,
$\mathcal{O}(b^4)$ or $\mathcal{O}(m_\pi^2 b^2) \sim
\mathcal{O}(b^4)$. This was explained to be a general feature of the
two-meson systems at this order, including the non-zero momentum
states~\cite{Chen:2006wf}.  There are additional lattice-spacing
corrections due to the hairpin interactions present in mixed-action
theories, but for domain-wall valence propagators calculated on the
asqtad improved MILC gauge configurations, these contributions are
completely calculable without additional counterterms at NLO.  They
depend only upon the valence meson and staggered taste-identity meson
mass splitting~\cite{Chen:2005ab,Chen:2006wf} which has been
computed~\cite{Aubin:2004fs}.  This allows for a precise determination
of the mixed-action corrections to the scattering lengths at the
various pion masses.  In two-flavor MA$\chi$-PT (i.e. including finite
lattice-spacing corrections) the chiral expansion of the scattering
length at NLO is~\cite{Chen:2006wf}
\begin{eqnarray}
m_\pi\ a_{\pi\pi}^{I=2} =  
-{m_\pi^2\over 8\pi f_\pi^2}
 \Biggl\{
1 + {m_\pi^2\over 16\pi^2 f_\pi^2}\ \Biggl[
3 \log\left({m_\pi^2\over\mu^2}\right) - 1 - l_{\pi\pi}^{I=2}(\mu) -
{\tilde\Delta_{ju}^4\over 6 m_\pi^4}\ \Biggr]
\Biggr\} \ ,
\label{eq:su2chiPT}
\end{eqnarray}
where it is understood that $m_\pi$ and $f_\pi$ are the lattice-physical parameters~\cite{Chen:2006wf}
and
\begin{eqnarray}
\tilde{\Delta}_{ju}^2 &\equiv \tilde{m}_{jj}^2 - m_{uu}^2
                = 2 B_0 (m_j- m_u) + b^2 \Delta_I +\dots\, ,
\label{eq:su2chiPTB}
\end{eqnarray}
contains the leading lattice-spacing artifacts.
$u$ denotes a valence quark and $j$ denotes a sea-quark, and 
isospin-symmetry is assumed in both the sea and valence sectors.
$\tilde{m}_{jj}$
($m_{uu}$) is the mass of a meson composed of two sea (valence) quarks
of mass $m_j$ ($m_u$) and the dots denote higher-order corrections to
the meson masses.  Clearly Eq.~(\ref{eq:su2chiPT}), which contains all
$\mathcal{O}(m_\pi^2 b^2)$ and $\mathcal{O}(b^4)$ lattice artifacts,
reduces to the continuum expression for the scattering
length~\cite{Gasser:1983yg} in the QCD limit where
$\tilde{\Delta}_{ju}^2\rightarrow 0$~\footnote{The counterterm
$l_{\pi\pi}^{I=2}(\mu)$ is the same counterterm that
appears in continuum $\chi$-PT.}.  
The three-flavor MA$\chi$-PT expression for $K^+K^+$ and $K^+\pi^+$ scattering
are somewhat more complicated, for obvious reasons and we refer the reader to Ref.~\cite{Chen:2007ug}.

The fact that baryons do not have the same special status with respect to
chiral symmetry as the pseudo-Goldstone bosons means that the field
redefinitions that parametrically suppress finite-lattice spacing
contributions to $\pi^+\pi^+$ scattering, do not exist for baryons.
Therefore one expects to see finite lattice spacing contributions to the
scattering lengths for nucleon-nucleon, and hyperon-nucleon scattering at
${\cal O}(b^2)$ (and residual mass-type  contributions).  
In order to achieve high precision calculations of baryon-baryon scattering,
calculations on lattices with a small lattice spacing will be required.

\subsection{Potentials From Lattice Calculations}
\noindent
One remarkable feature of nuclear physics is that one can understand
and compute to reasonable accuracy the properties and interactions of
nuclei working with an energy-independent two-nucleon potential alone.
Phenomenologically, one finds that the three-nucleon interaction is
required to improve agreement with experiment, as is the inclusion of
meson exchange currents into electroweak matrix elements.  But the
fact remains that the two-nucleon potential is the dominant
interaction in nuclei.

There is a burning desire to construct a nucleon-nucleon
potential~\footnote{In this context, the word {\it potential} means an
energy-independent potential.  }  directly from QCD, and hence from
Lattice QCD. One can extract an energy-dependent and
source/sink-dependent potential, defined at one energy (the energy of
the two nucleons determined in the finite volume), however, this
contains no more information than the phase-shift at the energy
determined with L\"uscher's method.  One such potential was calculated
in quenched QCD in Ref~\cite{Ishii:2006ec} but, for the reasons mentioned above, it 
cannot be used as an input in nuclear calculations and it cannot be meaningfully
compared to traditional nucleon-nucleon potentials.

A nucleon-nucleon potential may be defined from Lattice
QCD calculations in the same way that potentials are determined from
experimental measurements of the elastic scattering cross-section.  A
large number of calculations are performed, producing values for
the phase-shift, along with an uncertainty, over a wide range of
low-energies, and a potential is defined that minimizes the
$\chi^2/dof$ in a global fit to the lattice calculations.  At present,
there is no practical program underway to perform such an analysis due
to limited computational power.

\subsection{Statistical Errors Associated with Calculations of Observables in
  the Nucleon and Systems of Nucleons}
\label{sec:staterrors}
\noindent
Lattice QCD calculations of quantities are performed by a Monte-Carlo
evaluation of the path integral. As such, an important aspect of any
such calculation is its statistical error and it is useful to consider
a discussion presented by Lepage~\cite{Lepage:1989hd}.  Consider an
observable that is extracted from the correlation function $\langle
\theta(t)\rangle$, such as a correlation function resulting from a
pion source at time $t=0$ and a pion sink at time $t$.  In any
calculation there are a finite number of configurations on which to
perform measurements and a finite number of lattice sites on each
lattice.  For $N$ statistically independent measurements of
$\theta(t)$, the noise to signal ratio of this correlator behaves as
$1/\sqrt{N}$ in the limit of large $N$.  However, the time-dependence
of this ratio depends upon the specific correlation function that is
being calculated.

Consider the case where $\langle\theta(t)\rangle$ is the correlation function associated with $n$
pions, each with a source of the form 
$\pi^+({\bf x},t)=\overline{u}({\bf  x},t)\gamma_5 d({\bf x},t)$, or one that
is smeared over neighboring lattice sites,
\begin{eqnarray}
\langle \theta(t) \rangle & = & \langle 
\left(\sum_x \pi^-({\bf x},t)
\right)^n
\left( 
\phantom{\sum_x\hskip -0.2in}
\pi^+({\bf 0},0)
\right)^n
\rangle\ \rightarrow\ A_0 \ e^{-n m_\pi t}
 \ \ ,
\label{eq:Gfun}
\end{eqnarray}
where the interactions between the pions have been neglected in the large time behavior.
The variance of this correlator is estimated to be 
\begin{eqnarray}
N \sigma^2 & \sim &\langle \theta(t)^\dagger  \theta(t)\rangle  - \langle \theta(t) \rangle^2
\nonumber\\
& = & 
\langle 
\left(\sum_x \pi^-({\bf x},t) \right)^n
\left(\sum_y \pi^+({\bf y},t)  \right)^n
\left( 
\phantom{\sum_x\hskip -0.2in}
\pi^+({\bf 0},0) \right)^n
\left( 
\phantom{\sum_x\hskip -0.2in}
\pi^-({\bf 0},0)  \right)^n
\rangle\ \ -\ \langle \theta(t) \rangle^2
\nonumber\\
&& \rightarrow\ \left(A_2 - A_0^2 \right) \ e^{-2 n m_\pi t}
 \ \ ,
\label{eq:GGdaggerfun}
\end{eqnarray}
where the large time behavior of the variance is dictated by the lightest
intermediate state that can be formed from the propagators emerging from the sources
associated with $\theta(t)^\dagger  \theta(t)$ and coupling to the sinks (annihilation
diagrams are not included).
It then follows that the ratio of noise to signal behaves as 
\begin{eqnarray}
{\sigma\over\overline{x}} & = & {\sigma (t)\over \langle \theta(t) \rangle }
\sim \ {\sqrt{\left(A_2 - A_0^2 \right)} \ e^{-n m_\pi t}\over \sqrt{N} A_0\
e^{-n m_\pi t} }
\sim {1\over \sqrt{N}}
 \ \ ,
\label{eq:NtoSpi}
\end{eqnarray}
which is independent of time. (Here ${\sigma/\overline{x}}$ denotes the ratio of standard deviation
to mean.) That is to say that for the single pion
correlator and correlators involving arbitrary numbers of pions the errors are time-independent, and importantly do not
exponentially grow with time.

The situation is, unfortunately, not so pleasant for systems involving any number
of baryons.
For the case of a single proton, the correlation function has the form
\begin{eqnarray}
\langle \theta^{ii}(t) \rangle & = & 
\sum_{\bf x}\ \langle p^i({\bf x},t) \overline{p}^{i} ({\bf 0},0)\rangle
\ \rightarrow\ A^{ii}_{p0} \ e^{-m_p t}
 \ \ ,
\label{eq:Gfunproton}
\end{eqnarray}
where an interpolating field that has non-vanishing overlap with the proton is
$p^i \sim u^{a,T}C\gamma_5 d^b u^{i,c} \epsilon_{abc}$,
where $a,b,c$ are color indices and $i$ is a spin-index.
The variance of this correlation function is
\begin{eqnarray}
N \sigma^2 & \sim & \langle \theta^{ii\dagger}(t)  \theta(t)^{ii}\rangle  - \langle \theta^{ii}(t) \rangle^2
\ = \ \sum_x 
\langle p^i({\bf x},t) \overline{p}^{i}({\bf x},t) p^i({\bf 0},0)
\overline{p}^{i}({\bf 0},0) \rangle\ \ -\ \langle \theta^{ii}(t) \rangle^2
\nonumber\\
&& \rightarrow\ A_{p2}\  e^{-3 m_\pi t}- A_{p0}^2\  e^{-2m_p t}
\ \ \rightarrow \ A_{p2}\  e^{-3 m_\pi t}
 \ \ ,
\label{eq:GGdaggerfunproton}
\end{eqnarray}
and therefore the noise to signal ratio behaves as
\begin{eqnarray}
{\sigma\over\overline{x}} & = & {\sigma (t)\over \langle \theta(t) \rangle }
\sim   {1\over \sqrt{N}} \ e^{\left( m_p - {3\over 2} m_\pi\right) t} 
 \ \ .
\label{eq:NtoSproton}
\end{eqnarray}
More generally, for a system of $A$ nucleons, the noise to signal ratio behaves
as 
\begin{eqnarray}
{\sigma\over\overline{x}} & & 
\sim   {1\over \sqrt{N}} \ e^{A \left( m_p - {3\over 2} m_\pi\right) t}
 \ \ .
\label{eq:NtoSnucleus}
\end{eqnarray}
Therefore, in addition to the signal itself falling as $G\sim e^{-A m_p t}$,
the noise associated with the correlator grows exponentially as in Eq.~(\ref{eq:NtoSnucleus}).


\section{Our Current Techniques and Resources}

\begin{figure}[!ht]
\vspace*{4pt}
\centerline{\psfig{file=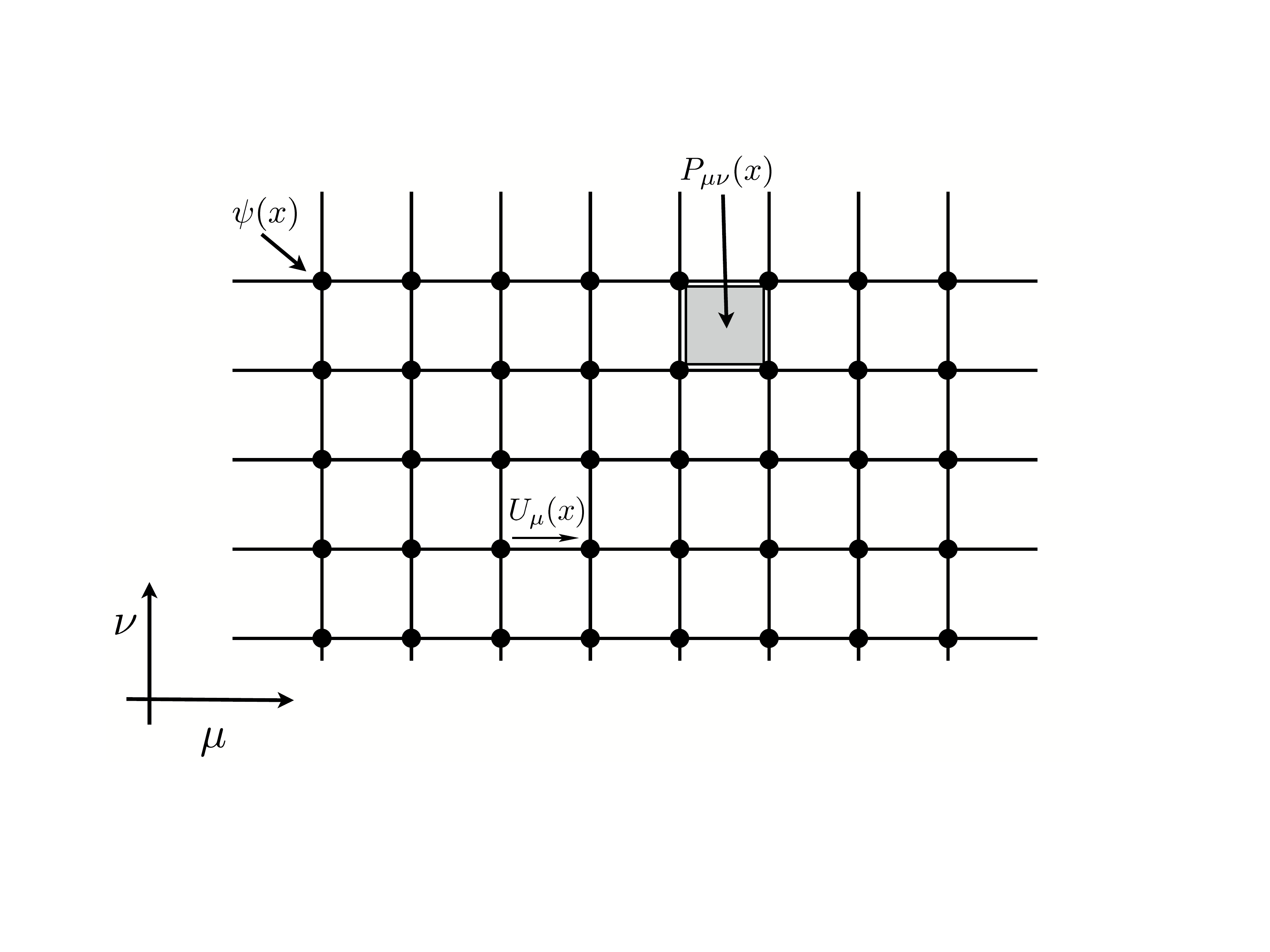,width=17.0cm}}
\vspace*{8pt}
\caption{A two dimensional slice of the four dimensional  lattice.
}
\label{fig:lattice}
\end{figure}
\noindent 
The Lattice formulation of QCD is the perfect tool for evaluating the correlation functions
required to 
extract physical observables, such as hadron masses and phase shifts. 
It both provides an ultraviolet  regulator of the continuum field theory and 
converts functional integrals into regular integrals of very high dimension.
In the continuum, the QCD path integral is
 \begin{equation}
 {\cal Z} = \int {\cal D} A_\mu {\cal D} \bar \psi {\cal D}  \psi \; e^{\int d^4 x \left(-\frac{1}{4} F^a_{\mu\nu} F^{a\mu\nu} - 
 \bar{\psi} \left[ D_\mu \gamma_\mu + m\right] \psi\ +\ {\cal L}_{G.F.}\right)}
\end{equation}
where $A_\mu$ is the gauge field representing the gluons,
$F^a_{\mu\nu}$ is the gauge field strength and $\bar \psi$, $\psi$ are
the fermion fields representing the quarks.  $D_\mu$ is the covariant
derivative which ensures gauge invariance and $\gamma_\mu$ are
matrices satisfying the Clifford algebra.  The
physical quantities in this theory can be calculated from correlation
functions of operators $\cal O$ that are functions of the quantum
fields (quarks and gluons).
\begin{equation}
\langle {\cal O}\rangle = \frac{1}{\cal Z} \int {\cal D} A_\mu {\cal D} \bar
\psi {\cal D}  \psi \; {\cal O}\; e^{\int d^4 x \left(-\frac{1}{4} F^a_{\mu\nu}
    F^{a \mu\nu} - 
 \bar{\psi} \left[  D_\mu \gamma_\mu + m\right] \psi\ +\ {\cal L}_{G.F.}\right)}
 \end{equation}
We can now discretize the continuum path integral introducing a discrete space time.
In order to preserve gauge invariance the gauge fields are discretized
as special unitary matrices, $SU(3)$, living on the links of the
lattice (see Figure~\ref{fig:lattice}). The discrete gauge action is
given by the sum over all plaquettes $P_{\mu\nu}(x)$ which are the
product of the links $U$ going around the elementary plaquettes of the
lattice.
\begin{equation}
S_g(U) ={\beta} \sum_{x\mu\nu}\left( 1 - \frac{1}{3}
{\rm Re Tr}{P_{\mu\nu}(x)} \right)  
\end{equation}
with
 \begin{equation}
 P_{\mu\nu} = U_\mu(x)U_\nu(x+\hat\mu)U^\dagger_\mu(x+\hat\nu)U^\dagger_\nu(x)\,,
 \label{eq:probability}
 \end{equation}
 and $\beta$ is the lattice gauge coupling.  Taking the lattice
 spacing to zero, the above action becomes the continuum gauge action
 $\ -\int d^4 x \frac{1}{4}\left( F^a_{\mu\nu}(x)\right)^2 $. This is the
 well known Wilson gauge~\cite{Wilson:1974sk} action.  This
 discretization is not unique but it is the simplest. One can modify
 this discrete action by adding larger loops with coefficients
 appropriately chosen in order to achieve better convergence to the
 continuum limit, which is the ultimate goal of the calculation.
 
The fermions, which live on the vertices of the lattice, present a
more challenging problem. Naive discretization results in the so called
fermion doubling problem, i.e. lattice fermions come naturally in
sixteen copies, too many for describing real QCD which has three
light quarks (up, down and strange) and three heavy quarks (top,
bottom and charm). The doublers can be avoided by several ingenious
formulations of lattice fermions. Wilson fermions, which were
introduced first~\cite{Wilson:1974sk}, eliminate the doublers by
adding irrelevant dimension five operators in the action that lift the
masses of the doublers, leaving only the light fermion in the
spectrum. The price to pay is breaking of chiral symmetry and the
introduction of lattice artifacts that scale as $O(b)$. Kogut-Susskind
fermions~\cite{Kogut:1974ag} provide another way to remove some of the
doublers and re-interpret the remaining four as four degenerate
flavors.  In this approach a $U(1)$ chiral symmetry still remains
unbroken and lattice artifacts scale as $O(b^2)$.  Kogut-Susskind
fermions become problematic when the required number of flavors is not
a multiple of four (as is the case for QCD).  In addition, the broken
flavor and chiral symmetries introduce large lattice artifacts,
although they scale as $O(b^2)$. Finally, the so called domain wall
fermions~\cite{Kaplan:1992bt,Shamir:1993zy,Furman:1994ky} and overlap
fermions~\cite{Narayanan:1994gw,Neuberger:1997fp} are fermionic
actions that both preserve chiral symmetry at finite lattice spacing
and are doubler free. Unfortunately, such formulations are
computationally significantly more expensive.
In all cases the lattice fermion action is of the form
\begin{equation}
S_f = \bar \psi D(U) \psi
\end{equation}
where $\psi$ is the fermion "vector" and $D(U)$ is a sparse
matrix~\footnote{In certain cases such as the overlap fermions the
matrix is not sparse but has sparse like properties i.e. the matrix
vector multiplication is a "cheap" operation.}  acting on this vector,
that depends on the gauge field $U$.

The partition function in the case of two quark flavors is
\begin{eqnarray}
{\cal Z} &=& \int \prod_{\mu,x}dU_\mu(x)\prod_x 
d\bar{\psi}d\psi
\;\; e^{-S_g(U)-S_f(\bar{\psi},\psi,U)} \nonumber\\
&=&  \int \prod_{\mu,x}dU_\mu(x)\;\;
 {\rm det}\left(D(U)^\dagger D(U)\right)  \;\; e^{-S_g(U)}\,.
 \end{eqnarray}
The integration over the quark fields, which are represented by
Grassman numbers, can be done exactly.  In addition, the quark matrix
$D(U)$ represents one flavor, however since ${\rm det} D(U)^\dagger =
{\rm det} D(U)$, the determinant ${\rm det}\left(D(U)^\dagger
D(U)\right) $ represents two flavors. In the case of correlation functions, integrating out
the quarks gives the following expression
\begin{equation}
\langle{\cal O}\rangle = \frac{1}{\cal Z}
 \int \prod_{\mu,x}dU_\mu(x)\;\; {\cal O}(\frac{1}{D(U)},U)\;\;
 {\rm det}\left(D(U)^\dagger D(U)\right)  \;\; e^{-S_g(U)}\, ,
 \label{eq:CorFunc}
 \end{equation}
resulting in operators ${\cal O}$ that depend on the inverse of the
quark matrix.  The above manipulation is only valid in the case of two
flavors of quarks (the up and the down) which both have the same mass,
which is a good approximation to the low energy physics of QCD. A
strange quark can be easily added by including ${\rm
det}\left(D(U)^\dagger D(U)\right)^{1/2}$ in the partition function.

The computation of Eq.~(\ref{eq:CorFunc}) is the main numerical task
faced in Lattice QCD calculations. The integral in
Eq.~(\ref{eq:CorFunc}) over the gauge fields is of extremely large
dimensionality. Considering that we have discretized QCD, which has a
fundamental scale of $\sim1~{\rm fm}$ ($10^{-13}~{\rm cm}$), we need
to work with a lattice that has a physical size much larger than $1~{\rm fm}$
in order to control finite volume effects, and a
lattice spacing much smaller than $1~{\rm fm}$ in order to control the
continuum limit.  With moderate choices for the volume and the lattice
spacing, a typical lattice size of $32^4$ is arrived at.  Counting the
color, flavor and spin degrees of freedom, the calculation involves
$\approx 10^8$ degrees of freedom. The only way this computation can
be done is by using Monte Carlo integration. Fortunately, the
combination of the quark determinant and the gauge action,
\begin{equation}
 {\cal P}(U) =\frac{1}{\cal Z} {\rm det}\left(D(U)^\dagger D(U)\right)  \;\; e^{-S_g(U)}\,,
 \end{equation}
is a positive definite quantity which can be interpreted as a
probability and hence importance sampling can be employed.  The basic
algorithm is to produce $N$ gauge field configurations $\{U\}$ with
probability distribution ${\cal P}(U)$ and then evaluate
\begin{equation}
\langle {\cal O} \rangle = \lim_{N\rightarrow \infty} \frac{1}{N}\sum_{i=1}^N {\cal O}(U_i,\frac{1}{D(U_i)}) \ .
\label{eq:average}
\end{equation}
At finite $N$, the estimate of ${\cal O}$ is approximate, with an error
that converges to zero as ${\cal O}(1/\sqrt{N})$.  Both for the gauge
field configuration generation and the evaluation of
Eq.~(\ref{eq:average}), the linear system of equations
\begin{equation}
 {D^\dagger(U)[m]D(U)[m]}\chi =  \phi\,,
 \label{eq:linearsyst}
\end{equation}
needs to be solved where the dependence of the quark matrix on the
quark mass $m$ is made explicit.  Since the quark matrix is
sparse, iterative solvers such as conjugate gradient can be used.  The
condition number of the quark matrix is inversely proportional to
the quark mass.  Since the physical quark masses for the up and down
quarks are quite small, the quark matrix has a large condition
number. With current computer resources this linear system cannot be
solved exactly at the physical quark mass. For that reason the
calculation is performed at heavier quark masses and then extrapolated
to the physical point.  The vast majority of the computer time used in
these calculations is devoted to the solution of this linear system
both in the context of gauge field generation and in the later stage
of the calculation of physical observables through
Eq.~(\ref{eq:average}).

Realistic lattice calculations require quark masses that result in
pion masses below 400 MeV, allowing chiral effective field theories to
be used with some reliability. In addition, a dynamical strange quark
is required in order to guarantee that the low energy constants of the
EFT match those of the physical theory.  Although this task seems
formidable, in the last several years there have been developments that
make phenomenologically interesting calculations now possible.

\subsection{Domain Wall and Staggered Fermions}
\noindent
The emergence of fermions that respect chiral
symmetry~\cite{Kaplan:1992bt,Shamir:1993zy,Furman:1994ky,Narayanan:1994gw,Neuberger:1997fp}
on the lattice was one of the major recent developments in Lattice
QCD. These formulations of lattice fermions allow us to reduce the
lattice spacing errors and approach the continuum limit in a smoother
manner. However, the cost of calculating with these fermions is an
order of magnitude larger than any other variant of lattice fermions.
In addition, the development of improved Kogut-Susskind fermion
actions~\cite{Orginos:1999cr,Orginos:1998ue} that significantly reduce
the $O(b^2)$ errors, allowed for cheap inclusion of quark loop effects
in the QCD correlation functions computed on the lattice.  With this
formulation, volumes with spatial extent as large as $L\sim 3.5~{\rm fm}$ are possible with
light-quark masses as low as 1/10th of the strange quark mass
depending on available computing resources. However the fact that
Kogut-Susskind fermions represent four flavors of quarks complicates
calculations when two or one flavors are needed. From the operational
point of view the problem is solved by introducing into the path
integral the Kogut-Susskind determinant raised to the $n_f/4$ power
(rooted), where $n_f$ is the desired number of flavors. The
non-integer power of the quark determinant introduces non-localities
in the lattice action. It has been argued, however, that the long
distance physics that survives the continuum limit is not affected by
such
non-localities~\cite{Shamir:2006nj,Bernard:2006ee,Bernard:2007ma,Bernard:2007eh,Durr:2006ze,Durr:2004ta}. In
addition, at finite lattice spacing, the pathologies arising in the
Kogut-Susskind fermion formulation can be dealt with in staggered
$\chi$-PT~\cite{Bernard:2006ee,Bernard:2007ma,Aubin:2003mg,Aubin:2003uc,Bernard:2006zw,Lee:1999zxa,Bernard:2006vv}.  Although
no rigorous proof exists, empirical evidence indicates that
Kogut-Susskind fermions do describe the correct physics as long as the
continuum limit is taken before the chiral limit~\cite{Durr:2004ta}.
It should be noted that there are some members of the lattice
community who believe that the rooted-staggered action is
fundamentally flawed and its continuum limit does not correspond to
QCD (for a summary of these arguments, see Ref.~\cite{Creutz:2007rk}).
We disagree with these arguments, 
however we acknowledge that there is no proof that the
continuum limit of the rooted-Kogut-Susskind action corresponds to
QCD. All of the work
that we present in this review based upon mixed-action calculations on
the MILC lattice ensembles assumes that the continuum limit is, in
fact, QCD.

In our calculations we use Kogut-Susskind fermions to represent the QCD
vacuum polarization effects associated with the two light flavors
(up/down quarks) and the somewhat heavier strange quark. This is done
by using gauge configurations generated with the appropriate
Kogut-Susskind fermion determinants incorporated into the probability
distribution that enters the path integral. Since this part of the
computation is completely disconnected from the calculation of
correlation functions, we can use gauge fields generated by other
collaborations. In our case we use the gauge configurations generated
by the MILC collaboration~\cite{Bernard:2001av}.

For all external quarks we use domain wall fermions. Because of the
chiral symmetry that domain wall fermions satisfy, all our correlation
functions satisfy chiral Ward identities, ensuring that the leading
order chiral behavior is continuum-like.  The small corrections
appearing due to Kogut-Susskind fermions in the vacuum loops can be
taken care of systematically in $\chi$-PT~\cite{Chen:2007ug,Chen:2005ab,Chen:2006wf}. 
Compared to calculations with
Kogut-Susskind fermions in the valence sector, this formulation results
in better control of the chiral behavior and possibly smaller
discretization errors. This approach was first introduced by the LHP
collaboration for the
study of nucleon structure~\cite{Edwards:2006zza,Renner:2007pb,Hagler:2007xi,Edwards:2006qx,Edwards:2005ym}.

\subsection{The Lattice Actions}
\noindent
The gauge configurations  used in our work  were generated by the MILC collaboration using
the one loop tadpole improved gauge action~\cite{Alford:1995hw} where both
${\cal O}(b^2)$ and ${\cal O}(g^2b^2)$ errors are removed. This action is defined as
\begin{equation}
S_G[U] = - \frac{\beta}{3} \left(  c_0 \sum_{x;\mu<\nu} P[U]_{x,\mu\nu}
        + c_1 \sum_{x;\mu\neq\nu} R[U]_{x,\mu\nu} 
        + c_2 \sum_{x;\mu<\nu<\sigma}  C[U]_{x,\mu\nu\sigma}
  \right)
\end{equation}
where $R[U]_{x,\mu\nu}$ and $C[U]_{x,\mu\nu\sigma}$ denote the real
part of the trace of the ordered product of SU(3) gauge links along 
$1\times 2$ rectangles in the $\mu,\nu$ plane and the
$\mu,\nu,\sigma,-\mu,-\nu,-\sigma$ paths, respectively.  The
coefficients $c_0$, $c_1$, and $c_2$ are determined  in one loop  tadpole improved
perturbation theory~\cite{Alford:1995hw}, and $\beta = 10/g^2_0$ where
$g_0$ is the bare gauge coupling. For the fermions in the vacuum the
Asqtad improved Kogut-Susskind action~\cite{Orginos:1999cr,Orginos:1998ue,Toussaint:1998sa,Lagae:1998pe,Lepage:1998vj,Orginos:1999kg} is used. This action
is the Naik action~\cite{Naik:1986bn} ($O(b^2)$ improved  Kogut-Susskind action),
with smeared links for the one link terms so that couplings to
gluons with any of their momentum components equal to $\pi/b$ are set to zero. The form
of the action is the following:
\begin{eqnarray}
S_{Asqtad} &=&  \frac{1}{2} \left[\sum_{x,y}\sum_{\mu=1,4} \bar\chi(x)\eta_\mu(x) \left( V_\mu(x) \delta_{y,x+\hat\mu} - V^\dagger_\mu(x-\hat\mu) \delta_{y,x+\hat\mu} \right)\chi(y) -\right.\nonumber\\
&-& \left.\sum_{x,y}\sum_{\mu=1,4} \bar\chi(x) \frac{1}{24 u_0^2} \eta_\mu(x) \left( W_\mu(x) \delta_{y,x+\hat\mu} - W^\dagger_\mu(x-\hat\mu) \delta_{y,x+\hat\mu} \right)\chi(y)\right] +\nonumber\\
&+& \sum_x m \bar\chi(x)\chi(x)
\end{eqnarray}
where the $V_\mu$ and $W_\mu$ links are 
\begin{eqnarray}
V_\mu(x) &=& \frac{5}{8} U_\mu(x)  + 
\sum_{\nu\ne\mu} \frac{1}{16 u_0^2}S_{\mu\nu}(x) + \nonumber\\ 
&+& \sum_{\nu\ne\mu}\sum_{\rho\ne\mu,\rho\ne\nu} \frac{1}{64 u_0^4}S_{\mu\nu\rho}(x) + \nonumber\\ 
 &+&
 \sum_{\nu\ne\mu}\sum_{\rho\ne\mu,\rho\ne\nu}
 \sum_{\sigma\ne\mu,\sigma\ne\nu,\sigma\ne\rho} \frac{1}{384 u_0^6}S_{\mu\nu\rho\sigma}(x) \nonumber\\ 
 &-& \sum_{\nu\ne\mu} \frac{1}{16 u_0^4} L_{\mu\nu}(x) \\
 W_\mu(x) &=&  U_\mu(x)U_\mu(x+\hat\mu)U_\mu(x+2\hat\mu) \\ 
\end{eqnarray}
and
\begin{eqnarray}
 S_{\mu\nu}(x)& =& U_\nu(x)U_\mu(x+\hat\nu)U^\dagger_\nu(x+\hat\mu) + U^\dagger_\nu(x -\hat\nu)U_\mu(x-\hat\nu)U_\nu(x-\hat\nu+\hat\mu) \\
 L_{\mu\nu}(x) &=& U_\nu(x)U_\nu(x+\hat\nu)U_\mu(x+2\hat\nu)U^\dagger_\nu(x+\hat\nu +\hat\mu)U^\dagger_\nu(x+\hat\mu) +  \nonumber \\
 &+&  U^\dagger_\nu(x -\hat\nu) U^\dagger_\nu(x -2\hat\nu)U_\mu(x-2\hat\nu)U_\nu(x-2\hat\nu+\hat\mu) U_\nu(x-\hat\nu+\hat\mu) \\
  S_{\mu\nu\rho}(x)& =& U_\nu(x)S_{\mu\rho}(x+\hat\nu)U^\dagger_\nu(x+\hat\mu) + 
  \nonumber\\
  &+& U^\dagger_\nu(x -\hat\nu)S_{\mu\rho}(x-\hat\nu)U_\nu(x-\hat\nu+\hat\mu)  \\
   S_{\mu\nu\rho\sigma}(x)& =& U_\nu(x)S_{\mu\rho\sigma}(x+\hat\nu)U^\dagger_\nu(x+\hat\mu) + \nonumber \\
   &+& U^\dagger_\nu(x -\hat\nu)S_{\mu\rho\sigma}(x-\hat\nu)U_\nu(x-\hat\nu+\hat\mu)\;. 
 \end{eqnarray}
In all of the above $u_0$ is the tadpole coefficient which was determined self consistently
by MILC, $\eta_\mu(x) = (-1)^{\sum_{\rho<\mu} x_\rho}$ and $x$ the four integers describing the lattice coordinates.
\newcommand{\Pp}{\frac{1+\gamma_5}{2}}
\newcommand{\Pm}{\frac{1-\gamma_5}{2}}

For the valence sector we use the five dimensional Shamir domain wall fermion action~\cite{Shamir:1993zy,Furman:1994ky}
\begin{eqnarray}
  S_{DW} &=&  -\sum_{x,x'}\sum_{s=0}^{L_s-1}
\left[
\bar{\Psi}(x,s) \left[ D_w(x,x') + 1\right]\Psi(x',s)  \right]- \nonumber \\
& - & \left[
\bar{\Psi}(x,s)\Pm\Psi(x',s+1) +
\bar{\Psi}(x,s)\Pp\Psi(x',s-1)\right]
 + \nonumber \\
&+& m \left[
  \bar{\Psi}(x,0    ) \Pp\Psi(x',L_s-1) +
  \bar{\Psi}(x,L_s-1)\Pm\Psi(x',0    )
\right]
\ ,
\label{eq:DWFaction}
\end{eqnarray}
with $D_w(x,x')$ the regular four dimensional Wilson fermion action,
\begin{equation}
  D_w(x,x') = (4+M_5)\delta_{x,x'} - \sum_\mu \left[
      \frac{1-\gamma_\mu}{2} U_\mu(x) \delta_{x+\hat\mu,x'} +
      \frac{1+\gamma_\mu}{2} U^\dagger_\mu(x') \delta_{x,x'+\hat\mu}\right] \ ,
\label{eq:Dw}
\end{equation}
and $L_s$ is the extent of the 5th dimension.  Hypercubic-smeared
(HYP-smeared)~\cite{Hasenfratz:2001hp,DeGrand:2002vu,DeGrand:2003in,Durr:2004as}
gauge links were used in Eq.~(\ref{eq:DWFaction}) and (\ref{eq:Dw}) to
improve chiral symmetry.  The physical four dimensional quark fields
appear as boundary modes at the surface of the five dimensional space when
$M_5$ lies in the interval $(-2,0)$.  The physical quark
fields ($\bar q(x)$ and $q(x)$ are related to the underlying 5D
fermions by
\begin{eqnarray}
  q(x) &=& \Pm\Psi(x,0)  +  \Pp\Psi(x,L_s-1) \nonumber \\
  \bar{q}(x) &=&\bar{\Psi}(x,L_s-1)\Pm + \bar{\Psi}(x,0)\Pp \;.
\end{eqnarray}
The parameter $m$ in Eq.~(\ref{eq:DWFaction}) is related to the physical 
quark mass as it introduces in the effective action a $m \bar q q$ term.
Domain wall fermions in the infinite $L_s$ limit poses an exact chiral symmetry when
$m$ vanishes. This symmetry transformation is
\begin{eqnarray}
  \Psi(x,s)  &\rightarrow& e^{i\Gamma_5(s) \theta(x)} \Psi(x,s) \\
  \bar\Psi(x,s)  &\rightarrow& \bar \Psi(x,s) e^{-i\Gamma_5(s) \theta(x)} 
\end{eqnarray}
where $\Gamma_5(s) = sign( \frac{L_s - 1}{2} - s ) $.

However, at finite $L_s$ this chiral symmetry is explicitly broken by
the coupling of left handed and right handed modes in the middle of
the 5th dimension. As a result one can construct the following
partially conserved axial vector current
\begin{equation}
  {\cal A}_\mu(x) = -\sum_{s=0}^{L_s-1} \Gamma_5(s) j_\mu(x,s) 
\end{equation}
where $j_\mu$ is the four dimensional conserved vector current that corresponds to the 4D Wilson fermion action. 
This current satisfies a Ward-Takahashi identity which in the flavor non-singlet case
takes the form~\cite{Furman:1994ky}:
\begin{eqnarray}
  \Delta_\mu \langle {\cal A}^a_\mu(x) O(y) \rangle &= &
        2m \langle \bar q(x)\tau^a \gamma_5 q(x) O(y) \rangle +\nonumber \\
        &+& 2 \langle  \bar q_{mp}(x)\tau^a \gamma_5 q_{mp}(x) O(y)
        \rangle + i \langle \delta^a O(y) \rangle \;\;
\label{eq:ward_tak_id}
\end{eqnarray}
where 
\begin{eqnarray}
  q_{mp}(x) &=& \Pm\Psi(x,\frac{L_s}{2})  +  
                \Pp\Psi(x,\frac{L_s}{2}-1)
 \nonumber \\
 \bar{q}_{mp}(x) &=& \bar{\Psi}(x,\frac{L_s}{2}-1)\Pm+
                    \bar{\Psi}(x,\frac{L_s}{2})\Pp 
\end{eqnarray}
are four dimensional fields constructed at the midpoint of the the 5th
dimension.  The Ward-Takahashi identity of Eq.~(\ref{eq:ward_tak_id})
is the same as the continuum counterpart with just an additional term
$2 \langle \bar q_{mp}(x)\tau^a \gamma_5 q_{mp}(x) O(y) \rangle$. This
term is there only at finite $L_s$~\footnote{For the flavor singlet
current this term survives the infinite $L_s$ limit and gives rise to
the anomaly.} and it is a measure of the explicit chiral symmetry
breaking. At long distances this term is proportional to $2 \langle
\bar q(x)\tau^a \gamma_5 q(x) O(y) \rangle$.  Using the pseudo-scalar
density as a probe operator $O(y)$ the residual mass is defined as
\begin{equation}
{\rm m}_{res} = \frac{1}{t_{max}-t_0}\sum_{t_0}^{t_max}\frac{\langle  \bar q_{mp}(t)\tau^a \gamma_5 q_{mp}(t)  \bar q(0)\tau^a \gamma_5 q(0)\rangle}{\langle  \bar q(t)\tau^a \gamma_5 q(t)  \bar q(0)\tau^a \gamma_5 q(0)\rangle}\;,
\end{equation}
where $t_0, t_{max}$ is the time interval where only the ground state pion contributes
to the two correlators in the ratio.

\subsection{Mixed Action Tuning}
\noindent
Because the valence and sea quark actions are different, the
calculation is inherently partially quenched. In other words, the
calculation violates unitarity.  
Unlike conventional  partially quenched calculations, which become unitary when the
valence quark mass is tuned to the sea quark mass,
unitarity cannot be restored by tuning the valence quark mass.
The next best option is to tune the valence quark mass in such
a way that the resulting pions have the same mass as those made of the
sea Kogut-Susskind fermions. In this case unitarity should be restored
in the continuum limit, where the $n_f=2$ staggered action has an
$SU(8)_L\otimes SU(8)_R\otimes U(1)_V$ chiral symmetry due to the
four-fold taste degeneracy of each flavor, and each pion has 15
degenerate additional partners.  At finite lattice spacing this
symmetry is broken and the taste multiplets are no longer degenerate,
but have splittings that are ${\cal O}(\alpha^2
b^2)$~\cite{Orginos:1999cr,Orginos:1998ue,Toussaint:1998sa,Orginos:1999kg,Lee:1999zxa}.
The domain wall fermion mass is tuned to give valence pions that match
the Goldstone Kogut-Susskind pion~\footnote{This is the only Goldstone boson that becomes
massless in the chiral limit at finite lattice spacing.}.  This choice
gives pions that are as light as possible, resulting in better
convergence of the $\chi$-PT  needed to extrapolate
the lattice results to the physical quark mass point.  This tuning was
also done by LHPC
collaboration~\cite{Renner:2004ck,Edwards:2005kw,Edwards:2006zza,Renner:2007pb,Hagler:2007xi,Edwards:2006qx}.

\subsection{Method of Contractions}
\noindent
Throughout our work, gauge invariant Gaussian smeared quark
propagators centered around a single point were used. In order to
facilitate the complicated Wick contractions of many body
interpolating fields, the following strategy was adopted. Noticing
that the permutations needed for Wick contractions are in effect
"scalar"~\footnote{In parallel computing language "scalar" refers to
operations that are not data parallel.}, all the contractions at the
annihilation operator point (sink) were performed and all the color
and spin indices were left open at the creation operator point
(source). The resulting data are Fourier transformed (space indices at
the annihilation operator point) and saved on disk. All the two-body
(and N-body) correlation functions can then be constructed by
appropriate contractions of the source spin and color indices on
scalar machines (such as simple work stations).

Code was constructed to automatically perform all permutations, keeping track of the
signs associated with fermion exchanges, allowing the construction of
complicated diagrams in a relatively simple and efficient
manner.  This approach works well when quark annihilation diagrams are
absent. For this reason such processes (such as the
I=0 $\pi\pi$ channel), have not been explored in our work to date.  
Another reason for avoiding such processes is that the mixed action approach has large
artifacts in these channels. In the future we plan to address such
processes using the so called all-to-all propagator
approach~\cite{Foley:2005ac}.

\subsection{Data Analysis}
\label{subsec:analysis}
\noindent
 Since Monte-Carlo integration is used to compute the relevant
 correlation functions, the statistical uncertainty 
must be carefully determined.  The main observables
 extracted in all calculations presented in this review are energy
 levels and energy level differences. These results contain
 information about phase shifts, scattering lengths and the three body
 interaction as discussed above.  The extraction of energy levels is
 done by fitting the relevant correlation functions to a sum of
 exponentials (or appropriately hyperbolic cosine functions when
 anti-periodic boundary conditions in time are imposed). We performed
 correlated $\chi^2$ minimization fits that take into account the time
 correlations in the lattice data.  In particular, the relevant
 parameters, such as the masses and the amplitude each state
 contributes to the correlation function, are determined as those that
 minimize
 \begin{equation}
\chi^2(A) = \sum_{ij} \left[\bar G(t_i) - F(t_i, A)\right]C^{-1}_{ij} \left[\bar G(t_j) - F(t_j, A)\right]
\end{equation}
where $\bar G(t)$ are the lattice two point correlation functions,
$F(t,A)$ are the fitting functions used, with $A$ denoting the set of
fitting parameters over which $\chi^2(A)$ is minimized, and $C_{ij}$
is the covariance matrix.  The lattice two point correlation functions
are determined as averages over $N$ Monte-Carlo samples $G_k(t)$:
\begin{equation}
\bar G(t) = \frac{1}{N} \sum_{k=1}^N G_k(t)\; 
\end{equation}
and 
\begin{equation}
C_{ij} = \frac{1}{N(N-1)}  \sum_{k=1}^N \left[G_k(t_i) - \bar G(t_i)\right] \left[G_k(t_j) - \bar G(t_j)\right] \ .
\end{equation}
The (standard) errors on the fitted parameters are determined by the boundaries of the
error ellipsoid, which is defined by the locus of points where $\chi^2
= \chi^2_{min} + 1$. (For a pedagogical presentation of fitting see the
TASI lectures by D. Toussaint~\cite{DeGrand:1990ss}.)

In computing scattering lengths, the procedure for determining the 
statistical errors is a little more involved due to the highly
non-linear relation between the scattering length and the energy
levels of the two particle system. First one is interested in the
energy differences between the energy levels of the two particle
system and the sum of the masses of the two free particles (similarly
for the case of more than two particles).  These energy differences
can be determined in two ways. The simplest is the one described in
the previous sections where ratios of correlators are constructed in
such way so that these ratios are a sum over exponentials parametrized
by the desired energy splittings (see for example
Eq.~(\ref{eq:energieshift})). 
In this case Jackknife is used to determine
the covariance matrix and then a correlated $\chi^2$-fit is performed as
described above. 
For a  single elimination Jackknife, the
covariance matrix of a ratio of correlators is
\begin{equation}
C_{ij} = \frac{N-1}{N}  \sum_{k=1}^N \left[R_k(t_i) - \bar R(t_i)\right] \left[R_k(t_j) - \bar R(t_j)\right]
\end{equation}
where $R_k$ is the desired ratio computed with the $k$th sample omitted from
the full ensemble and $\bar R$ is the ratio computed on the full ensemble.

Fitting correlation functions to extract the ground state energy
requires fitting ranges that start at time separations from the source
that are large enough so that excited states have negligible
contributions. The determination of the minimum time separation that
can be included in the fit is sometimes subjective. Hence a systematic
error due to the choice of the minimum time separation in the fit is
included. This error is determined by observing the variation of the
extracted results as a function of the choice of fitting interval.  The
final errors include both systematic and statistical errors combined
in quadrature.

 One way to reduce the systematic errors due to coupling to excited
 states is to use appropriate interpolating fields that reduce these
 couplings. This can be achieved by using smeared quark sources and
 sinks. We have used gauge invariant Gaussian smeared sources 
for the quark propagators
which were implemented as
 \begin{equation}
\chi(\vec x) =  \left (1 + \frac{\omega}{4 N}  \nabla^2 \right)^N \delta(\vec x_0)
\end{equation}
where $\delta(\vec x)$ is a 3D delta function, $\nabla^2 $ is the
covariant 3D Laplacian, and $\omega$ is a parameter controlling the width
of the smearing.  In the limit of $N\rightarrow\infty$ this becomes an
exponential of the Laplace operator. While the value of $\omega$ was
chosen to optimize the coupling to the nucleon, it also worked
very well for the mesons.  In addition, we used both point and
smeared sinks, both of which produced comparable results.

All of our lattice calculations are performed away from the physical
pion mass.  For that reason we have to rely on $\chi$-PT
to extrapolate to the physical point. Since results from
ensembles with different sea quark masses are un-correlated (i.e. the
covariance matrix is diagonal), we performed un-correlated $\chi^2$
fits of our data to the $\chi$-PT  formulas. In all
cases we estimate a systematic error due to these fits and
extrapolation.  Finally, except for $\pi^+\pi^+$ and $K^+K^+$
scattering lengths where the leading order lattice spacing effects are
calculable, lattice spacing errors are estimated by dimensional
analysis.

In order to avoid systematic errors due to scale setting,
dimensionless quantities (appropriate products or ratios of
dimensional quantities) are used wherever possible.  In the few cases
we need to quote results in MeV we use the scale determined by the
MILC collaboration and confirmed by our own determination using the
pion decay constant~\cite{Beane:2005rj}.

An essential parameter for all of our calculations using domain wall
fermions in the valence sector is the residual mass ${\rm m}_{res}$
that measures the degree of explicit chiral breaking. The results of
the residual mass calculation are presented in
Table~\ref{tab:MILCcnfs}.

\begin{table}[!ht]
\caption{The parameters of the MILC gauge configurations and
   domain-wall propagators used in this work. The subscript $l$
   denotes light quark (up and down), and  $s$ denotes the strange
   quark. The superscript $dwf$ denotes the bare-quark mass for the
   domain-wall fermion propagator calculation. }
{\begin{tabular}{cccccc}
 Ensemble        
&  $b m_l$ &  $b m_s$ & $b m^{dwf}_l$ & $ b m^{dwf}_s $ & $10^3 \times b
m_{res}$\\
\hline 
2064f21b676m007m050 &  0.007 & 0.050 & 0.0081 & 0.081  & 1.604  \\
2064f21b676m010m050 &  0.010 & 0.050 & 0.0138 & 0.081  & 1.552  \\
2064f21b679m020m050 &  0.020 & 0.050 & 0.0313 & 0.081  & 1.239   \\
2064f21b681m030m050 &  0.030 & 0.050 & 0.0478 & 0.081  & 0.982  \\
2896f2b709m0062m031 & 0.0062 & 0.031 & 0.0080 & 0.0423 & 0.380\\
\end{tabular}\label{tab:MILCcnfs}}
\end{table}

\subsection{Resources}
\noindent
The computational resources needed for the calculations presented here
were obtained from several sources including the USQCD clusters at
JLab and FNAL (pentium clusters with infiniband interconnect),
Tungsten at NCSA (pentium infiniband cluster) and Mare-Nostrum in
Barcelona, Spain (Power PC 970MP IBM with Myrinet
interconnect)~\footnote{We thank Andrew Pochinsky and Balint Joo for
essential help in optimizing our code for this machine.}.  The total
computer power that went into our results reviewed here is about
1.2Tflop-years, an amount of time that is rather small by current
standards. This amount of time was sufficient to achieve precision
results in the meson sector.

The results presented here are snapshots of an ongoing effort. Not all
of them were obtained with the same statistics on a given
ensemble. For details on exactly what went into these calculations the
reader should refer to the original publications.  We typically
computed several propagators per configuration by shifting the location
of the source, and have cases with as many as 24 propagators per
configuration.  We carefully monitored the variance reduction as more
propagators were added, and concluded that correlation functions were
almost statistically independent.  
The residual correlations were
taken care of 
by averaging correlation functions on a given configuration (blocking) 
and
proceeding with
the statistical analysis of data discussed in Section~\ref{subsec:analysis}.
Further blockings over multiple configurations did not change the results.

\section{Two-Body Physics}

\subsection{$\pi\pi$ Scattering}

\subsubsection{Introduction}

\noindent Pion-pion scattering at low energies is the simplest and
best-understood hadron-hadron scattering process.  Its simplicity and
tractability follow from the fact that the pions are identified as the
pseudo-Goldstone bosons associated with the spontaneous breaking of
the approximate chiral symmetry of QCD.  For this reason, the
low-momentum interactions of pions are strongly constrained by the
approximate chiral symmetries, more so than other hadrons.  The
scattering lengths for $\pi\pi$ scattering in the s-wave are uniquely
predicted at LO in $\chi$-PT~\cite{Weinberg:1966kf}:
\begin{eqnarray}
m_\pi a_{\pi\pi}^{I=0} \ = \ 0.1588 \ \ ; \ \ m_\pi a_{\pi\pi}^{I=2} \ = \
-0.04537 
\ \ \ ,
\label{eq:CA}
\end{eqnarray}
at the charged pion mass.
Subleading orders in the chiral expansion of the $\pi\pi$ amplitude
give rise to perturbatively-small deviations from the tree level, and
contain both calculable non-analytic contributions and analytic terms
with new coefficients that are not determined by chiral symmetry
alone~\cite{Gasser:1983yg,Bijnens:1995yn,Bijnens:1997vq}.  In order to
have predictive power at subleading orders, these coefficients must be
obtained from experiment or computed with Lattice QCD.

Recent experimental efforts have been made to compute the s-wave
$\pi\pi$ scattering lengths, $a_{\pi\pi}^{I=0}$ and
$a_{\pi\pi}^{I=2}$: E865~\cite{Pislak:2001bf,Pislak:2003sv} ($K_{e4}$
decays), CERN DIRAC~\cite{Adeva:2005pg} (pionium lifetime) and CERN
NA48/2~\cite{Batley:2005ax} ($K^\pm\rightarrow\pi^\pm\pi^0\pi^0$).
Unfortunately, these experiments do not provide stringent constraints
on $a_{\pi\pi}^{I=2}$.  However, a theoretical determination of s-wave
$\pi\pi$ scattering lengths which makes use of experimental data has
reached a remarkable level of
precision~\cite{Colangelo:2001df,Leutwyler:2006qq}:
\begin{eqnarray}
m_\pi a_{\pi\pi}^{I=0} \ = \ 0.220\pm 0.005 \ \ ; \ \ m_\pi a_{\pi\pi}^{I=2} \ = \ -0.0444\pm 0.0010
\ \ \ .
\label{eq:roy}
\end{eqnarray}
These values result from the Roy
equations~\cite{Roy:1971tc,Basdevant:1973ru,Ananthanarayan:2000ht},
which use dispersion theory to relate scattering data at high energies
to the scattering amplitude near threshold. In a striking recent
result, this technology has allowed a model-independent determination
of the mass and width of the resonance with vacuum quantum numbers
(the $\sigma$ meson) that appears in the $\pi\pi$ scattering
amplitude~\cite{Caprini:2005zr}. Several low-energy constants of
one-loop $\chi$-PT are critical inputs to the Roy equation analysis.
One can take the values of these low-energy constants computed with
Lattice QCD by the MILC
collaboration~\cite{Aubin:2004fs,Bernard:2006wx} as inputs to the Roy
equations, and obtain results for the scattering lengths consistent
with the analysis of Ref.~\cite{Colangelo:2001df}.

The first lattice calculations of $\pi\pi$ scattering were
performed in quenched
QCD~\cite{Sharpe:1992pp,Gupta:1993rn,Kuramashi:1993ka,Kuramashi:1993yu,Fukugita:1994na,Gattringer:2004wr,Fukugita:1994ve,Fiebig:1999hs,Aoki:1999pt,Liu:2001zp,Liu:2001ss,Aoki:2001hc,Aoki:2002in,Aoki:2002sg,Aoki:2002ny,Juge:2003mr,Ishizuka:2003nb,Aoki:2005uf,Aoki:2004wq,Li:2007ey,Sasaki:2008sv},
and 
the first partially-quenched QCD calculation of $\pi\pi$ scattering (the scattering length
and phase-shift) was carried
through by the CP-PACS collaboration, who exploited the finite-volume
strategy to study $I=2$, s-wave scattering with two flavors ($n_f=2$)
of
improved Wilson fermions~\cite{Yamazaki:2004qb}, with pion masses in
the range $m_\pi\simeq 0.5-1.1~{\rm GeV}$.  
The first fully-dynamical
calculation of the $I=2$ $\pi\pi$ scattering length with three flavors ($n_f=2+1$)
of light quarks was performed by NPLQCD  using domain-wall
valence quarks on asqtad-improved staggered sea quarks at four pion
masses in the range $m_\pi\simeq 0.3-0.5~{\rm GeV}$ at a single
lattice spacing, $b\sim 0.125~{\rm fm}$~\cite{Beane:2005rj}.

\subsubsection{Data Analysis and Chiral and Continuum Extrapolation}
\label{sec:Extrapolatepipi}

\begin{figure}[!ht]
\centering                  
\includegraphics*[width=0.7\textwidth,viewport=2 5 700 240,clip]{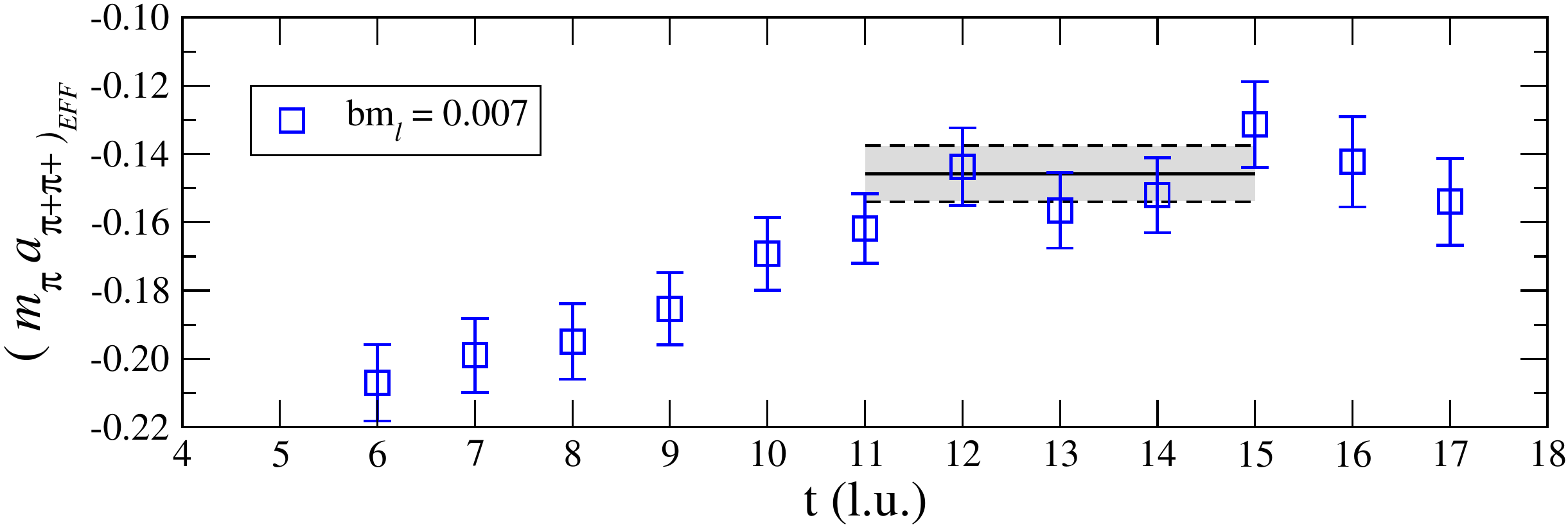}
\hfill
\includegraphics*[width=0.7\textwidth,viewport=2 5 700 240,clip]{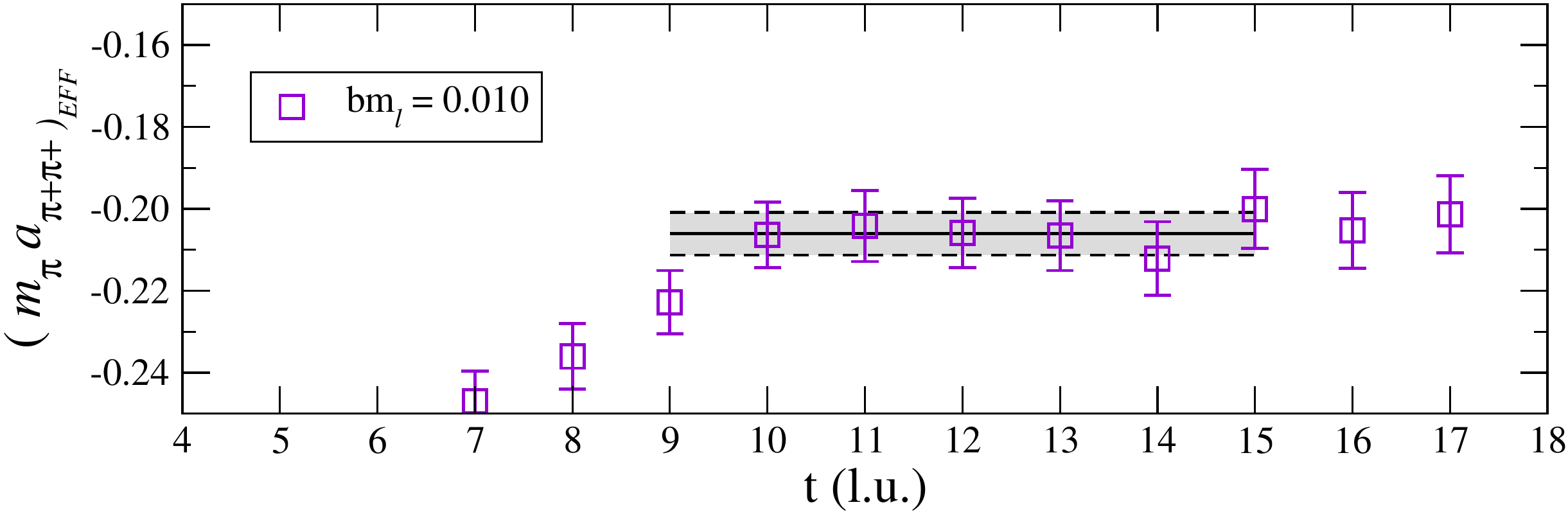}
\hfill
\includegraphics*[width=0.7\textwidth,viewport=2 5 700 240,clip]{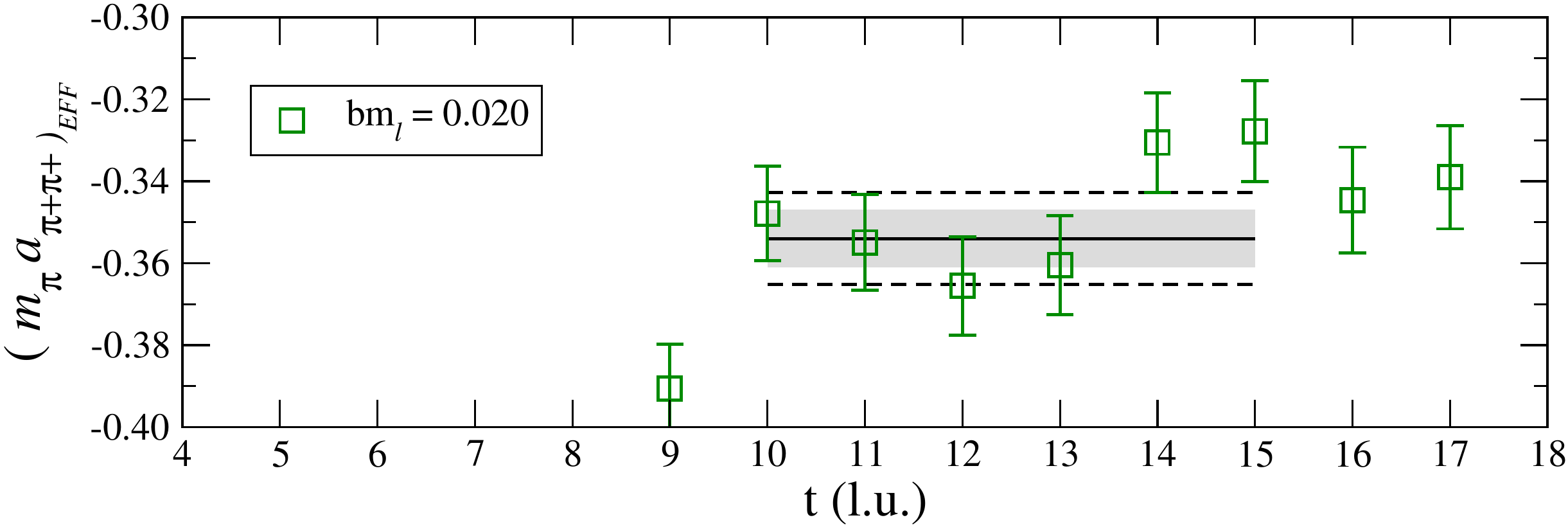}
\hfill
\includegraphics*[width=0.7\textwidth,viewport=2 5 700 240,clip]{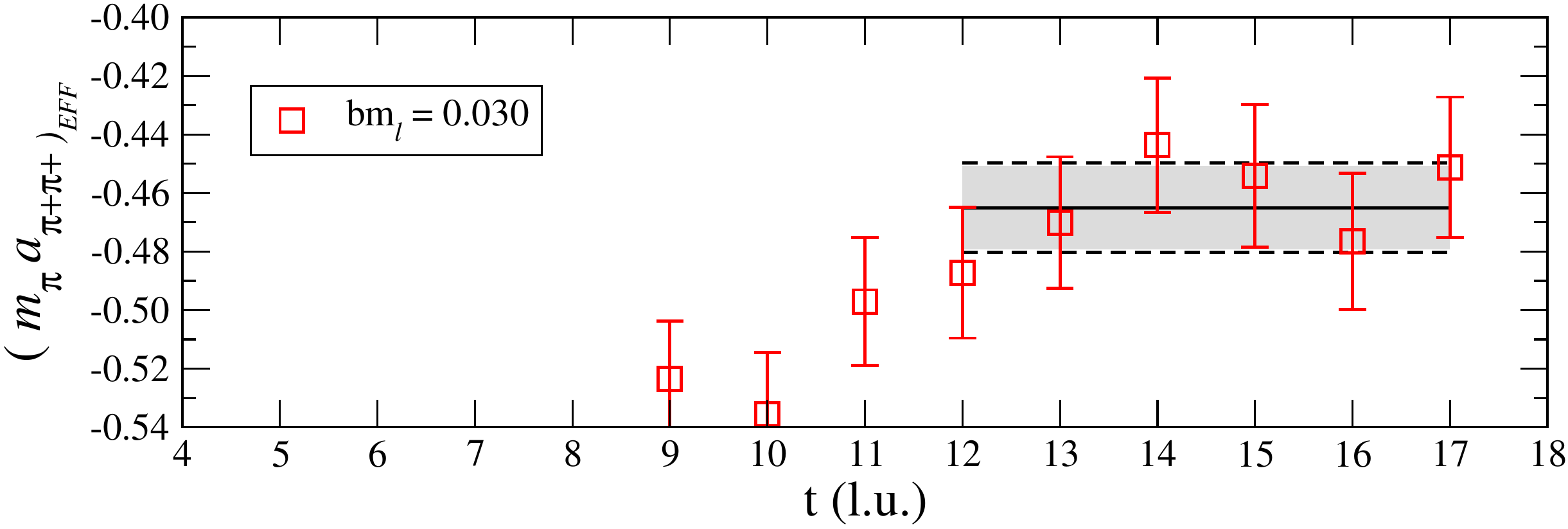}
\caption{ 
The effective $\pi^+\pi^+$ scattering length times the effective $\pi$ mass
as a function of time-slice arising from smeared sinks. The solid black lines and
shaded regions are fits with 1-$\sigma$ statistical 
uncertainties; the dashed lines are estimates of the systematic uncertainty due to fitting.}
\label{fig:SSPPplots}
\end{figure}

\noindent It is convenient to present the results of the lattice
calculation in ``effective scattering length'' plots, simple variants
of the effective-mass plots discussed above.  The effective energy splitting is formed
from the ratio of correlation functions given in Eq.~(\ref{ratio_correlator}),
\begin{eqnarray}
\Delta E_{\pi^+ \pi^+}(t) & = & \log\left({ G_{\pi^+ \pi^+}(0,t)\over  G_{\pi^+ \pi^+}(0,t+1)}\right)
\  \ ,
\label{eq:effene} 
\end{eqnarray}
which in the limit of an infinite number of gauge configurations would
become a constant at large times that is equal to the lowest energy of
the interacting $\pi^+\pi^+$ system in the volume relative to the
non-interacting two-pion energy.  
At each time-slice, $\Delta
E_{\pi^+ \pi^+}(t)$ is inserted into Eq.~(\ref{eq:energies}) (or Eq.~(\ref{eq:e0})),
to give a scattering length at each time
slice, $a_{\pi^+\pi^+}(t)$.  It is customary to consider the
dimensionless quantity given by the pion mass times the scattering
length, $m_\pi\; a_{\pi^+\pi^+}$, where $m_\pi(t)$ is the pion
effective mass, in order to remove scale-setting uncertainties.  
For each of the MILC ensembles that were analyzed, the effective scattering
lengths are shown in Fig.~\ref{fig:SSPPplots}, and the values of the pion masses, decay constants and
$\pi\pi$ energy-shifts can be found in
Refs.~\cite{Beane:2005rj,Beane:2007xs}.
The mixed-action corrections for the $I=2\ \pi\pi$ scattering length
have been determined in Ref.~\cite{Chen:2005ab} and have been discussed above.
With domain-wall fermion masses tuned to match the staggered Goldstone
pion~\cite{Renner:2004ck,Edwards:2005kw}, one finds (in lattice units)
$\tilde{\Delta}_{ju}^2 = b^2 \Delta_I =
0.0769(22)$~\cite{Aubin:2004fs} on the coarse MILC lattices with
$b\sim 0.125~{\rm fm}$.  
For each ensemble,  $m_\pi\ a_{\pi\pi}^{I=2}$ was determined and the chiral
extrapolation formula 
in Eq.~(\ref{eq:su2chiPT}) was used 
to extract a value of the counterterm
$l_{\pi\pi}^{I=2}(\mu=f_\pi)$. 
The results of the two-flavor
extrapolation to the continuum are given in Ref.~\cite{Beane:2007xs}.

Fitting to lattice data at the lightest accessible values of the quark
masses will optimize the convergence of the chiral expansion.  While
there are only four different quark masses in the data set, 
with pion and kaon masses of approximately
$(m_\pi,m_K)\sim (290,580) , (350,595) , (490, 640)$ and $(590, 675)~{\rm MeV}$,
fitting all four data sets and then ``pruning'' the heaviest data set
and refitting provides a useful measure of the convergence of the
chiral expansion. Hence, in ``fit A'', we fit the
$l_{\pi\pi}^{I=2}(f_\pi)$'s extracted 
from all four lattice ensembles 
($m_\pi\sim 290, 350, 490,\ {\rm and }\  590~{\rm MeV}$)
to a constant, 
while in ``fit B'', we
fit the $l_{\pi\pi}^{I=2}(f_\pi)$'s from the lightest three lattice
ensembles 
($m_\pi\sim 290, 350,\ {\rm and }\  490~{\rm MeV}$). 
In ``fit C'', we fit the $l_{\pi\pi}^{I=2}(f_\pi)$'s from the lightest two lattice
ensembles ($m_\pi\sim 290 \ {\rm and }\  350~{\rm MeV}$). 
Results are given in Table~\ref{tab:FitResultsNLOsu2}.
\begin{table}[!ht]
 \caption{Results of the fits in two-flavor Mixed-Action $\chi$-PT.
The values of $m_\pi\ a_{\pi\pi}^{I=2}$ correspond to the extrapolated values at the physical point.
The first uncertainty is statistical and the second is a comprehensive systematic uncertainty.}
{\begin{tabular}{cccc}
FIT &  $l_{\pi\pi}^{I=2}(\mu=f_\pi)$ &  $m_\pi\ a_{\pi\pi}^{I=2}$
(extrapolated) & $\chi^2$/dof \\
\hline
A & $6.43\pm 0.23\pm 0.26$  &  $-0.043068\pm 0.000076\pm 0.000085$  & $1.17$  \\
B & $5.97\pm 0.29\pm 0.42$  &  $-0.043218\pm 0.00009\pm 0.00014$  & $0.965$  \\
C & $4.89\pm 0.64\pm 0.68$  &  $-0.04357\pm 0.00021\pm 0.00022$  & $0.054$  \\
\end{tabular} \label{tab:FitResultsNLOsu2}}
\end{table}
Taking the range of parameters spanned by fits A-C
one finds:
\begin{eqnarray}
l_{\pi\pi}^{I=2}(\mu=f_\pi) & = & 5.4\pm 1.4
\nonumber\\
m_\pi\ a_{\pi\pi}^{I=2} & = & -0.04341\pm 0.00046 \ .
\end{eqnarray}

\begin{figure}[!ht]
\vskip 0.95cm
\centering                  
\centerline{{\epsfxsize=4.5in \epsfbox{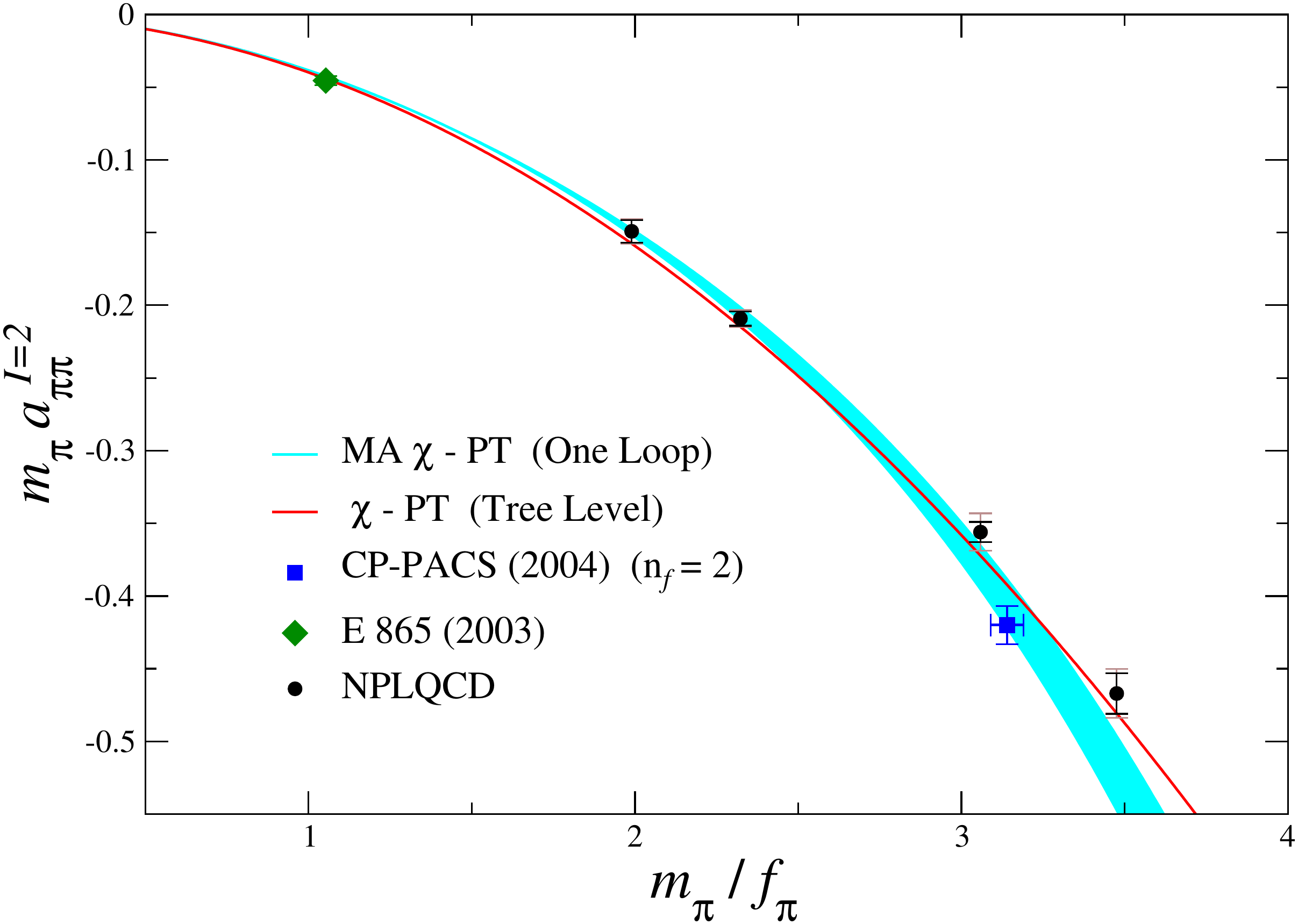}}}
\caption{$m_\pi \ a_{\pi\pi}^{I=2}$ vs. $m_\pi/f_\pi$ (ovals) with
statistical (dark bars) and systematic (light bars) uncertainties.
Also shown are the experimental value from
Ref.~\protect\cite{Pislak:2003sv} (diamond) and the lowest quark
mass result of the $n_f=2$ dynamical calculation of
CP-PACS~\protect\cite{Yamazaki:2004qb} (square).  The blue band
corresponds to a weighted fit to the lightest three data points (fit
B) using the one-loop MA$\chi$-PT formula in Eq.~(\protect\ref{eq:su2chiPT})
(the shaded region corresponds only to the statistical
uncertainty). The red line is the tree-level $\chi$-PT result. The
experimental data is not used in the chiral extrapolation fits.}
\label{fig:CAplot}
\end{figure}

In Fig.~\ref{fig:CAplot} we show the results of our calculation, along
with the lowest mass $n_f=2$ point from CP-PACS (not included in the
fit). The tree-level prediction and the results of the
two-flavor fit described in this section are also shown.  The experimental point
shown in Fig.~\ref{fig:CAplot} is not included in the fit and
extrapolation.  It is interesting that the lattice calculations indicate
little deviation from the tree level $\chi$-PT curve. The significant
deviation of the extrapolated scattering length from the tree-level
result is largely a consequence of fitting to MA$\chi$-PT at one-loop
level.

An important check of the systematic uncertainties involved
in the chiral extrapolation is to perform the same analysis using
three-flavor MA$\chi$-PT~\cite{Chen:2005ab,Chen:2006wf} as both the
real world and our lattice calculation have three active light
flavors.  In addition to the computations needed for the two-flavor
analysis, it is necessary to determine masses and decay constants for
the kaon and the $\eta$.  
The Gell-Mann--Okubo mass-relation
among the mesons is used to determine the $\eta$ mass, which is not computed
in this lattice calculation due to the enormous computer resources
(beyond what is available) required to compute the disconnected
contributions. This procedure is consistent to this order in the chiral
expansion.

The chiral expansion of the $\pi^+\pi^+$ scattering length in
three-flavor mixed-action $\chi$-PT as well as the numerical values for
the various ensembles are given in Ref.~\cite{Beane:2007xs}.  For the
three-flavor analysis, the pruning analysis
gives the results shown in
Table~\ref{tab:FitResultsNLOsu3}.
\begin{table}[!ht]
 \caption{Results of the NLO fits in three-flavor Mixed-Action $\chi$-PT.
The values of $m_\pi\ a_{\pi\pi}^{I=2}$ correspond to the extrapolated values at the physical point.
The first uncertainty is statistical and the second is a comprehensive systematic uncertainty.}
{\begin{tabular}{cccc}
FIT &  $32(4\pi)L_{\pi\pi}^{I=2}(\mu=f_\pi)$ &  $m_\pi\ a_{\pi\pi}^{I=2}$
(extrapolated) & $\chi^2$/dof \\
\hline
D & $7.09\pm 0.23\pm 0.23$  &  $-0.042992\pm 0.000076\pm 0.000077$  & $0.969$  \\
E & $6.69\pm 0.29\pm 0.39$  &  $-0.04312\pm 0.00009\pm 0.00013$  & $0.803$  \\
F & $5.75\pm 0.63\pm 0.64$  &  $-0.04343\pm 0.00021\pm 0.00021$  & $0.073$  \\
\end{tabular} \label{tab:FitResultsNLOsu3}}
\end{table}
Taking the range of parameters spanned by fits D-F
one finds:
\begin{eqnarray}
32(4\pi)L_{\pi\pi}^{I=2}(\mu=f_\pi) & = & 6.2\pm 1.2
\nonumber\\
m_\pi\ a_{\pi\pi}^{I=2} & = & -0.04330\pm 0.00042
\ \ \ .
\end{eqnarray}

\subsubsection{Systematic Uncertainties}
\label{sec:Systerrors}

\noindent There are many sources of systematic uncertainty to
be quantified; there are lattice-spacing artifacts that arise at 
$\mathcal{O}(m_\pi^4 b^2)$; there are exponentially-suppressed
finite-volume effects; there are effects due to residual chiral
symmetry breaking; there are generic two-loop effects due to
the truncation of the chiral expansion at one loop; and finally,
there are range corrections that enter at ${\cal O}\left(L^{-6}\right)$ 
in Eq.~(\ref{eq:e0}). All of these effects have been taken
into account~\cite{Beane:2007xs}.  It is noteworthy that the residual mass
turns out to be one of the leading systematic errors in our analysis
of $\pi^+\pi^+$ scattering.

It is worth emphasizing that the calculations have exact isospin
symmetry, as do the extrapolation formulas used to analyze the
results.  The conventional discussion of the scattering length is in
the unphysical theory with $e=0$ and $m_u=m_d=m$, with
$m_\pi=m_{\pi^+}=139.57018\pm 0.00035~{\rm MeV}$ and $f_\pi=f_{\pi^+}
= 130.7\pm 0.14\pm 0.37~{\rm MeV}$.  Hence
$m_{\pi^+}/f_{\pi^+}=1.0679\pm 0.0032$, where the statistical and
systematic uncertainties have been combined in quadrature.  The
results of the lattice calculations are extrapolated to this value.
The leading contribution to isospin breaking in $\pi\pi$ scattering is
due to the electromagnetic interaction, and this has been studied
extensively~\footnote{For discussions of the contributions of virtual
photons to $\pi\pi$ scattering see
Ref.~\cite{Maltman:1996nw,Meissner:1997fa,Knecht:1997jw,Knecht:2002gz},
and for very recent work on the electromagnetic contributions to the
extraction of $\pi\pi$ scattering lengths from kaon decays see
Ref.~\cite{BD_kaon2007} and Ref.~\cite{JG_kaon2007} }. Such
contributions must be removed from the experimentally-determined
scattering amplitude in order to make a comparison with the
strong-interaction calculations.  Isospin breaking due to the
difference in mass of the light quarks occurs at next-to-leading order
in the chiral expansion, and is expected to be small, as is its
contribution to $m_{\pi^+}^2-m_{\pi^0}^2$.

\subsubsection{Discussion}
\label{sec:Concludepipi}

\begin{figure}[!ht]
\vskip 0.5in
\centerline{{\epsfxsize=3in \epsfbox{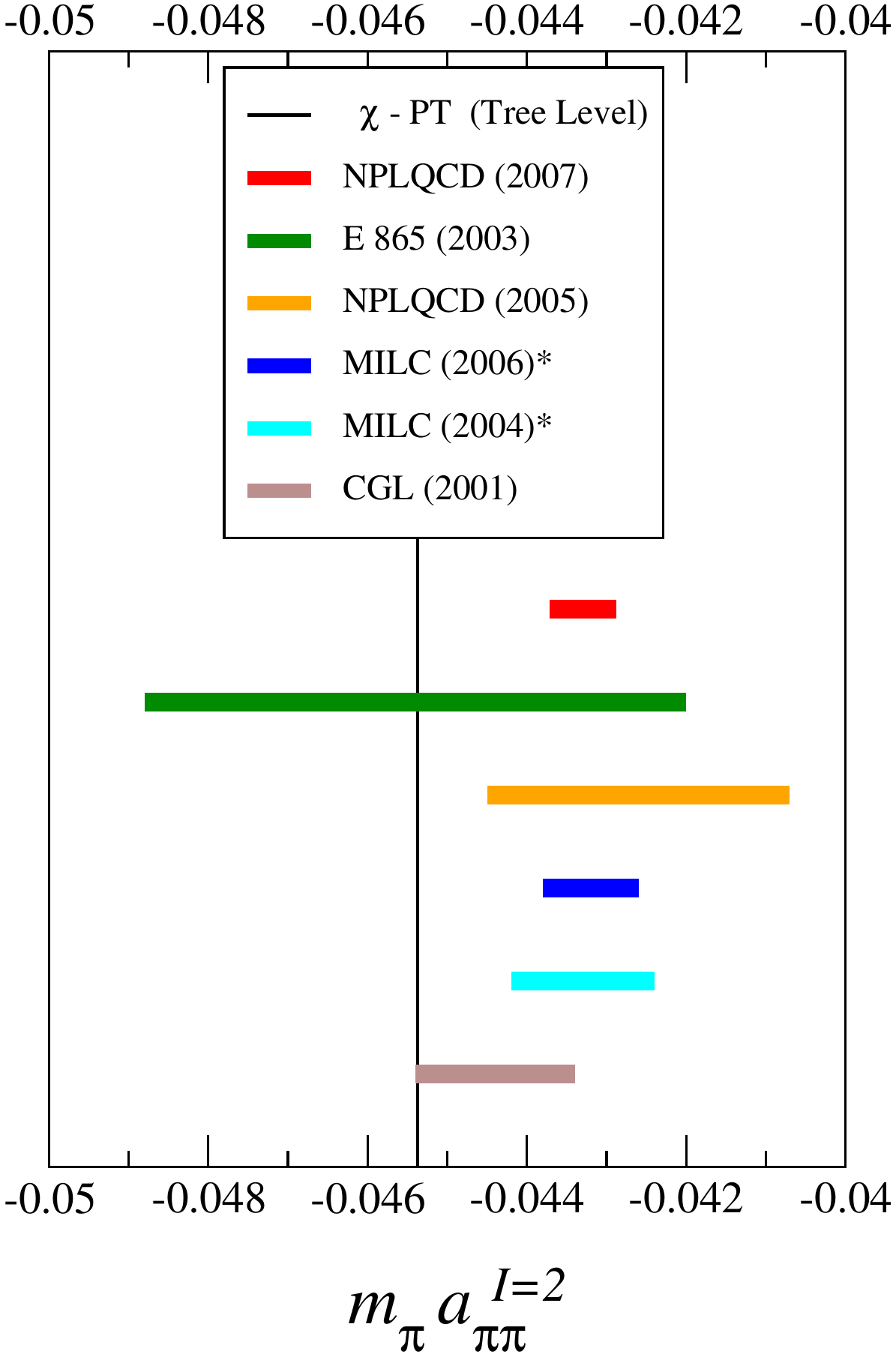}}}
\caption{ Bar chart of the various determinations of 
the $I=2$ $\pi\pi$ scattering length tabulated in
Table~\protect\ref{tab:vardet}. 
See footnote~\protect\ref{fn:milcstar}.}
\label{fig:barchart}
\end{figure}

\noindent The prediction for the physical value of the $I=2$ $\pi\pi$
scattering length is $m_\pi a_{\pi\pi}^{I=2} = -0.04330 \pm 0.00042$,
which agrees within uncertainties with the (non-lattice) determination
of CGL~\cite{Colangelo:2001df}.
In  Table~\ref{tab:vardet} and Fig.~\ref{fig:barchart} we offer a comparison 
between various determinations~\footnote{The stars on the
MILC results indicate that these are not lattice calculations of the
$I=2$ $\pi\pi$ scattering length but rather a hybrid prediction which
uses MILC's determination of various low-energy constants together
with the Roy equations), and the Roy equation determination of Ref.~\protect\cite{Colangelo:2001df}
(CGL (2001).\label{fn:milcstar}}.
\begin{table}[!ht]
\caption{ A compilation of the 
various calculations and predictions for the $I=2$ $\pi\pi$ scattering
length.}
{\begin{tabular}{cc}
{} &  $m_\pi\ a_{\pi\pi}^{I=2}$ \\
\hline
$\chi$-PT (Tree Level) & $-0.04438$ \\
NPLQCD (2007) & $-0.04330 \pm 0.00042$ \\
E 865 (2003) & $-0.0454\pm0.0031$ \\
NPLQCD (2005) & $-0.0426\pm 0.0018$ \\
MILC (2006)* & $-0.0432\pm0.0006$ \\
MILC (2004)* & $-0.0433\pm0.0009$ \\
CGL (2001) & $-0.0444 \pm 0.0010$ \\
\end{tabular} \label{tab:vardet}}
\end{table}
%

\subsection{K$\pi$ Scattering}

\subsubsection{Introduction}

\noindent In hadronic atoms, nature has provided a relatively clean
environment in which to explore the low-energy interactions of charged
hadrons.  The electromagnetic interaction allows for
oppositely-charged, long-lived hadrons to form Coulomb bound states.
The locations of the energy-levels of these systems are perturbed by
the strong interactions, while the lifetimes of the ground states are
dictated by the strong interactions that couple the charged hadrons to
lighter neutral ones.

Study of the low-energy interactions between kaons and pions with
$\pi^-K^+$ bound-states allows for an explicit exploration of the
three-flavor structure of low-energy hadronic interactions, an aspect
that is not directly probed in $\pi\pi$ scattering.  Experiments have
been proposed by the DIRAC collaboration~\cite{DIRACprops} to study
$\pi K$ atoms at CERN, J-PARC and GSI, the results of which would
provide direct measurements or constraints on combinations of the
scattering lengths.  In the isospin limit, there are two isospin
channels available to the $\pi K$ system, $I={1\over 2}$ and
$I={3\over 2}$.  The width of a $\pi^-K^+$ atom depends upon the
difference between scattering lengths in the two channels, $\Gamma\sim
(a_{1/2}-a_{3/2})^2$, (where $ a_{1/2}$ and $a_{3/2}$ are the
$I={1\over 2}$ and $I={3\over 2}$ scattering lengths, respectively)
while the shift of the ground-state depends upon a different
combination, $\Delta E_0\sim 2 a_{1/2}+a_{3/2}$.  Recently, the
Roy-Steiner equations (analyticity, unitarity and crossing-symmetry)
have been used to extrapolate high-energy $\pi K$ data down to
threshold~\cite{Buettiker:2003pp}, where it is found that
\begin{eqnarray}
m_\pi \left(a_{1/2}-a_{3/2}\right) & = & 0.269\pm 0.015
\ \ ,\ \ 
m_\pi \left(a_{1/2}+ 2 a_{3/2}\right) \ = \ 0.134\pm 0.037
\ \ \ ,
\label{eq:roypik}
\end{eqnarray}
which can be decomposed to $m_\pi a_{1/2}=0.224\pm 0.022$ and $m_\pi
a_{3/2}=-0.0448\pm 0.0077$. (See also
Refs.~\cite{Ananthanarayan:2001uy,Ananthanarayan:2000cp} and
Ref.~\cite{Jamin:2000wn} for a similar
approach.) In addition, three-flavor $\chi$-PT has been used to predict these
scattering lengths out to next-to-next-to-leading order (NNLO) in the
chiral expansion.  At NLO~\cite{Bernard:1990kw,Bernard:1990kx,Kubis:2001bx},
\begin{eqnarray}
m_\pi \left(a_{1/2}-a_{3/2}\right) & = & 0.238\pm 0.002
\ \ ,\ \ 
m_\pi \left(a_{1/2}+ 2 a_{3/2}\right) \ = \ 0.097\pm 0.047
\ \ \ ,
\label{eq:nlo}
\end{eqnarray}
while at NNLO~\cite{Bijnens:2004bu} $m_\pi a_{1/2}=0.220$ and $m_\pi
a_{3/2}= -0.047$~\footnote{ At tree-level,
Weinberg~\cite{Weinberg:1966kf} determined that $m_\pi a_{1/2}=0.137$
and $m_\pi a_{3/2}= -0.0687$.}.  One must be cautious in assessing the
uncertainties in these theoretical calculations, as one can only make
estimates based on power-counting for the contribution of higher-order
terms in the chiral expansion. There has been one determination of the
$\pi^+ K^+$ scattering length in quenched QCD~\cite{Miao:2004gy},
however, the chiral extrapolation of the scattering length did not
include the non-analytic dependences on the light quark masses that
are predicted by $\chi$-PT.

As discussed
above, recent work has identified in a model-independent way the
lowest-lying resonance in QCD which appears in $\pi\pi$
scattering~\cite{Caprini:2005zr}. Crucial to this development has been
the accurate determination of the low-energy $\pi\pi$ scattering
amplitude, including the recent Lattice QCD determination of the $I=2$
scattering length~\cite{Beane:2005rj}. A similar analysis has very
recently been carried out for $\pi K$ scattering in the
$I=\frac{1}{2}$ s-wave in order to determine the lowest-lying strange
resonance~\cite{DescotesGenon:2006uk}. Improved accuracy in the
low-energy $\pi K$ scattering amplitude should be welcome to this
endeavor.

\subsubsection{Analysis and Chiral Extrapolation}

\begin{figure}[!ht]
\centering                  
\includegraphics*[width=0.7\textwidth,viewport=2 5 700 240,clip]{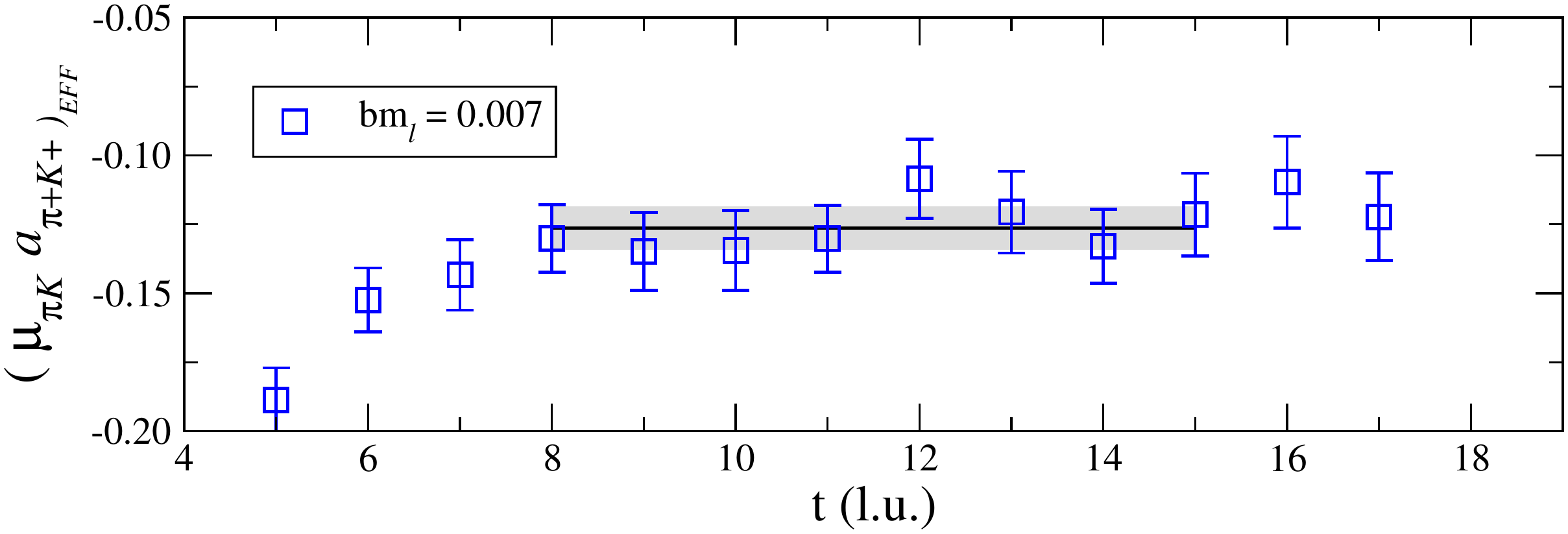}
\hfill
\includegraphics*[width=0.7\textwidth,viewport=2 5 700 240,clip]{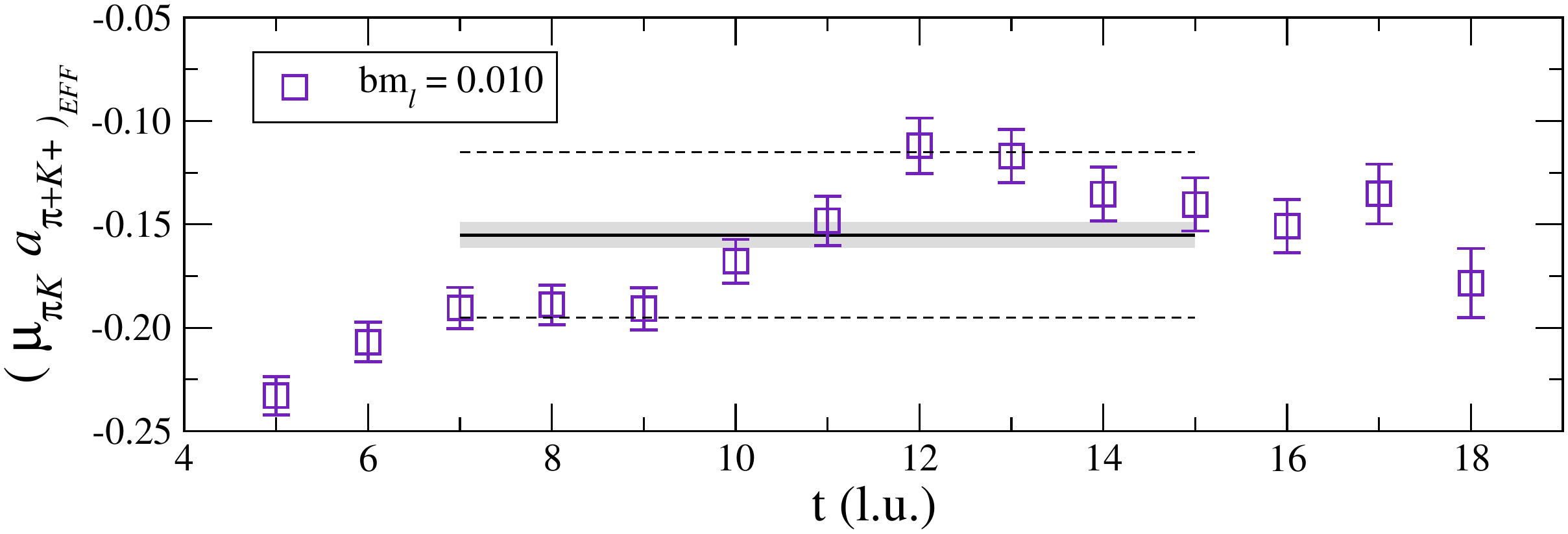}
\hfill
\includegraphics*[width=0.7\textwidth,viewport=2 5 700 240,clip]{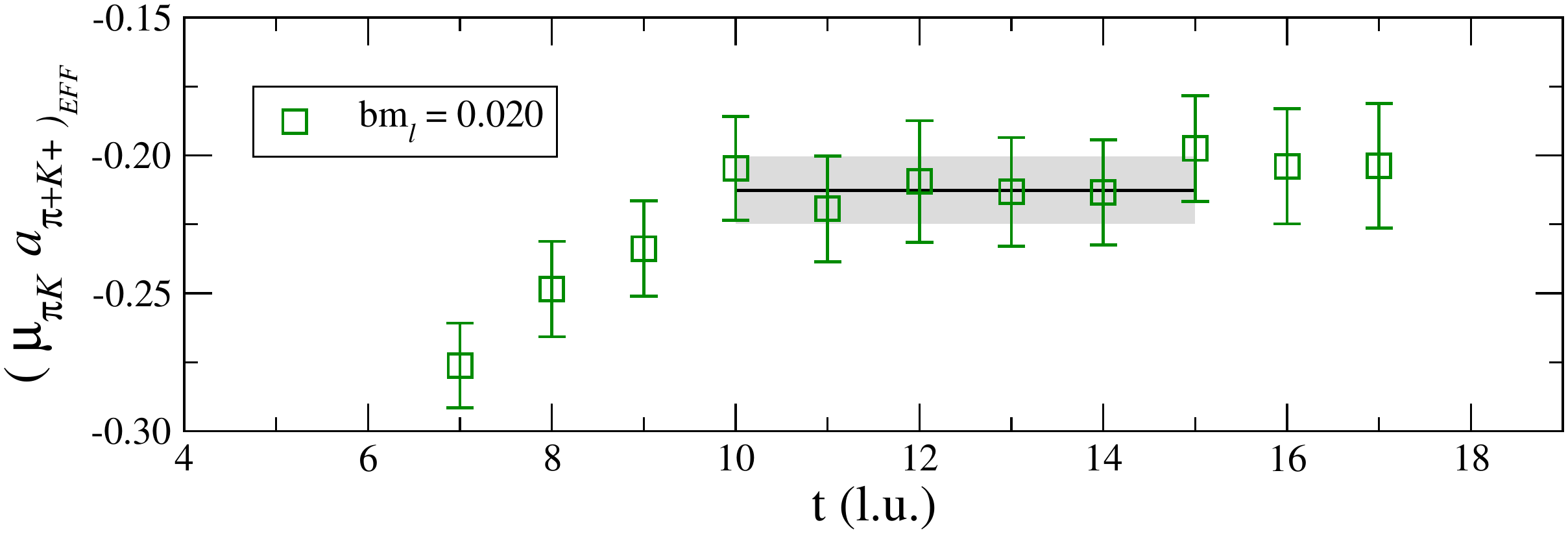}
\hfill
\includegraphics*[width=0.7\textwidth,viewport=2 5 700 240,clip]{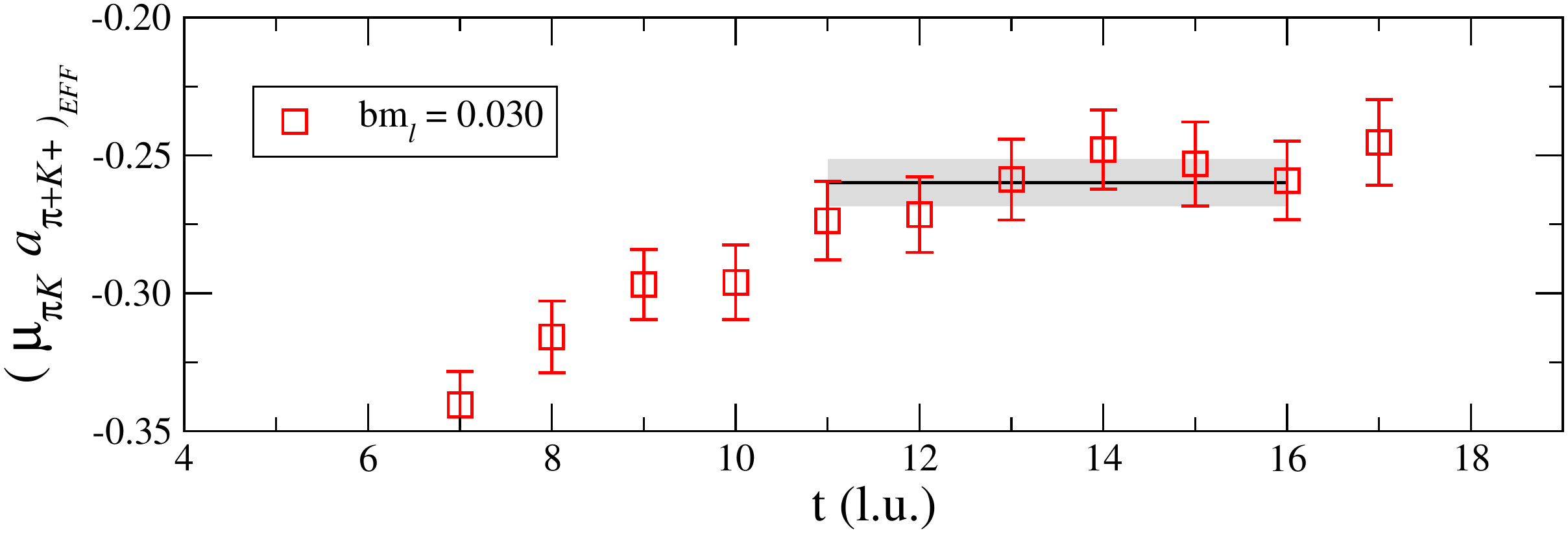}
\caption{ 
The effective $\pi^+K^+$ scattering length times the reduced mass, $ \mu_{\pi K}\  a_{\pi^+K^+}(t)$
as a function of time slice arising from smeared sinks. The solid black lines and
shaded regions are fits with 1-$\sigma$ errors.
The dashed lines on the $m_\pi\sim 350~{\rm MeV}$ ensemble plot are an estimate of the systematic error due
to fitting.}
\label{fig:SSKPplots}
\end{figure}

\begin{figure}[!ht]
\centering
\includegraphics*[width=0.7\textwidth]{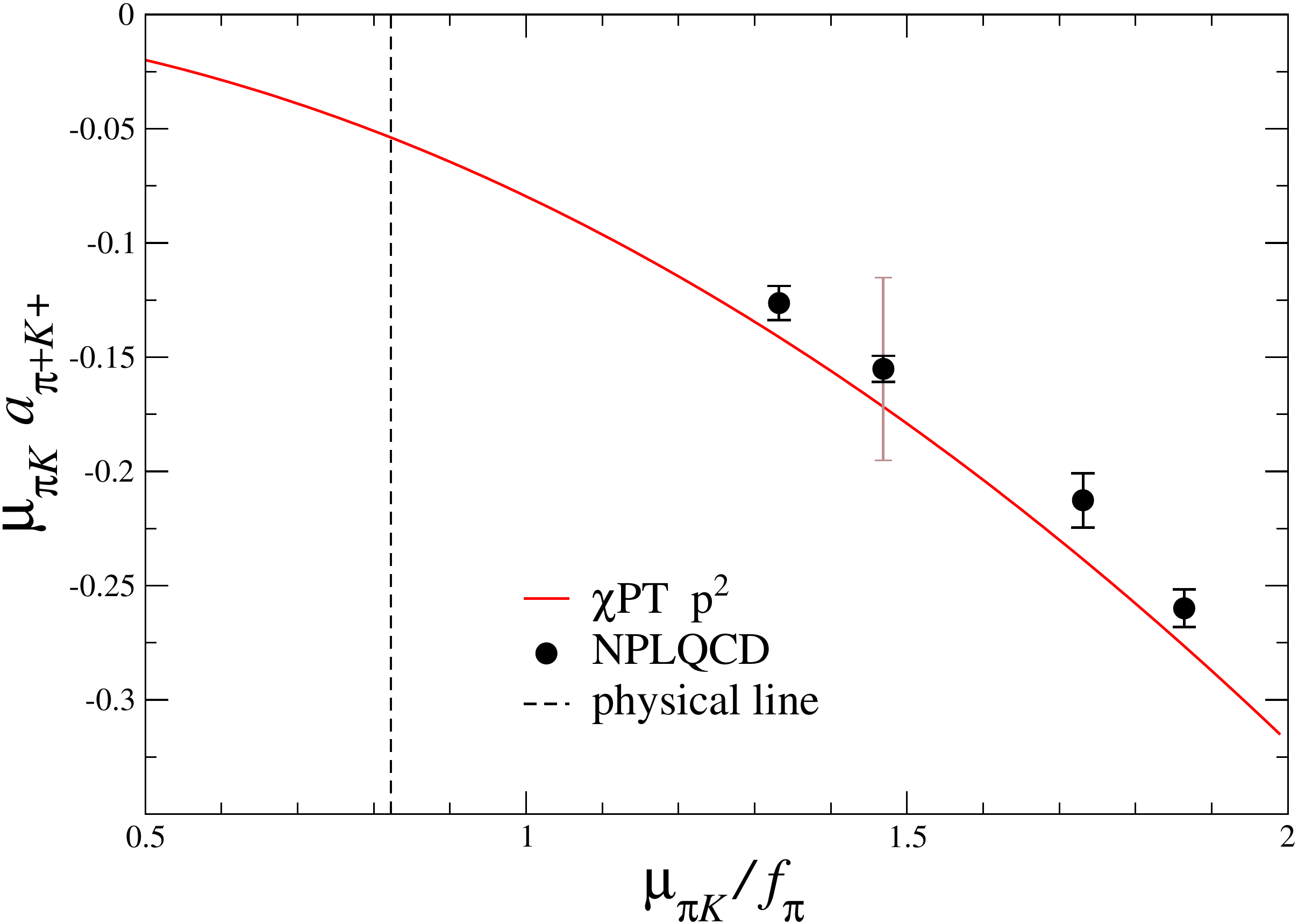}
\caption{$\mu_{\pi K} \ a_{\pi^+K^+}$ vs. $\mu_{\pi K}/f_\pi$.  The
data points are the results of this lattice calculation, while the
curve is the theoretical prediction at tree level in chiral
perturbation theory~\protect\cite{Weinberg:1966kf}.  The dark error
bar is statistical, while the lighter error bar corresponds to the
systematic error.  The vertical dashed line denotes the physical pion
and kaon masses.  }
\label{fig:muamu}
\end{figure}

\noindent It is useful to consider the dimensionless
quantity of the reduced mass times the scattering length, $\mu_{\pi
K}\ a_{\pi^+K^+}$, where $\mu_{\pi K}(t)$, the
``effective reduced mass'' is constructed from the effective mass of
the single particle correlators.  For each of the MILC ensembles, 
the effective scattering lengths are shown in
Fig.~\ref{fig:SSKPplots},
while the results of the lattice calculation of the 
decay constants, meson masses, $\pi^+K^+$ energy shifts and scattering lengths 
are given in Ref.~\cite{Beane:2006gj}.
The scattering lengths as a function of reduced mass are shown in Fig.~\ref{fig:muamu}.

\begin{figure}[!ht]
\centering
\includegraphics*[width=0.7\textwidth]{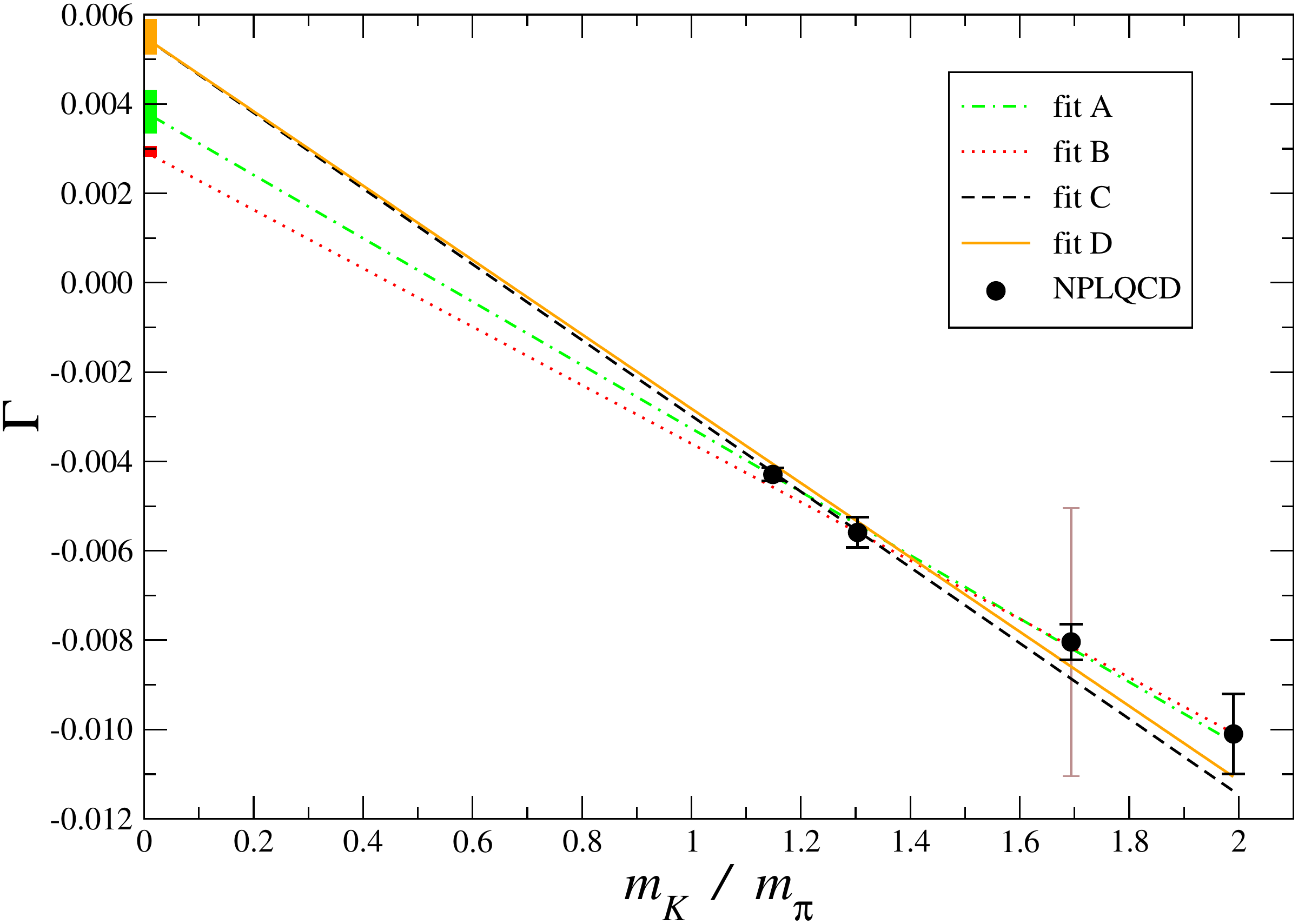}
\caption{${\Gamma}$ vs. $m_K/m_\pi$.
The dark error bar on the data points is statistical, while 
the lighter error bar corresponds to the systematic error.
The lines correspond to the four linear fits (A,B,C,D). The bars on
the y axis represent the 1-$\sigma$ errors in the determinations of $L_5=\Gamma(m_K/m_\pi =0)$
(see Ref.~\protect\cite{Beane:2006gj}). (At 95\% confidence level,
these determinations are in agreement.)}
\label{fig:G}
\end{figure}

In SU(3) $\chi$-PT~\cite{Gasser:1984gg,Gasser:1983yg,Gasser:1983ky} at NLO, the
expansion of the crossing even ($a^+$) and crossing odd ($a^-$)
scattering length times the reduced mass is
known~\cite{Bernard:1990kw,Bernard:1990kx,Kubis:2001bx} and reproduced
in Ref.~\cite{Beane:2006gj}. The counterterm $L_{\pi K}(\lambda)$ is a
renormalization scale, $\lambda$, dependent linear combination of the
Gasser-Leutwyler counterterms
\begin{eqnarray}
L_{\pi K} &\equiv& 2L_1 + 2L_2 + L_3 - 2L_4 - \frac{L_5}{2} + 2L_6 + L_8 
\ .
\end{eqnarray}
The $I={1\over 2}$ and $I={3\over 2}$ scattering lengths are related to 
crossing even and odd amplitudes by
\begin{eqnarray}
a_{1/2} & = & a^+ \ +\ 2 a^- 
\nonumber\\
a_{3/2} & = & a^+ \ -\ a^-\ =\ a_{\pi^+K^+} 
\ \  .
\end{eqnarray}
It is convenient to define the function ${\Gamma}$ via a subtraction of the
tree-level and one-loop contributions in order to 
isolate the counterterms, 
\begin{eqnarray}
{\Gamma} & \equiv & 
-
{f_\pi^2\over 16 m_\pi^2}
\left( {4\pi f_\pi^2\over\mu_{\pi K}^2} \left[ \mu_{\pi K}\ a_{\pi^+K^+} \right]
 + 1 + \chi^{(NLO,-)} - 2 {m_K m_\pi\over f_\pi^2} \chi^{(NLO,+)}
\right) 
\ ,
\label{eq:Gdef}
\end{eqnarray}
and at NLO this becomes
\begin{eqnarray}
{\Gamma} & = & 
L_5(f_\pi^{\rm phys})\ -\ 2\ {m_K\over m_\pi}\  L_{\pi K}(f_\pi^{\rm phys})
\ .
\label{eq:Gdefrhs}
\end{eqnarray}
The dependence of ${\Gamma}$ on $m_\pi$ and $m_K$ 
determines $L_5$ and $L_{\pi K}$ and, in turn, allows an
extraction of $a_{3/2}$ and $a_{1/2}$. The numerical values of
${\Gamma}$ and their errors are plotted in Fig.~\ref{fig:G}.
By fitting a straight line to the values of ${\Gamma}$ as a function of
$m_k/m_\pi$ the counterterms $L_5$ and $L_{\pi K}$ (renormalized at 
$f_\pi^{\rm phys}$) can be determined.

\begin{figure}[!ht]
\centering
\includegraphics*[width=0.7\textwidth]{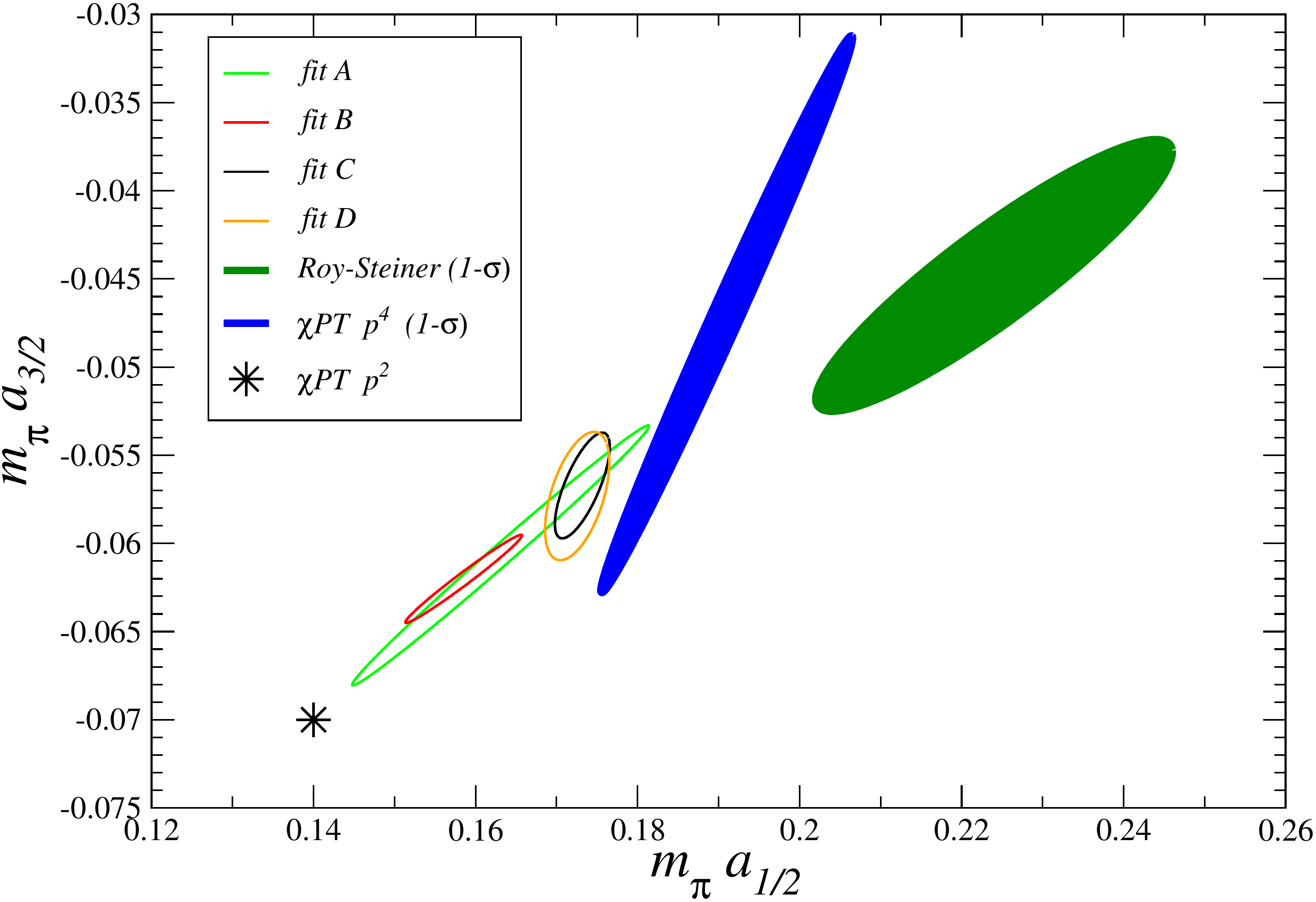}
\caption{Error ellipses for the four fits (A,B,C,D) at 95\% confidence
level.  The star corresponds
to the current-algebra predictions ($\chi$-PT $p^2$) from
Ref.~\protect\cite{Weinberg:1966kf}. The 1-$\sigma$ error
ellipses from a $\chi$-PT analysis at NLO~\protect\cite{Bernard:1990kw}
(denoted $\chi$-PT $p^4$) and from a fit using the Roy-Steiner
equations~\protect\cite{Buettiker:2003pp} are also shown.}
\label{fig:ellipses}
\end{figure}

As described in Section~\ref{sec:Extrapolatepipi}, the results of the lattice
calculations are pruned, to give ``fit A'' and ``fit B''.
\begin{table}[!ht]
 \caption{Results of the NLO fits.  The values of $m_\pi a_{3/2}$ and $m_\pi a_{1/2}$ 
correspond to their extrapolated values at the physical point.
}
{\begin{tabular}{cccccc}
FIT &  ${L}_5\times 10^3$ & $L_{\pi K}\times 10^3$ &  $m_\pi  a_{3/2}$ &  $m_\pi  a_{1/2}$  &$\chi^2$/dof \\
\hline
A & $3.83\pm0.49$  & $3.55\pm0.20$  & $-0.0607\pm0.0025$ & $0.1631\pm0.0062$  & $0.17$  \\
B & $2.94\pm0.07$  & $3.27\pm0.02$  & $-0.0620\pm0.0004$ & $0.1585\pm0.0011$  & $0.001$  \\ 
C & $\ \ 5.65\pm0.02^{+0.18}_{-0.54}$  
& $4.24 \pm 0.17$  & $-0.0567\pm0.0017$ & $0.1731\pm0.0017$ & $0.84$  \\ 
D & $\ \ {5.65\pm0.02^{+0.18}_{-0.54}}$ 
& $4.16 \pm0.18$  & $-0.0574\pm0.0016$ & $0.1725\pm0.0017$ & $0.90$  \\ 
\end{tabular} \label{tab:FitResultsNNLO}}
\end{table}
With this limited data set it is not practical to fit to the NNLO
expression~\cite{Bijnens:2004bu} for the scattering length.  However,
it is important to estimate the uncertainty in the values of the
scattering lengths extrapolated to the physical point that is
introduced by the truncation of the chiral expansion at NLO.  In our
work on $f_K/f_\pi$~\cite{Beane:2006kx} a value of $L_5$ was extracted
as it is the only NLO counterterm that contributes.  The numerical
value obtained is only perturbatively close to its true value, as it
is contaminated by higher-order contributions.  Therefore, by fixing
the $L_5$ that appears in Eq.~(\ref{eq:Gdefrhs}) to the value of $L_5$
extracted from $f_K/f_\pi$, an estimate of the uncertainty in both
$L_{\pi K}$ and in the extrapolated values of the scattering lengths
due to the truncation of the chiral expansion can be made.
Specifically, $L_5$ was sampled from a Gaussian distribution for a
range of $f_K/f_\pi$ values~\cite{Beane:2006kx} and then $L_{\pi K}$
was fit.  This fit is denoted ``fit C'', and the same fit but with the
$m_\pi\sim 590~{\rm MeV}$ data pruned is denoted ``fit D''. The
results of the four fits are given in Table~\ref{tab:FitResultsNNLO}
and plotted in Fig.~\ref{fig:G}. These lead to
\begin{eqnarray}
L_{\pi K} & = & 4.16 \pm 0.18 ^{+0.26}_{-0.91}
 \ \ \ ,
\end{eqnarray}
and a prediction of the scattering lengths extrapolated to the physical point of
\begin{eqnarray}
m_\pi \  a_{3/2} & = & -0.0574\pm 0.0016^{+0.0024}_{-0.0058} \nonumber \\ 
m_\pi \  a_{1/2} & = &\ \ 0.1725\pm 0.0017^{+0.0023}_{-0.0156} \ .
\end{eqnarray}
The central values and statistical errors were taken from
fit D and  the systematic error due to truncation of the
chiral expansion was set by taking the range of the various quantities allowed
by the four fits, including statistical and systematic errors. 
Fig.~\ref{fig:ellipses} shows the 95\% confidence-level error
ellipses associated with the four fits in the $m_\pi \ a_{1/2}$-$m_\pi
\ a_{3/2}$ plane. For purposes of comparison the
current-algebra point~\cite{Weinberg:1966kf} is included on the plot as well as
1-$\sigma$ error ellipses from analyses based on fitting
experimental data using $\chi$-PT at NLO~\cite{Bernard:1990kw} and
using Roy-Steiner equations~\cite{Buettiker:2003pp}. As 1-$\sigma$
error ellipses correspond to 39\% confidence level, one should be
careful in finding discrepancy between the various determinations of
the scattering lengths. 
The crossing-odd scattering length is of special interest as its corrections
are protected by SU(2) chiral symmetry and are therefore of order $m_\pi^4$
and expected to be small~\cite{Roessl:1999iu,Schweizer:2005nn}~\footnote{We thank Heiri Leutwyler
for emphasizing this point to us.}.

Given how well the lattice data fit the NLO continuum $\chi$-PT
formulas, it would seem that the $O(b^2)$ discretization errors are
comparable or smaller than the systematic error due to omitted
$O(m_q^3)$ effects in the chiral expansion. However, one should keep
in mind that our determinations of, for instance, the low-energy
constants $L_5$ and $L_{\pi K}$ are subject to $O(b^2)$ shifts.  In
contrast with the $\pi^+\pi^+$ and $K^+K^+$ scattering lengths, the
mixed-action quantity $\Delta_{\rm Mix}$ makes an explicit
contribution to the $K^+\pi^+$ scattering length~\cite{Chen:2005ab,donal}.
While this adds an additional unknown contribution to this process,  
a MA$\chi$-PT analysis of $\pi K$ scattering, including lattice  
data from the fine MILC lattices ($b\sim0.09~{\rm fm}$), as well as the  
determination of $\Delta_{\rm Mix}$~\cite{Orginos:2007tw}, will be  
able to address this source of systematic error quantitatively.

We anticipate that with improved statistics,
together with calculations on lattices with smaller lattice spacings,
the theoretically-predicted regions for $m_\pi a_{3/2}$ and $m_\pi
a_{1/2}$ can be further reduced beyond those shown in
Fig.~\ref{fig:ellipses}. These regions can then be compared with the
expected measurements from $K^+\pi^-$ atoms, to provide an exciting
test of hadronic theory.

We also note that (1) there are calculations underway to calculate the
scattering lengths in both isospin channels directly
with quenched QCD~\cite{Nagata:2007zz}; (2) an indirect method which
uses scalar form factors for exclusive semileptonic decays, calculated
with $n_f=2$ lattice data, gives
$m_\pi\   a_{1/2}  =   0.179\pm 0.017$~\cite{Flynn:2007ki,Flynn:2007rs}.

\subsection{KK Scattering}

\subsubsection{Introduction}

\noindent
Strange hadrons may play a crucial role in the properties and evolution of 
nuclear material under extreme conditions~\cite{Page:2006ud}.
The interior of neutron stars provide one such environment in which the
densities are high enough that it may be energetically favorable
to have  strange baryons present in significant
quantities, depending upon their interactions with non-strange hadrons.
Further, it may be the case that a kaon condensate forms due, in part, to strong
interactions between kaons and nucleons~\cite{Kaplan:1986yq}.  
Unfortunately, the theoretical analysis of both scenarios is somewhat 
plagued by the limited knowledge of the interactions of
strange hadrons with themselves and with non-strange hadrons.

Heavy-ion collisions, such as those at the BNL Relativistic Heavy Ion
Collider (RHIC), also produce nuclear material in an extreme
condition.  Recent observations suggesting the formation of a
low-viscosity fluid are quite exciting as they provide a first glimpse
of matter not seen previously.  The late-time evolution of such a
collision requires an understanding of the interaction between many
species of hadrons, not just those of the initial state, including the
interactions between strange mesons and baryons.  While pion
interferometry in heavy-ion collisions is a well-established tool for
studying the collision region (for recent theoretical progress, see
Refs.~\cite{Cramer:2004ih,Miller:2005ji,Miller:2007gh}), the STAR
collaboration has recently published the first observation of neutral
kaon ($K_s^0$) interferometry~\cite{Abelev:2006gu}.  In the analysis
of $K_s^0$-$K_s^0$ interferometry, the non-resonant contributions to
the final state interactions between the kaons were estimated using
three-flavor $\chi$-PT, the low-energy effective field theory of QCD.
Given the sometimes poor convergence of $SU(3)_L\otimes SU(3)_R$
$\chi$-PT due to the relatively large kaon mass compared to the scale
of chiral symmetry breaking ($\Lambda_\chi \sim 1$~GeV), particularly
in the baryon sector, it is important to be able to verify that the
non-resonant contributions to $KK$-scattering are indeed small, as
estimated in $\chi$-PT. To date, there have been no experimental
determinations of the $I=1\ KK$ scattering length, $a_{KK}^{I=1}$, but
recently it has been calculated at NLO in
$\chi$-PT~\cite{Chen:2006wf}.

%
%
\begin{figure}[!t]
\center
\begin{tabular}{c}
\includegraphics*[width=0.7\textwidth,viewport=2 5 700 240,clip]{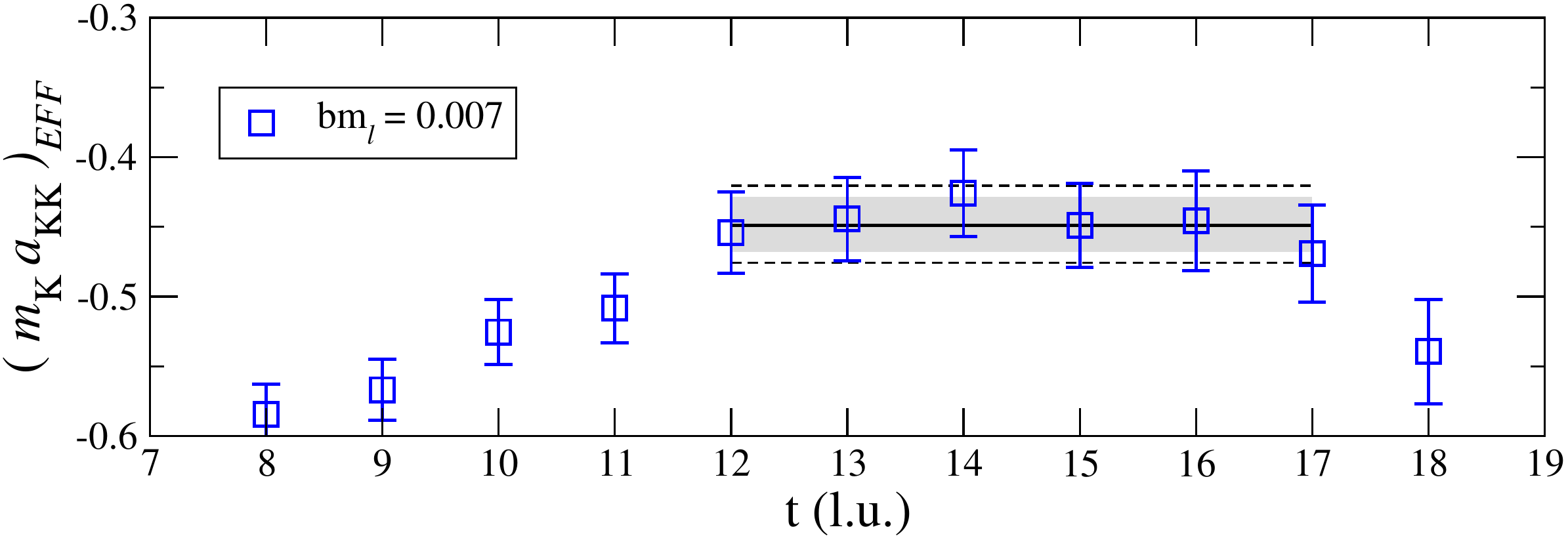}\\
\hfill
\includegraphics*[width=0.7\textwidth,viewport=2 5 700 240,clip]{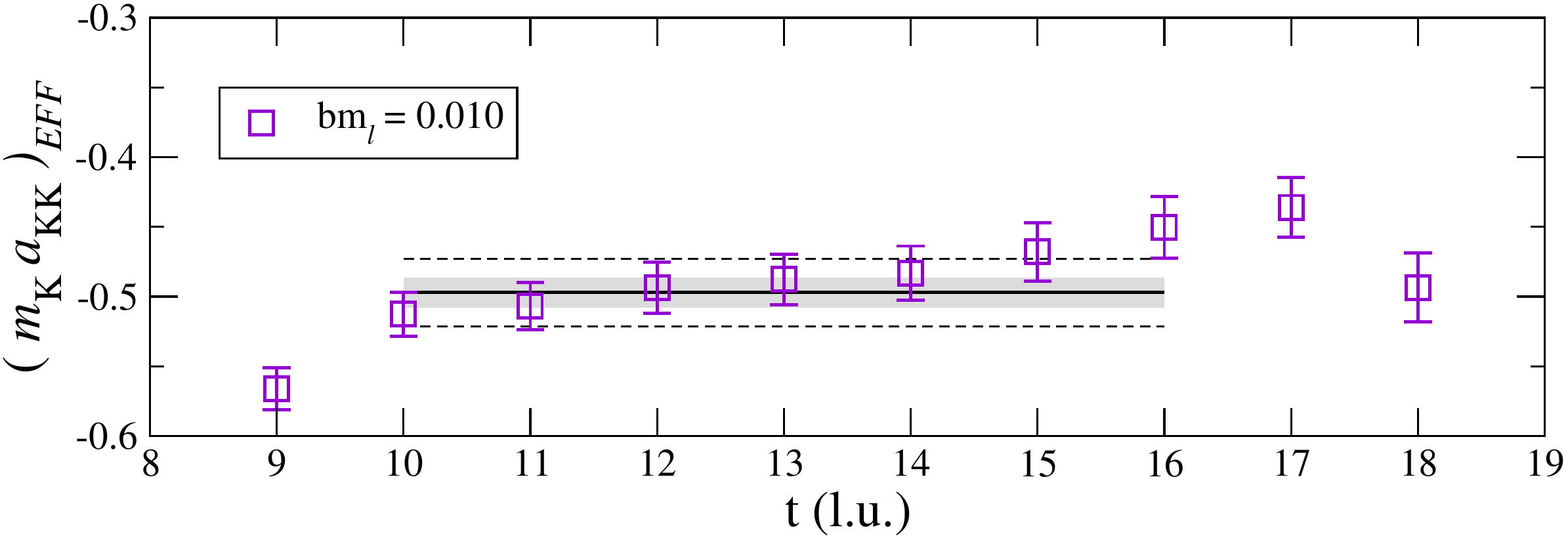}\\
\hfill
\includegraphics*[width=0.7\textwidth,viewport=2 5 700 240,clip]{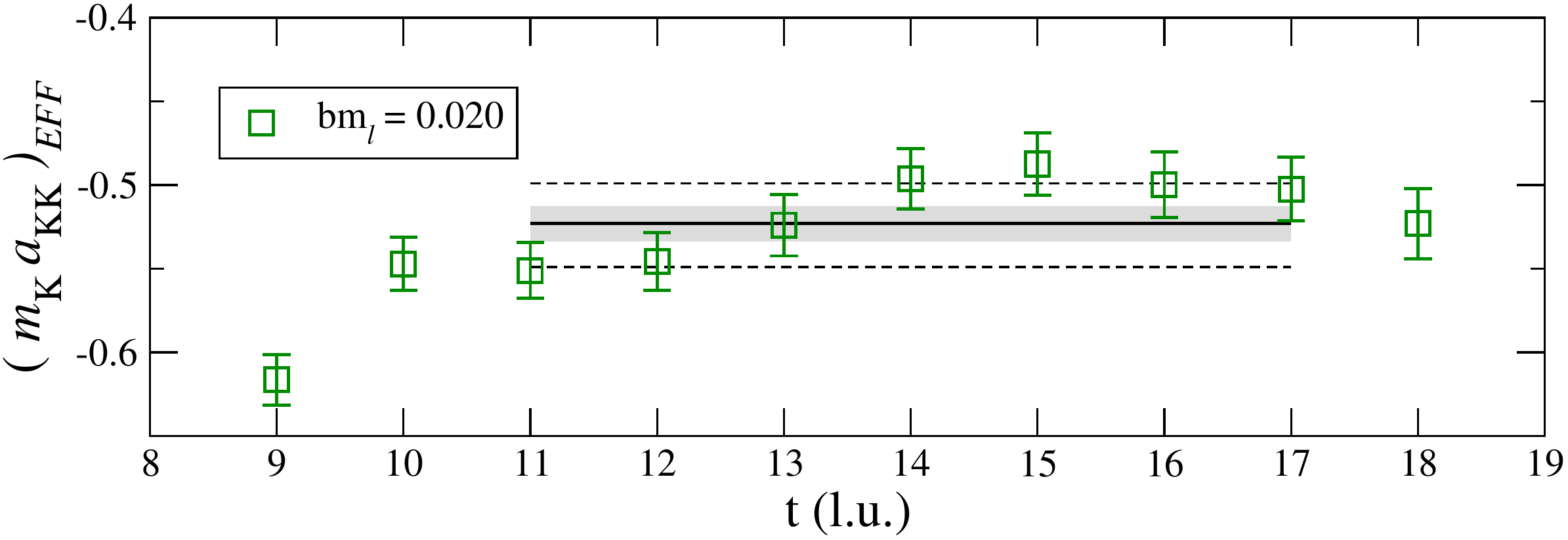}\\
\hfill
\includegraphics*[width=0.7\textwidth,viewport=2 5 700 240,clip]{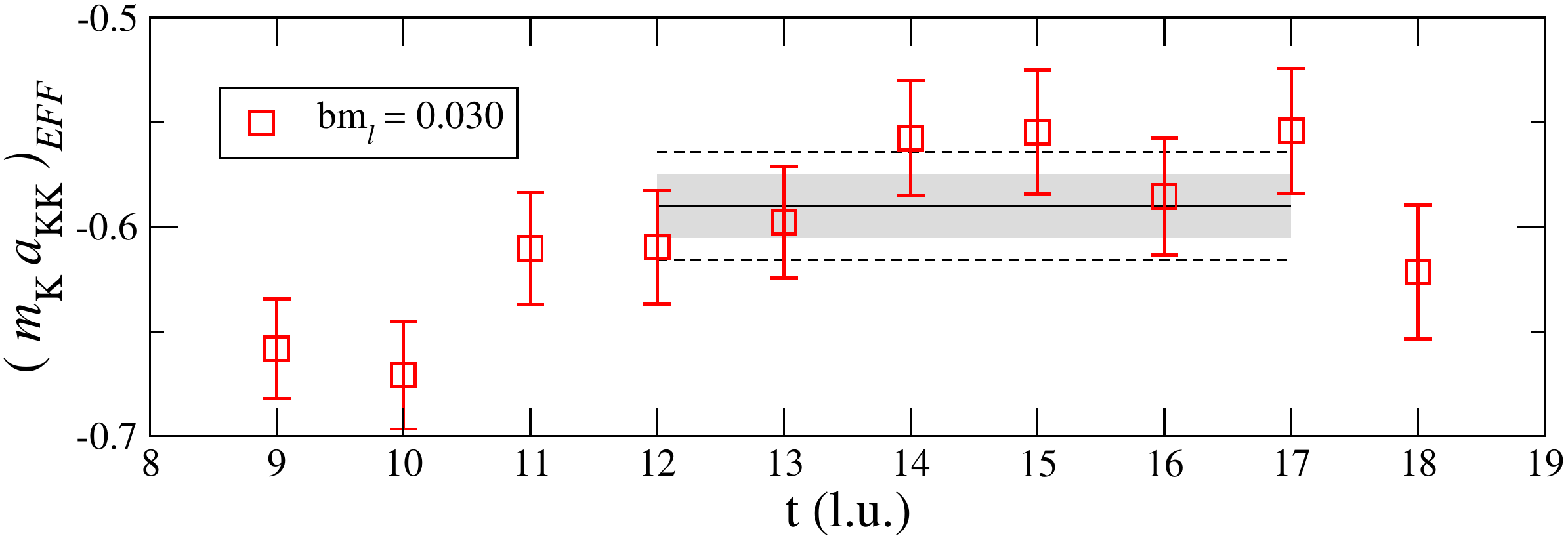}\\
\end{tabular}
\caption{\label{fig:effective_a} The effective $K^+K^+$
scattering length times the effective $m_{K^+}$ as a function of time
slice. The solid black lines and shaded
regions are fits with 1-$\sigma$ statistical uncertainties.
The dashed lines correspond to the statistical and systematic
uncertainties added in quadrature. (See Ref.~\protect\cite{Beane:2007uh} for data tables.)
}
\end{figure}

%
%
\subsubsection{Mixed-Action $\chi$-PT at One Loop \label{sec:ResultsB}}

\noindent
In Ref.~\cite{Chen:2006wf}, the expression for the $I=1\ KK$
scattering length was determined to NLO in $\chi$-PT, including
corrections due to mixed-action lattice artifacts.  As with the 
$I~=~2\ \pi\pi$ scattering length~\cite{Chen:2005ab}, it was demonstrated that
when the mixed-action extrapolation formula is expressed in terms of
the lattice-physical parameters there are no
lattice-spacing-dependent counterterms at $\mathcal{O}(b^2)$,
$\mathcal{O}(b^2 m_K^2)$ or $\mathcal{O}(b^4)$.  There are finite
lattice-spacing-dependent corrections, proportional to $b^2
\Delta_\mathrm{I}$, and therefore entirely determined to this order in
MA$\chi$-PT.  Again, as with the $I=2\ \pi\pi$ system, the NLO MA
formula for $m_K a_{KK}^{I=1}$ does not depend upon the mixed
valence-sea meson masses, and therefore does not require knowledge of
the mixed-meson masses~\cite{Orginos:2007tw}.  This
allows for a precise determination of the predicted MA corrections to
the scattering length.  At NLO in MA$\chi$-PT, the scattering length
takes the form
\begin{multline}\label{eq:mKaKKMA}
m_K a_{KK}^{I=1}(b\neq 0) =-\frac{m_{K}^{2}}{8 \pi f_{K}^{2}} \bigg\{ 1
        +\frac{m_{K}^{2}}{ (4\pi f_{K})^{2}} \bigg[ 
                C_\pi \ln \left( \frac{m_\pi^{2}}{\mu^2} \right) 
                +C_K \ln \left( \frac{m_K^{2}\,}{\mu ^2} \right) 
        \\
                +C_X \ln \left( \frac{\tilde{m}_X^{2}}{\mu^2} \right) 
                +C_{ss} \ln \left( \frac{m_{ss}^{2}}{\mu ^{2}} \right) 
                + C_0 
        - 32(4\pi)^2\, L_{KK}^{I=1}(\mu)
        \bigg] \bigg\}\, ,
\end{multline}
where the various coefficients, $C_i$, along with $\tilde{m}_X^{2}$ and $m_{ss}^{2}$,
can be
found in Appendix E of Ref.~\cite{Chen:2006wf}.  
To account for the predicted MA corrections, one can use Eq.~\eqref{eq:mKaKKMA} to
directly fit the results of the lattice calculation.
Further sources of systematic errors can be identified; higher-order effects in the chiral
expansion, $\Delta_{NNLO}(m_K a_{KK}^{I=1})$; exponentially-suppressed
finite-volume effects, $\Delta_{FV}(m_K a_{KK}^{I=1})$; residual
chiral symmetry breaking effects from the domain-wall action,
$\Delta_{m_{res}}(m_K a_{KK}^{I=1})$; and the error in truncating the
effective-range expansion with the inverse scattering length,
$\Delta_{range}(m_K a_{KK}^{I=1})$. For a detailed discussion of
these various sources of uncertainty, we refer the reader to Ref.~\cite{Beane:2007uh}.

\subsubsection{Extrapolation to the Physical Point
\label{sec:mKaKKPred}}

%
%
\begin{figure}[!t]
\vskip0.5in
\center
\begin{tabular}{c}
\includegraphics[width=0.7\textwidth]{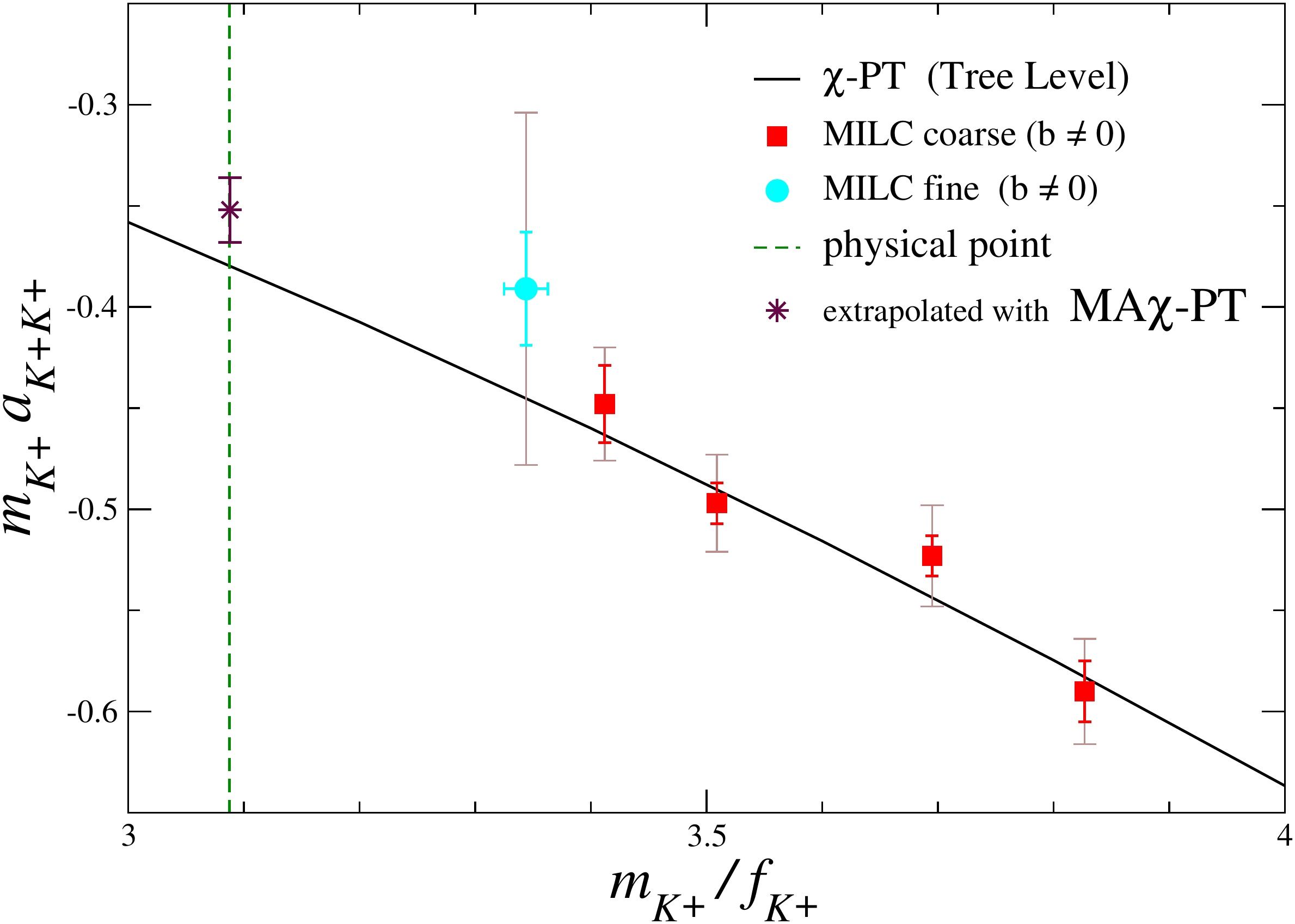} \\
\end{tabular}
\caption{\label{fig:mKaKKdataPhys} 
$m_{K^+} a_{K^+ K^+}$ versus $m_{K^+}/f_{K^+}$.
The points with error-bars are the results of this lattice calculation (not
extrapolated to the continuum) on both the coarse and fine MILC lattices.
The solid curve corresponds to the tree-level prediction of $\chi$-PT, and the
point denoted by a star and its associated uncertainty is the value
extrapolated to the physical meson masses and to the continuum.
}
\end{figure}

\noindent
The results of the lattice calculations were pruned (defined in
Section~\ref{sec:Extrapolatepipi})
to explore the convergence of the chiral expansion, 
giving fits A, B, and 
C~\cite{Beane:2005rj,Beane:2006gj,Beane:2007xs} for the low-energy constant
 $L_{KK}^{I=1}(\mu = f_K)$.
The results of these fits are collected in Table~\ref{tab:L_GL}.
%
%
\begin{table}[t]
\caption{
The results of  fitting three-flavor MA$\chi$-PT at NLO to the 
computed scattering lengths, as described in the text.
The values of $m_{K^+} a_{K^+ K^+}$ are those extrapolated to the physical
(isospin-symmetric) meson masses and to the continuum.}
{\begin{tabular}{c|ccc}
\ \ \ \ FIT\ \ \ \  & $32(4\pi)^2 L_{KK}^{I=1}(f_K)$ & $m_{K^+} a_{K^+ K^+}$ (extrapolated) & $\chi^2$/dof \\ \hline
A & 7.3(1)(4) & $-0.347 \pm 0.003 \pm 0.009$ & 0.22  \\
B & 7.3(2)(5) & $-0.347 \pm 0.004 \pm 0.011$ & 0.32  \\
C & 6.9(2)(6) & $-0.355 \pm 0.005 \pm 0.013$ & 0.14  \\ 
\end{tabular} \label{tab:L_GL}   }
\end{table}
The extracted values of $L_{KK}^{I=1}$ from each of the fits are consistent
with each other within the uncertainties.
In analogy with the comparison convention employed for $\pi^+\pi^+$,
the results of the lattice calculation are extrapolated to the physical values of 
$m_{\pi^+}/f_{K^+} = 0.8731 \pm 0.0096$, $m_{K^+}/f_{K^+} =
3.088 \pm 0.018$ and $m_\eta / f_{K^+} = 3.425 \pm 0.0019$ assuming isospin
symmetry, and the absence of electromagnetism.
Taking the range of values of $L_{KK}^{I=1}$
spanned by these fits, gives
\begin{eqnarray}
          m_{K^+} a_{K^+ K^+} \ =\  -0.352 \pm 0.016
 \ \ \ \ , \ \ \ 
32(4\pi)^2 L_{KK}^{I=1}(\mu=f_K) & = & 7.1 \pm 0.7
\, ,
\end{eqnarray}
where the statistical and systematic errors have been combined in quadrature.
The results are shown in Fig.~\ref{fig:mKaKKdataPhys}.

\subsection{Thoughts on Meson-Meson Scattering}

\noindent The lattice results for meson-meson scattering pose an
interesting puzzle.  In Fig.~\ref{fig:CAplot} one sees that the $I=2$
$\pi\pi$ scattering length tracks the current algebra result up to
pion masses that are expected to be at the edge of the chiral regime
in the two-flavor sector. While in the two flavor theory one expects
fairly good convergence of the chiral expansion and, moreover, one
expects that the effective expansion parameter is small in the channel
with maximal isospin, the lattice calculation clearly imply a cancellation
between chiral logs and counterterms (evaluated at a given scale). As
one sees in Fig.~\ref{fig:mKaKKdataPhys}, the same phenomenon occurs
in $K^+K^+$ where the chiral expansion is governed by the strange
quark mass and is therefore expected to be more slowly converging. The
$\pi^+K^+$ scattering length (see Fig.~\ref{fig:muamu}) indicates
similar behavior but one should keep in mind that this data 
is expected to contain larger 
lattice spacing corrections. This mysterious
cancellation between chiral logs and counterterms for the meson-meson
scattering lengths begs for an explanation.

\subsection{NN Scattering}

\noindent One of the ultimate goals of nuclear physics is to compute
the properties and interactions of nuclei directly from QCD.  The
first, pioneering study of NN scattering with Lattice QCD was
performed more than a decade ago by Fukugita~{\it et
al}~\cite{Fukugita:1994ve}.  This calculation was quenched and at
relatively large pion masses, $m_\pi\gsim 550~{\rm MeV}$.  A fully
dynamical Lattice QCD calculation of the NN scattering lengths in both
the $\si$-channel and $\siii-\diii$-coupled-channels at pion masses of
$m_\pi \sim 350~{\rm MeV}$, $490~{\rm MeV}$ and $590~{\rm MeV}$ in the
isospin-limit was performed within the last few
years~\cite{Beane:2006mx}.  The dependence of the NN scattering
lengths upon the light-quark masses has been determined to various
non-trivial orders in the EFT
expansion~\cite{Beane:2002vs,Beane:2002xf,Epelbaum:2002gb}, and is
estimated to be valid up to $m_\pi\sim 350~{\rm MeV}$.  The pion mass
dependence of nuclear observables may also be explored using a lattice
formulation of the low-energy EFT of nucleons~\cite{Lee:2008fa}.  The
results of the Lattice QCD calculation at the lightest pion mass and
the experimentally-determined scattering lengths at the physical value
of the pion mass are used to constrain the chiral dependence of the
scattering lengths from $m_\pi\sim 350~{\rm MeV}$ down to the chiral
limit.

For an arbitrary nucleus (or bound and continuum nucleons), of atomic
number $A$ and charge $Z$, there are naively $(A+Z)!\ (2A-Z)!$ contractions
that must be performed to produce the nuclear correlation 
function~\footnote{The sum over all contractions can be shown~\cite{DKprivate} to scale as
$A^3$, rather than $\sim (A!)^2$.}.
Therefore, in the $\si$ channel there are $48$ contractions, while in
the $\siii-\diii$ coupled channels system there are $36$.  
The two-nucleon correlator that projects onto the s-wave state 
in the continuum limit is
\begin{eqnarray}
& & C_{NN}^{IS}(t) \ =\  
X_{\alpha\beta\sigma\rho}^{ijkl}
\nonumber\\
& & \qquad
\sum_{\bf x , y}
\langle N^\alpha_i(t,{\bf x})N^\beta_j(t, {\bf y})
N^{\sigma\dagger}_k (0, {\bf 0})N^{\rho\dagger}_l(0, {\bf 0})
\rangle
\  , 
\label{NN_correlator} 
\end{eqnarray}
where $I$ denotes the isospin of the NN system and $S$ denotes its spin,
$\alpha,\beta,\sigma,\rho$ are isospin-indices and $i,j,k,l$ are
Dirac-indices.
The tensor $X_{\alpha\beta\sigma\rho}^{ijkl}$ has elements that produce the
correct spin-isospin quantum numbers of two-nucleons in an s-wave.
The summation over ${\bf x}$ (and ${\bf y}$) corresponds to summing over all
the spatial lattice sites, thereby projecting onto 
the momentum ${\bf p}={\bf 0}$ state.
The interpolating field for the proton is
$p_i(t,{\bf x}) = \epsilon_{abc} u^a_i(t, {\bf x}) \left( u^{b T}(t, {\bf x})
  C\gamma_5 d^c(t, {\bf x})\right)$, and similarly for the neutron.

At the pion masses used in these calculations the NN scattering
lengths are found to be of natural size in both channels, and are much
smaller than the $L\sim 2.5~{\rm fm}$ lattice spatial extent.  It is
noteworthy that the scattering lengths at the heaviest pion mass 
are not inconsistent with the lightest-mass quenched values of Ref.~\cite{Fukugita:1994ve}. However,
one should keep in mind the effects of quenching on the infrared properties of the
theory~\cite{Beane:2002nu}.

\begin{figure}[!ht]
\vskip 0.25in
\centerline{\includegraphics[width=4.0in]{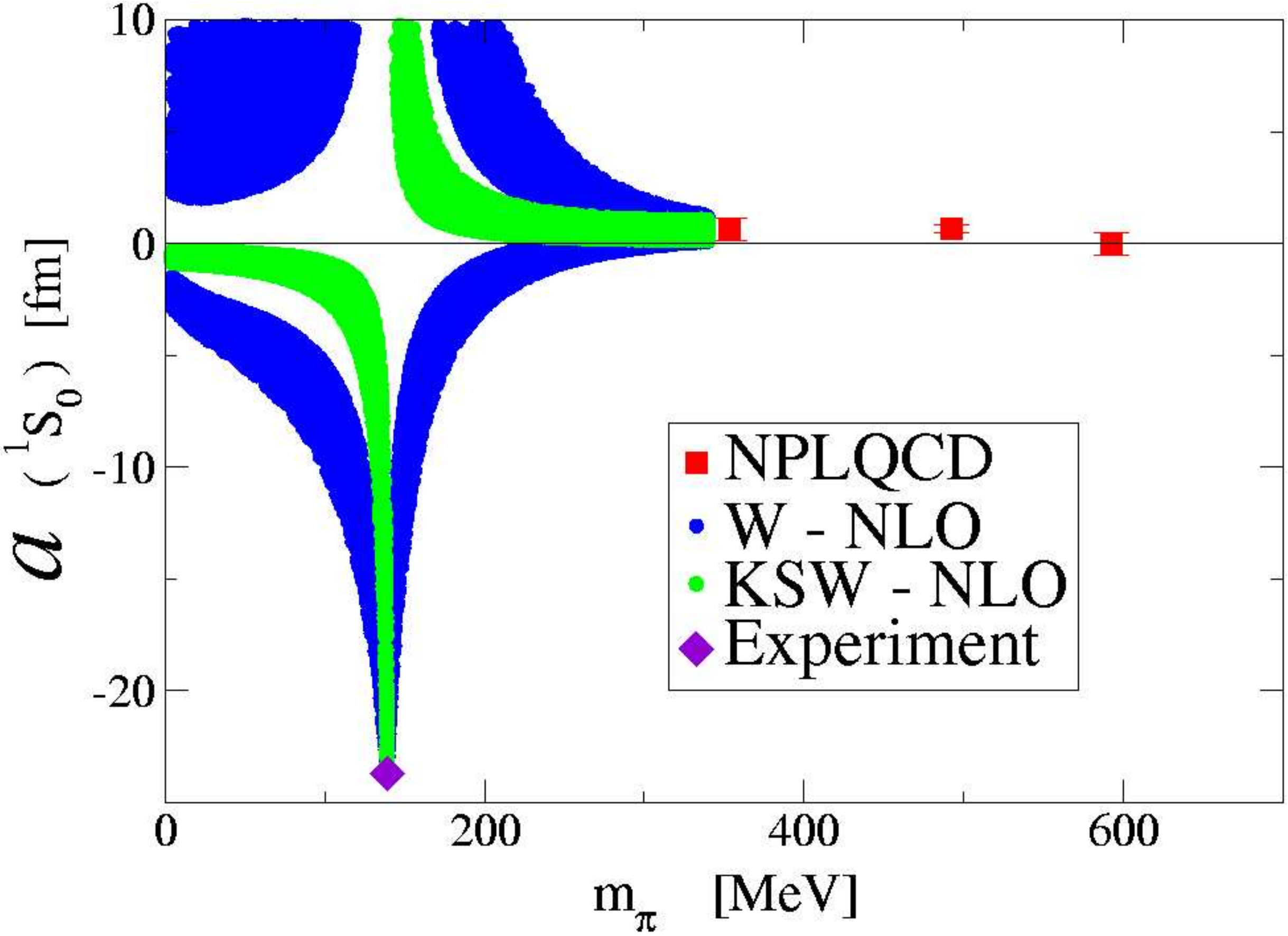}}
\noindent
\caption{Allowed regions for the scattering length in the $\si$ channel as a
  function of the pion mass. The experimental value of the scattering length 
  and NDA have been used to constrain the  extrapolation in both 
BBSvK~\protect\cite{Kaplan:1998we,Kaplan:1998tg,Beane:2001bc} and W~\protect\cite{Weinberg:1990rz,Weinberg:1991um,Ordonez:1995rz} 
power-countings at NLO.
}
\label{fig:ampi_sing}
\end{figure}

\begin{figure}[!ht]
\vskip 0.15in
\centerline{\includegraphics[width=4.0in]{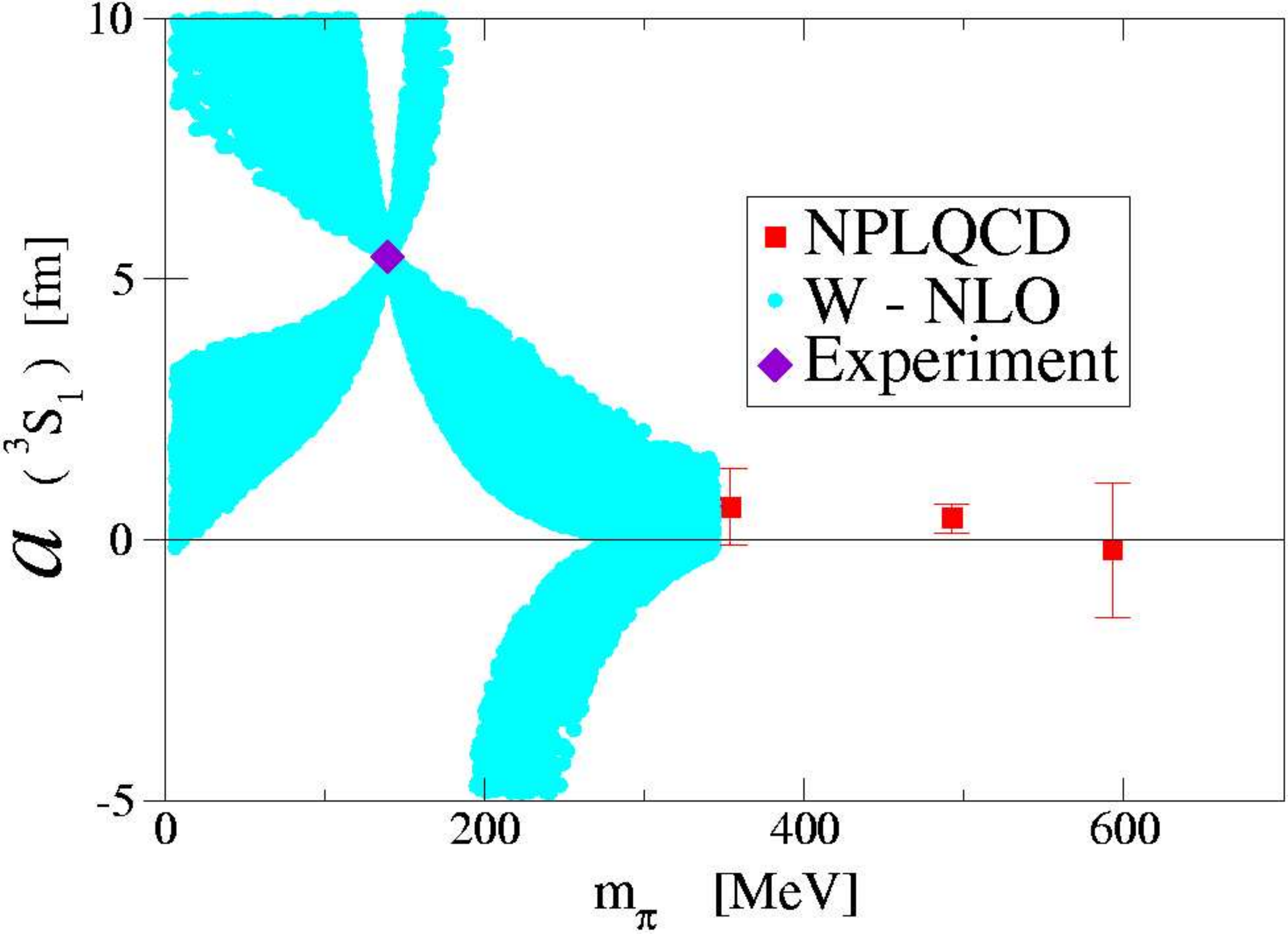}}
\noindent
\caption{Allowed regions for the scattering length in the $\siii-\diii$ 
coupled-channels as a function of the pion mass. 
The experimental value of the scattering length and NDA have been used to constrain the 
extrapolation in BBSvK~\protect\cite{Beane:2001bc} power-counting at NLO.}
\label{fig:ampi_trip}
\end{figure}

A pion mass of $m_\pi\sim 350~{\rm MeV}$ is expected to be at the
upper limit of where the EFT describing NN interactions will be
valid~\cite{Weinberg:1990rz,Weinberg:1991um,Ordonez:1995rz,Kaplan:1998we,Kaplan:1998tg,Beane:2001bc}.
The chiral extrapolation is performed with BBSvK
power-counting~\cite{Beane:2001bc} ($\equiv$KSW
power-counting~\cite{Kaplan:1998we,Kaplan:1998tg}) and W
power-counting~\cite{Weinberg:1990rz,Weinberg:1991um,Ordonez:1995rz}
in the $\si$-channel and BBSvK power-counting in the $\siii-\diii$
coupled channels.  The Lattice QCD determinations of the light-quark
axial-matrix element in the nucleon by LHPC~\cite{Edwards:2005ym} and
its physical value are used to constrain the chiral expansion of
$g_A$. The lattice calculations of the nucleon mass and pion decay
constant~\cite{Beane:2005rj} ---as well as their physical values---
are used to constrain their respective chiral expansions.  In addition
to the quark-mass dependence these three quantities contribute to the
NN systems, there is dependence on the quark masses at next-to-leading
order (NLO) from pion exchange, and from local four-nucleon operators
that involve a single insertion of the light-quark mass matrix,
described by the ``$D_2$''
coefficients~\cite{Beane:2002vs,Beane:2002xf,Epelbaum:2002gb}.  The
results of this Lattice QCD calculation constrain the range of allowed
values for the $D_2$'s, and consequently the scattering lengths in the
region between $m_\pi\sim 350~{\rm MeV}$ and the chiral limit, as
shown in Fig.~\ref{fig:ampi_sing} and Fig.~\ref{fig:ampi_trip}.  With
only one lattice point at the edge of the regime of applicability of
the EFT, a prediction for the scattering lengths at the physical pion
mass is not possible: the experimental values of the scattering
lengths are still required for an extrapolation to the chiral limit
and naive dimensional analysis (NDA) is still required to select only
those operator coefficients that are consistent with perturbation
theory. The regions plotted in the figures correspond to values of
$C_0$ -- the coefficient of the leading-order quark-mass independent
local operator -- and $D_2$ that fit the lattice datum and the
physical value, and are consistent with NDA.
In both channels the lightest lattice datum constrains the
chiral extrapolation to two distinct bands which are sensitive to both
the quark mass dependence of $g_A$ and the sign of the $D_2$
coefficient.  As the lattice point used to constrain the EFT is at the
upper limits of applicability of the EFT, we expect non-negligible
corrections to these regions from higher orders in the EFT expansion.
It is clear from Fig.~\ref{fig:ampi_sing} and Fig.~\ref{fig:ampi_trip}
that even a qualitative understanding of the chiral limit will require
lattice calculations at lighter quark masses.

Without the resources to perform similar Lattice QCD calculations in
different volumes, and observing that most energy-splitting are
positive, it is assumed that scattering states are found in each case.
Calculations in larger volumes will be done in the future to verify
the expected power-law dependence upon volume that scattering states
exhibit.  In addition to discriminating between bound and continuum
states, calculations in a larger volume would reduce the energy of the
lowest-lying continuum lattice states, and thus reduce the uncertainty
in the scattering length due to truncation of the ERE.  Further
improvement would result from measuring the energy of the first
excited state on the lattice, either with a single source or by using
the L\"uscher-Wolff~\cite{Luscher:1990ck} method.  The lattice spacing
effects in the present calculation appear at $\sim {\cal O}(b^2)$ (or
exponentially suppressed ${\cal O}(b)$ effects), and are expected to
be small.  However, the finite lattice spacing effects should be
determined by performing the same calculation on lattices with a finer
lattice spacing. A theoretical investigation of mixed-action EFT for
NN scattering which would allow a continuum extrapolation remains to
be carried through~\footnote{See Ref.~\cite{Chen:2007ug} for
preliminary work in this direction.
The masses and magnetic moments of the octet baryons have been
calculated at NLO in MA$\chi$-PT in Ref.~\cite{Tiburzi:2005is}.}.  In addition to more precise
Lattice QCD calculations through an increase in computing resources,
formal developments are also required.  In order to have a more
precise chiral extrapolation, calculations in the various relevant
EFTs must be performed beyond NLO.  Furthermore, it is clear that
lattice calculations at lower pion masses are essential for the
extrapolation to the chiral limit, and will ultimately allow for a
``prediction'' of the physical scattering lengths.

\subsection{YN Scattering}

\subsubsection{Introduction}

\noindent In high-density baryonic systems, the large value of the
Fermi energy may make it energetically advantageous for some of the
nucleons to become hyperons, with the increase in rest mass being more
than compensated for by the decrease in Fermi energy. This is
speculated to occur in the interior of neutron stars, but a
quantitative understanding of this phenomenon depends on knowledge of
the interactions among the hadrons in the medium.  In contrast to
NN interactions, where the wealth of experimental
data has allowed for the construction of high-precision potentials,
the hyperon-nucleon (YN) interactions are only very-approximately
known.  Experimental information about the YN interaction comes mainly
from the study of hypernuclei~\cite{hypernuclei-review,Hashimoto:2006aw}, the analysis
of associated $\Lambda$-kaon and $\Sigma$-kaon production in NN
collisions near threshold~\cite{Ba98,Bi98,Se99,Ko04,AB04,GHHS04}, and
hadronic atoms~\cite{Batty:1997zp}.  There are a total of 35
cross-sections measurements~\cite{NNonline} of the processes $\Lambda
p\rightarrow \Lambda p$, $\Sigma^-p\rightarrow\Lambda n$, $\Sigma^+
p\rightarrow\Sigma^+ p$, $\Sigma^- p\rightarrow\Sigma^- p$ and
$\Sigma^- p\rightarrow\Sigma^0 n$, and unsurprisingly, the extracted
scattering parameters are highly model dependent.  The theoretical
study of YN interactions is hindered by the lack of experimental
guidance.  The ``realistic'' potentials developed by the
Nijmegen~\cite{nij99,nij06} and J\"ulich~\cite{HHS89,RHKS96,HM05} groups are
just two examples of phenomenological models based on meson exchange.
These are soft-core potentials with one-boson exchange models of the
NN interaction.  Since $SU(3)$ flavor symmetry is broken by the
differences in the quark masses, the corresponding couplings are not
completely determined by the NN interaction and are instead obtained
by a fit to the available data.  In Ref.~\cite{nij99,nij06}, for example,
six different models are constructed, each describing the available YN
cross-section data equally well, but predicting different values for
the phase shifts.  The effective field theory
approach~\cite{savage-wise,KDT01,Hammer02,Beane:2003yx,PHM06} is less
developed and suffers from a large number of couplings that need to be
fit to the data.

In view of the large uncertainties in the YN scattering amplitudes and
their importance for modeling neutron stars and the study of
hypernuclei, a first-principles QCD calculation of YN scattering is
highly desirable.  The only way to achieve this is through numerical
calculations using Lattice QCD.  
Some of the present authors were part of a work~\cite{Beane:2003yx} that 
outlined a program to address this issue
with a combination of lattice calculations and the use of effective
field theories.  

We have computed the low-energy
s-wave phase shifts for YN scattering~\cite{Beane:2006gf} in the $\si$ channel and
$\siii-\diii$ coupled-channels at particular energies, using
L\"uscher's finite-volume
method~\cite{Luscher:1990ux,Hamber:1983vu,Beane:2003da}. 
No attempt was made to extrapolate to the physical pion mass as
it is likely that all but one of the data points lies outside the
regime of applicability of the YN EFTs.

An interpolating field of the form $n\times\Sigma^-$ (a simple product
of the single baryon interpolating fields) was used 
to determine the energy-eigenvalues of the s-wave
strangeness $=1$, isospin $={3\over 2}$ eigenstates in both spin
channels, and an interpolating field $n\times\Lambda$ to determine the
energy-eigenvalues of the s-wave strangeness $=1$, isospin $={1\over
2}$ eigenstates in both spin channels.

\subsubsection{Results}

\begin{figure}[!ht]
\vspace{0.2in}
\includegraphics[width = .32\textwidth,angle=0] {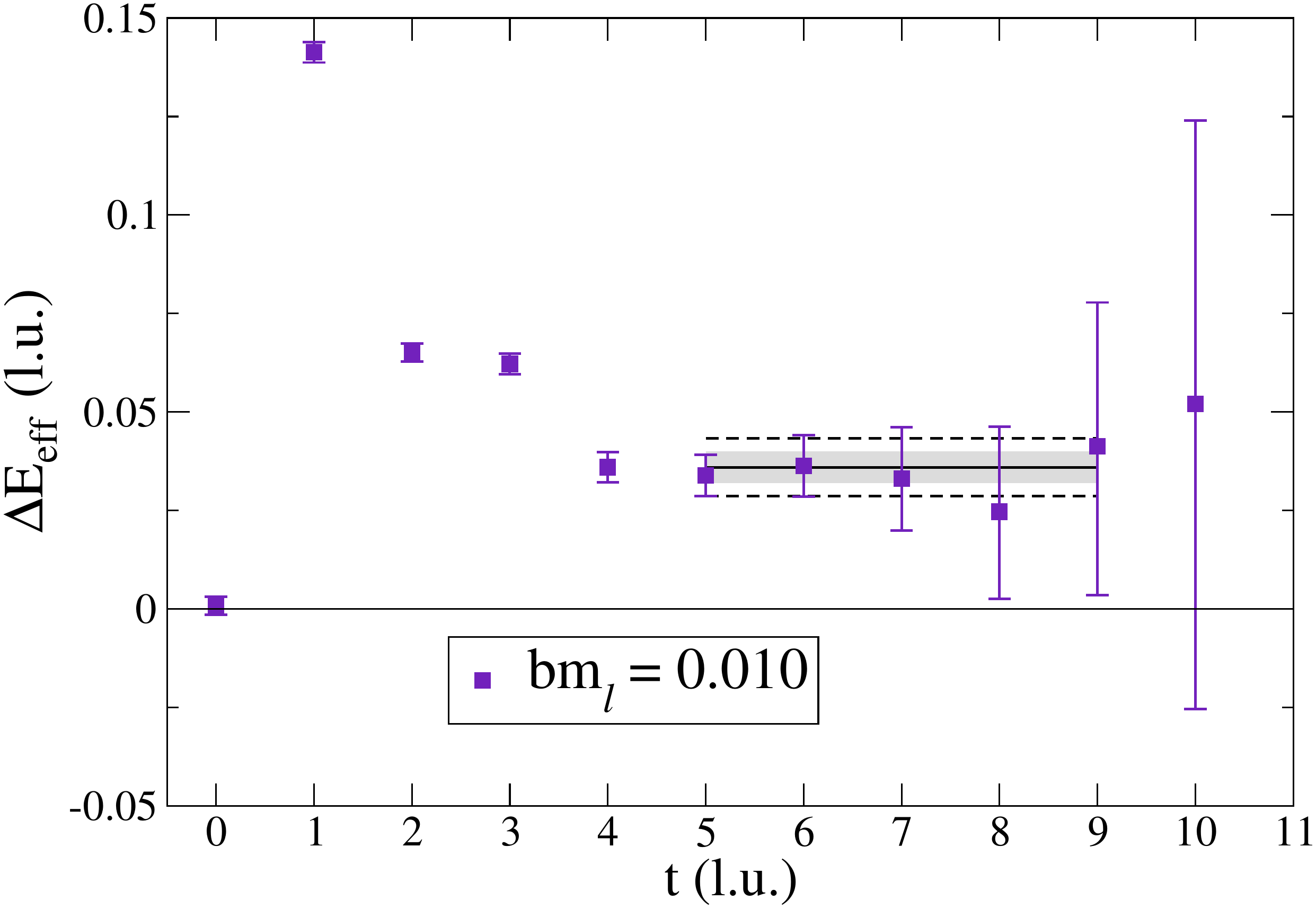} \hfill
\includegraphics[width = .32\textwidth,angle=0] {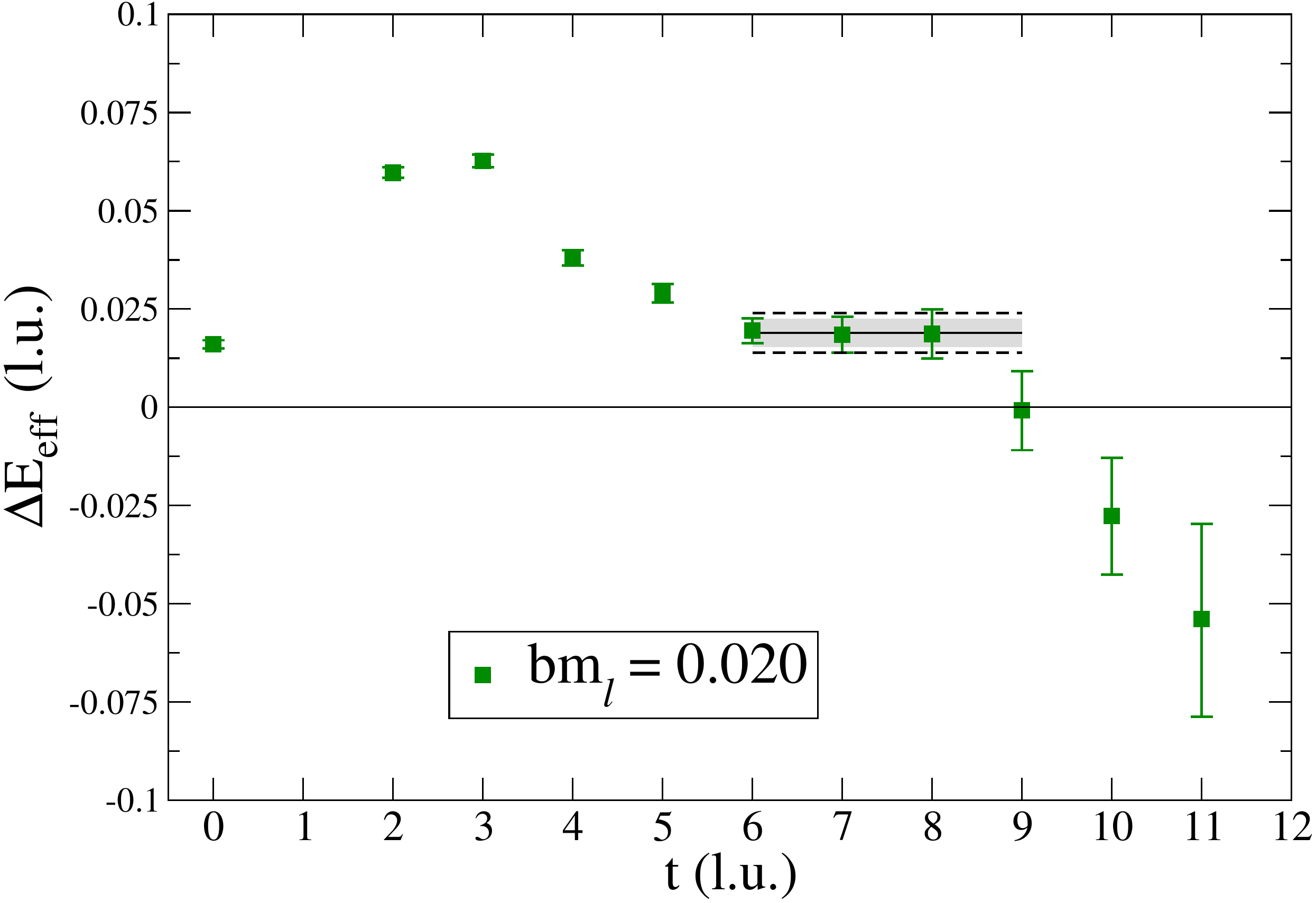} \hfill
\includegraphics[width = .32\textwidth,angle=0] {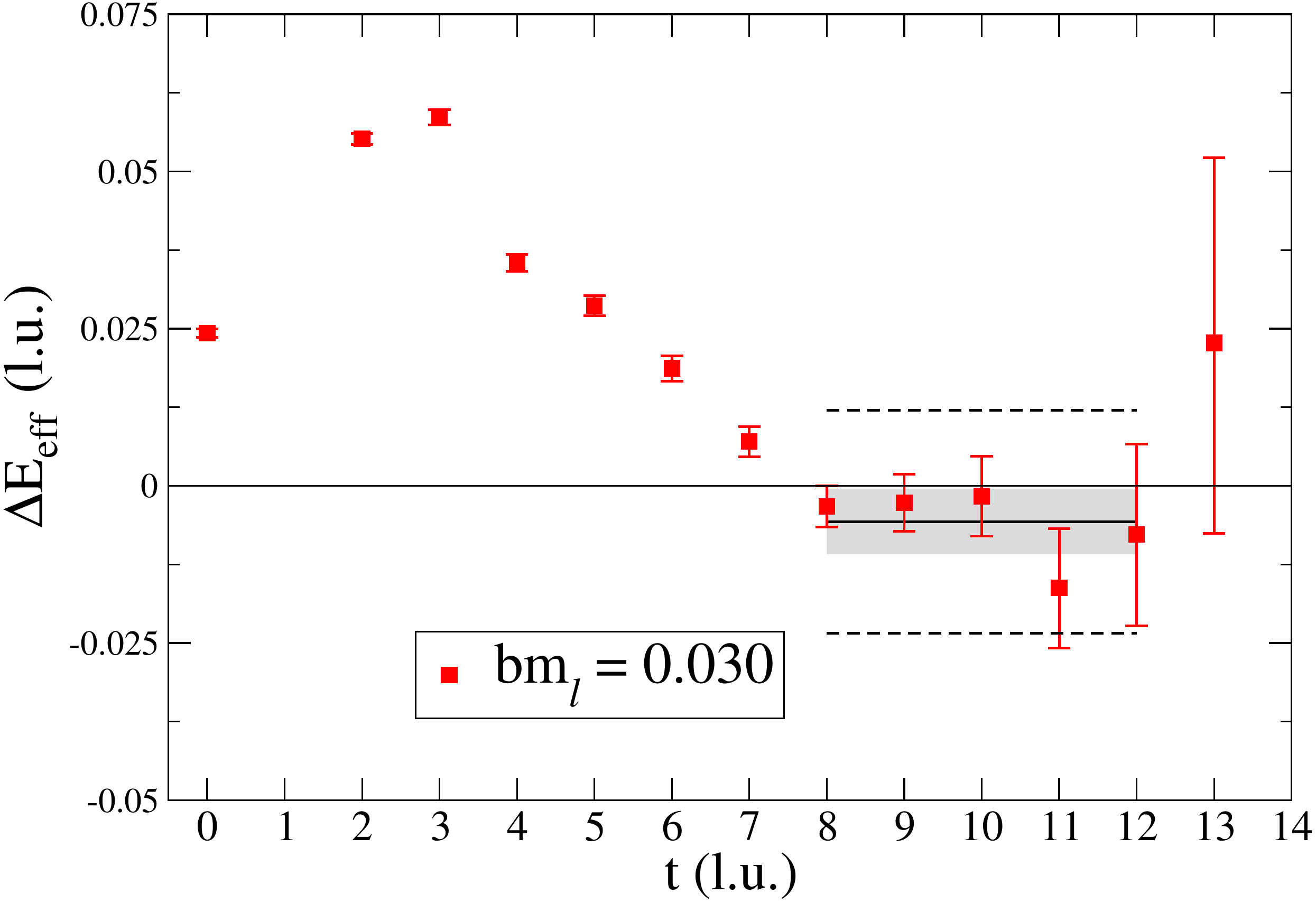}
\caption{
The effective mass plots for  $n\Lambda$ in the $\si$-channel
at pion masses of $m_\pi\sim 350~{\rm MeV}$ (left panel),
$m_\pi\sim 490~{\rm MeV}$ (center panel)
and $m_\pi\sim 590~{\rm MeV}$ (right panel).
The straight line and shaded region correspond to the extracted energy shift
and associated uncertainty. The dashed lines correspond to the statistical
and systematic errors added linearly.}
\label{fig:effmassNL1s0} 
\end{figure}

\noindent 
A typical example of the quality of the output of the lattice
calculation
can be seen  in the effective mass plots for $n\Lambda$ in the $\si$-channel, as
shown in Fig.~\ref{fig:effmassNL1s0}.
%
The plateaus in the effective energy plots persist for only a
small number of time-slices.  At small $t$ there is the usual
contamination from excited states whereas at larger $t$ the
signal-to-noise ratio degrades exponentially with $t$.  The Dirichlet boundary
at $t=22$ introduces a systematic uncertainty due to backward
propagating states.  However, in practice, the statistical noise
becomes a limiting factor at far earlier time slices and the boundary
at $t=22$ is not an issue for this calculation.  We obtained a
non-zero energy shift larger than the statistical error in ten of the
correlation functions.

It is not clear that we have been able to identify the ground states
in all of the correlation functions, e.g. $n\Sigma^-$ in the
$\si$-channel at $m_\pi\sim 490~{\rm MeV}$, and $n\Lambda$ in the
$\si$-channel at $m_\pi\sim 490~{\rm MeV}$, as the statistics are not
sufficient to determine whether the large-time behavior we observe is due
to noise or due to the presence of any states with lower energy than
those found. Indeed, it would be very exciting
if there were states with lower energy, as they would likely be bound
states (based on naturalness arguments and the exact L\"uscher
relation). 

\begin{figure}[!ht]
\vskip.1in
\includegraphics[width = .49\textwidth,angle=0] {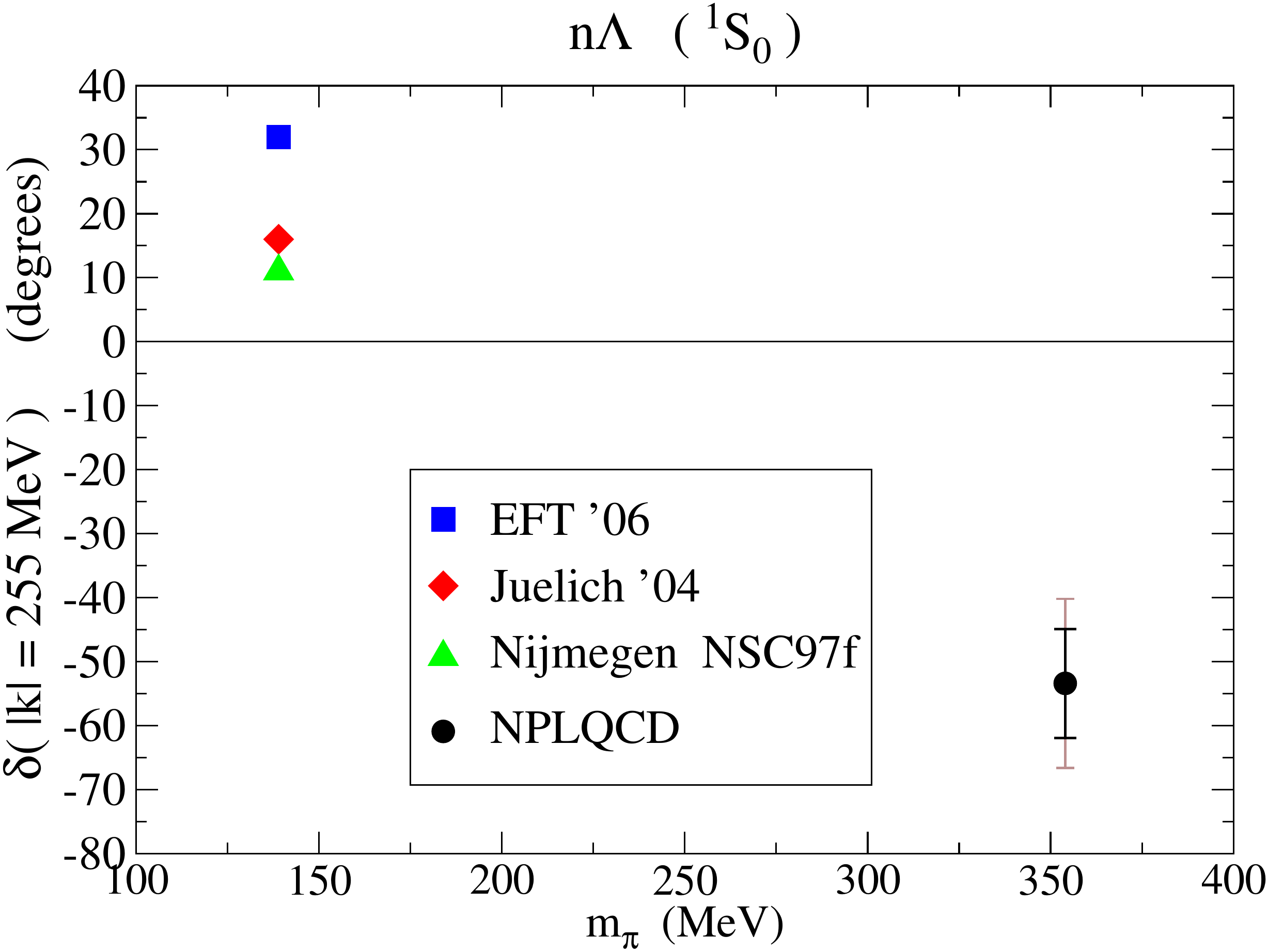} \hfill
\includegraphics[width = .49\textwidth,angle=0] {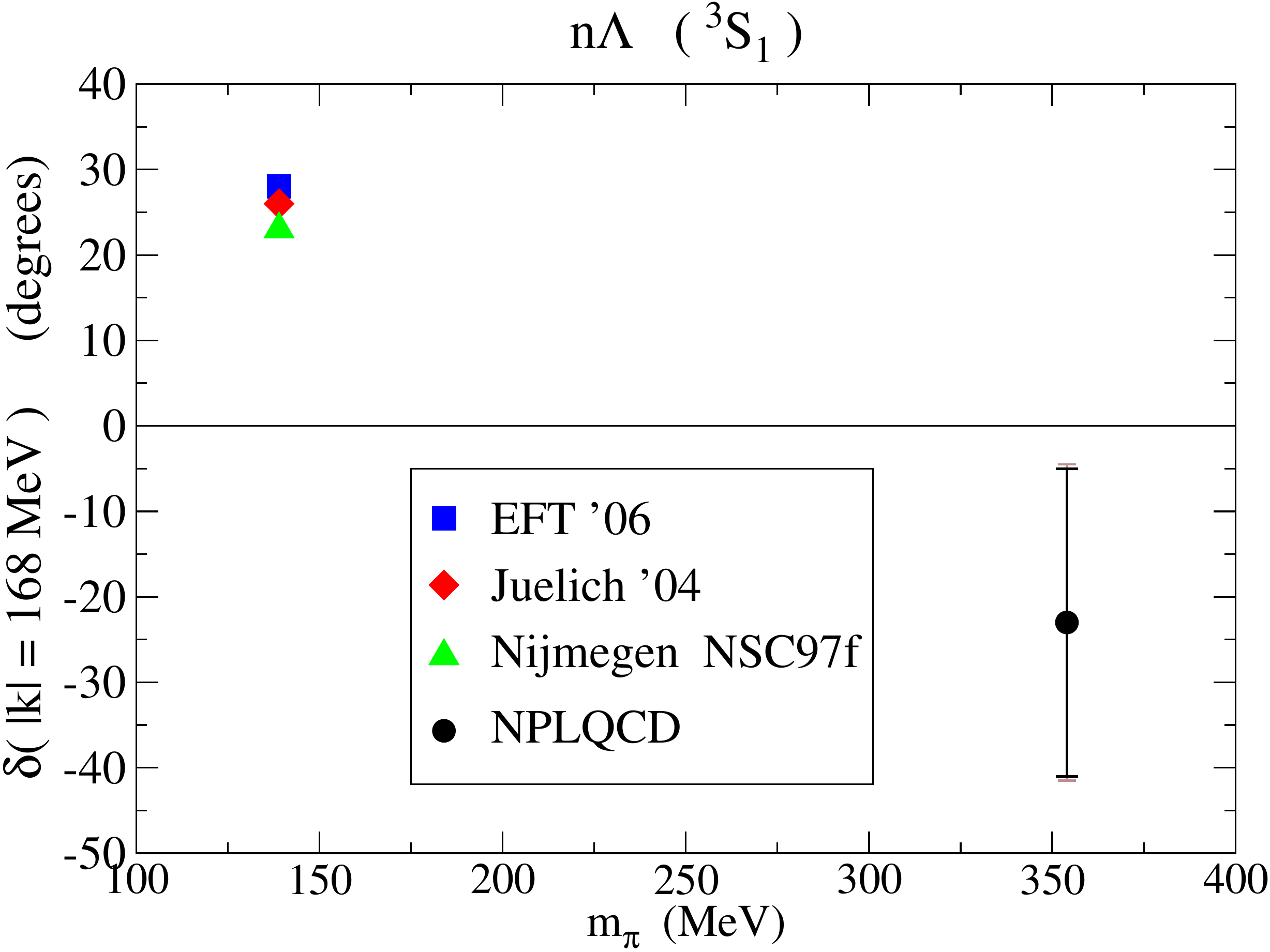}
\vskip.2in
\includegraphics[width = .49\textwidth,angle=0] {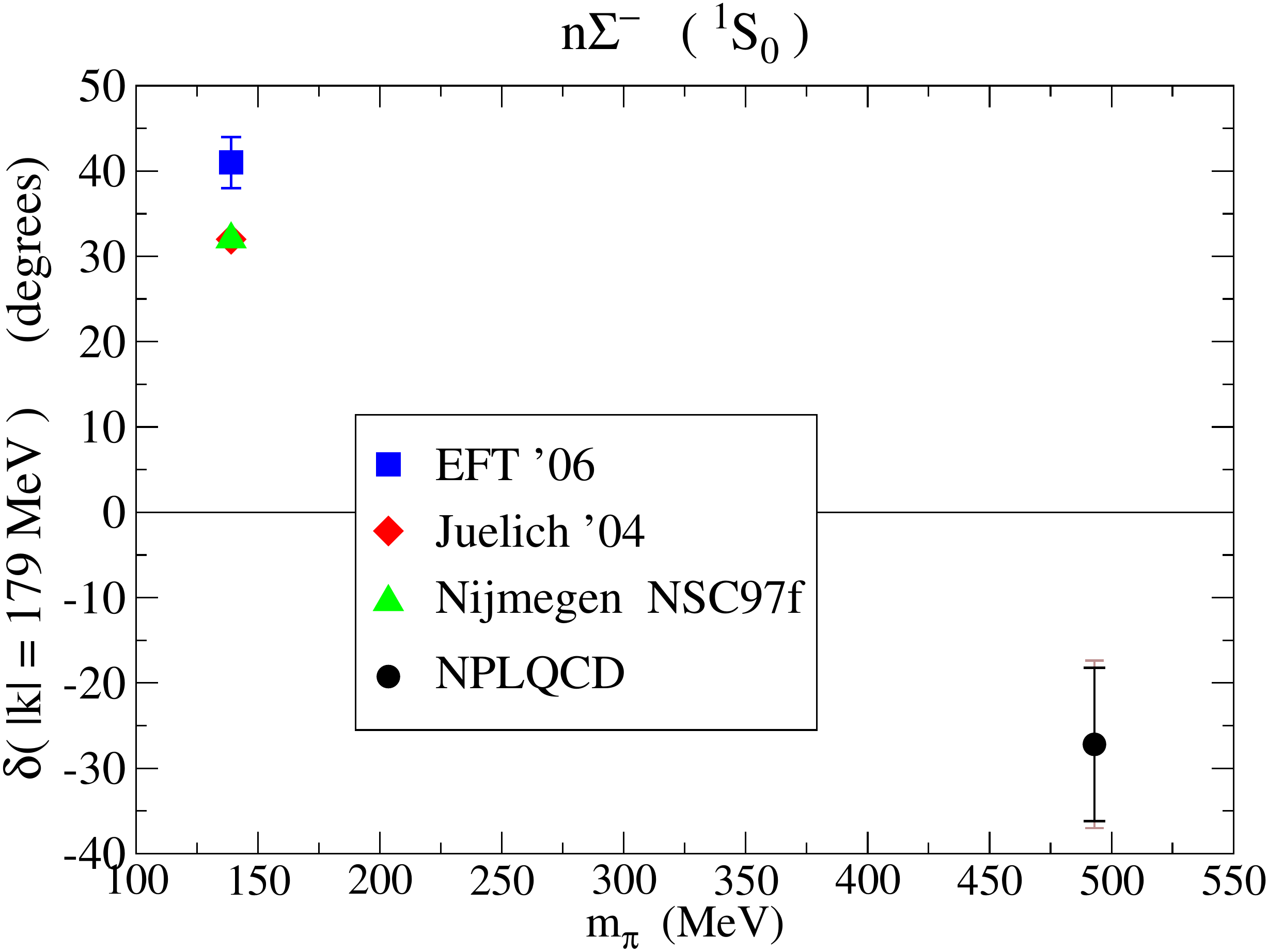} \hfill
\includegraphics[width = .49\textwidth,angle=0] {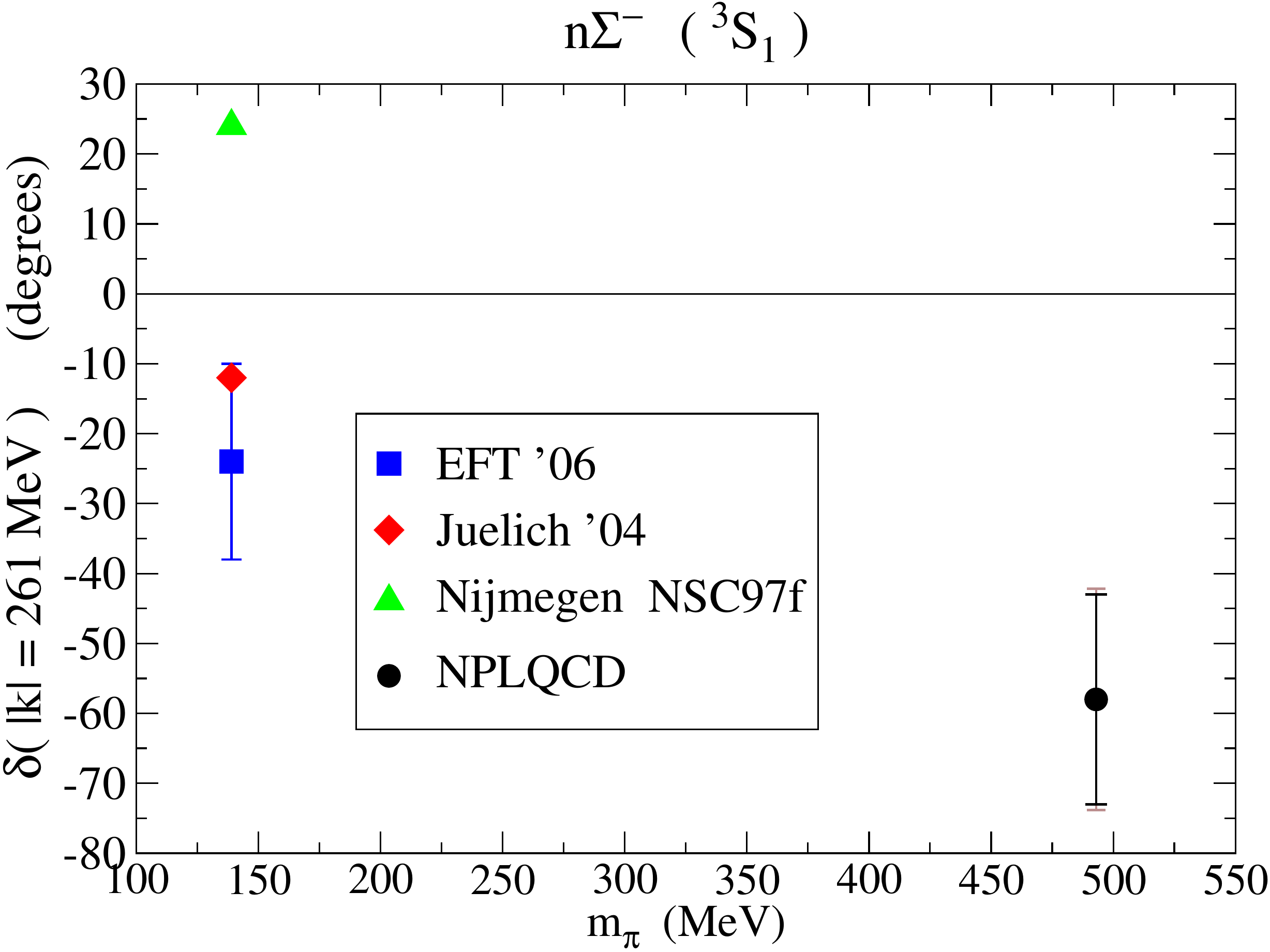}
\caption{Comparison of the lowest-pion-mass lattice results in each channel with
a recently developed YN EFT~\protect\cite{PHM06} (squares), and several potential
models: Nijmegen~\protect\cite{nij99} (triangles) and J\"ulich~\protect\cite{HM05}
(diamonds). The dark error bars on the lattice data are statistical and the light error bars
are statistical and systematic errors added in quadrature.}
\label{fig:phaseshcompare}
\end{figure}
%

\subsubsection{Discussion}

\noindent The lightest pion mass at
which a signal was extracted is at the upper limits of
the regime of applicability of the effective field theories that have
been constructed, thus precluding a chiral extrapolation.  However, this work does provide new rigorous
theoretical constraints on effective field theory, and potential model
constructions of YN interactions.  In Fig.~\ref{fig:phaseshcompare} we
compare the lattice values of the phase shifts to recent EFT
results~\cite{PHM06} (squares), and to several potential models:
Nijmegen~\cite{nij99} (triangles) and J\"ulich~\cite{HM05}. At
face value these results appear quite discrepant, however one should keep in
mind that extrapolation to the physical pion mass will seriously alter
individual contributions to the YN interaction.

While the measurements of the momenta and phase shifts are
unambiguous, their physical interpretation is not entirely resolved.
Each of the phase shifts at the lowest pion masses are negative.
Assuming that the observed state is the ground state in the lattice
volume, this implies that the interactions are all repulsive.  The
$n\Sigma^-$ interaction in the $\siii-\diii$ coupled channels is
strongly repulsive at $m_\pi\sim 490~{\rm MeV}$, while the interaction
in the $\si$-channel is only mildly repulsive.  The opposite is found
to be true for the $n\Lambda$ systems at $m_\pi\sim 350~{\rm MeV}$,
where the interaction in the $\si$-channel is found to be strongly
repulsive, while that in the $\siii-\diii$ coupled channels is mildly
repulsive.  However, there may be channels for which there exist
states of lower, negative energies, some of which may correspond to
bound states in the continuum limit.  If such states are present, then
we would conclude that the interaction is attractive, and that the
positive-shifted energy state we have identified corresponds to the
first continuum level.  Current statistics are sufficiently poor that
nothing definitive can be said about the existence of such
states. 

\subsection{Exploratory Quenched Calculations}

\subsubsection{BB Potentials}
\noindent
Energy-independent potentials can be rigorously defined and calculated
for systems composed of two (or more) hadrons containing a heavy quark
in the heavy-quark limit, $m_Q\rightarrow\infty$.  This is interesting
for more than academic reasons as the light degrees of freedom
(dof) in the B-meson have the same quantum numbers as the nucleon,
isospin-${1\over 2}$ and spin-${1\over 2}$.  As such, the EFT
describing the interactions between two B-mesons has the same form as
that describing the interactions between two nucleons, but the
counterterms that enter into each EFT are different.  Therefore, a deeper
understanding of the EFT description of nuclear physics can be gained
by Lattice QCD calculations of the potentials between B-mesons.  A
number of quenched explorations of these potentials have been
performed
previously~\cite{Michael:1999nq,Pennanen:1999xi,Green:1999mf,Fiebig:2001mr,Fiebig:2001nn,Cook:2002am,Takahashi:2006er,Doi:2006kx,Green:2004ia},
but there was little evidence for any potential simply due to limited
computational resources.  We returned to this
problem~\cite{Detmold:2007wk} and computed the potential between two
B-mesons in the four possible spin-isospin channels (neglecting
$B_d^0-\overline{B}_d^0$ mixing) in relatively small volume DBW2
lattices with $L\sim 1.6~{\rm fm}$, with a pion mass of $m_\pi\sim
403~{\rm MeV}$, and lattice-spacing of $b\sim 0.1~{\rm fm}$.  The
calculation was quenched and the naive Wilson action was used
for the quarks.  At this relatively fine lattice spacing, much finer
than previous calculations, were able to extract a non-zero potential,
but the small volume meant that the contributions to the potential
from image B-mesons (periodic BC's) were visible.

\begin{figure}[!ht]
\vspace*{4pt}
\centerline{\psfig{file=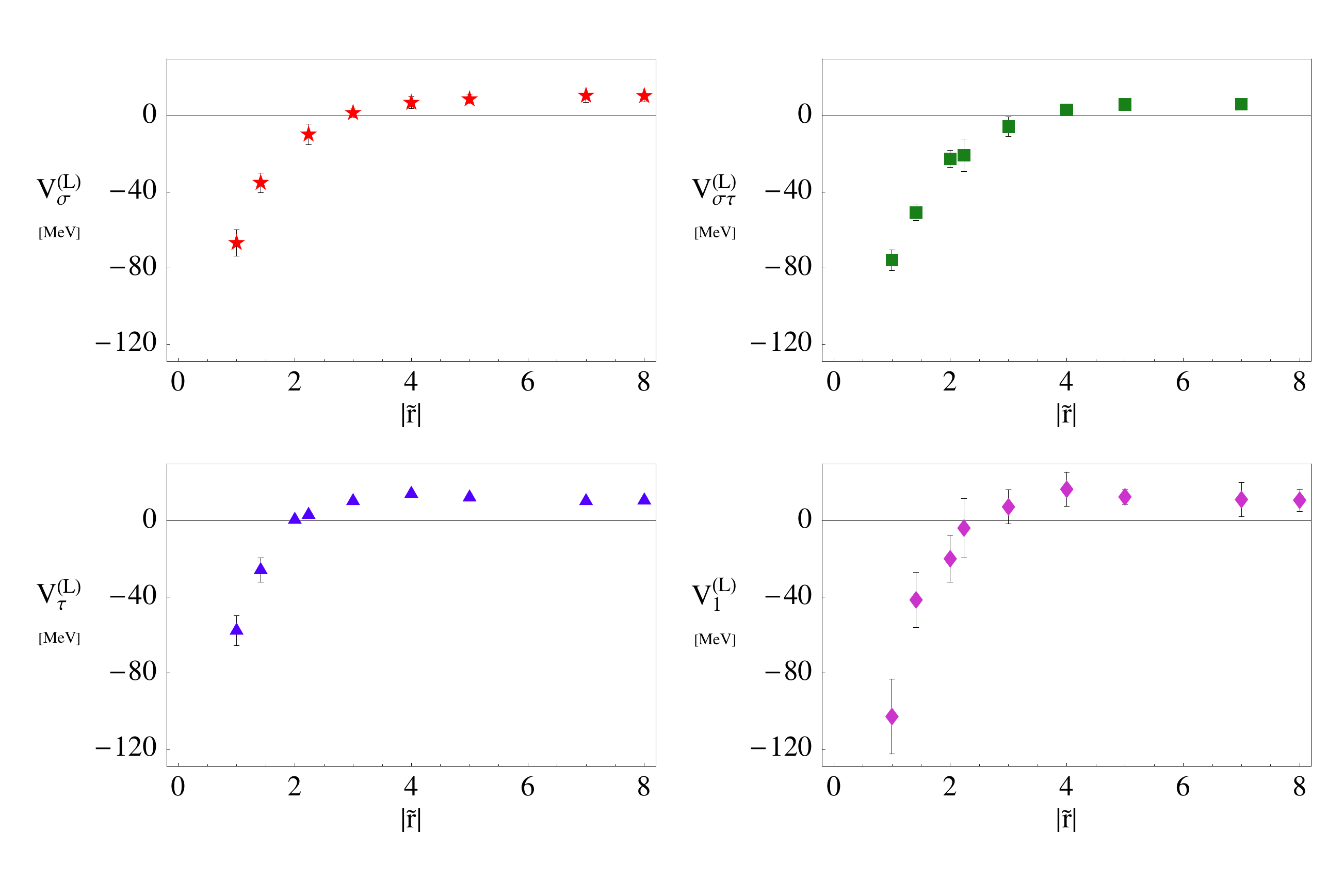,width=15.0cm}}
\vspace*{8pt}
\caption{The t-channel potentials between two B-mesons in the finite lattice volume. 
$ V_\sigma^{(L)}, V_\tau^{(L)}, V_{\sigma\tau}^{(L)}$, and $V_1^{(L)} $ 
correspond to the potentials in the exchange-channels with spin-isospin of
$(J,I) = (1,0)$, $(0,1)$, $(1,1)$ and $(0,0)$.
}
\label{fig:BBpots}
\end{figure}

Constructing the t-channel potentials, defined via the quantum
numbers of the exchange particles, in keeping with nuclear physics tradition,
isolated statistical fluctuations into the channel associated with the
``$\sigma$''-meson, leaving the channels with the quantum numbers of the $\pi$,
$\rho$ and $\omega$ with relatively small statistical errors.  The potentials
are shown in Fig.~\ref{fig:BBpots}.

Given the uncertainties in the potentials, and the number of counterterms that
appear in the EFT describing the long- and medium- distance interactions
between the B-mesons, it was possible to make only a parameterization of each
potential beyond the leading light-meson contribution.
\begin{figure}[!ht]
\vspace*{4pt}
\centerline{\psfig{file=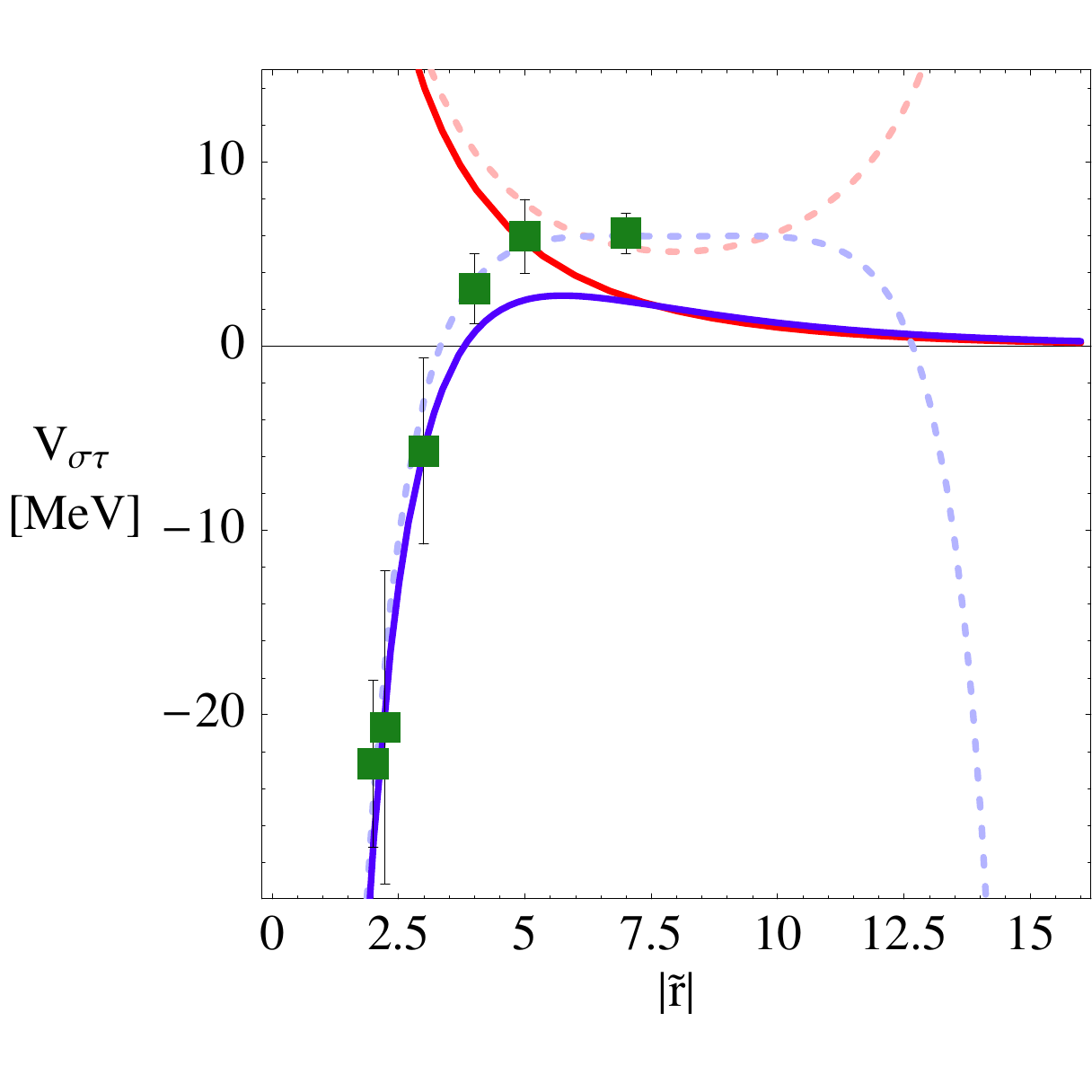,width=6.0cm}\qquad\psfig{file=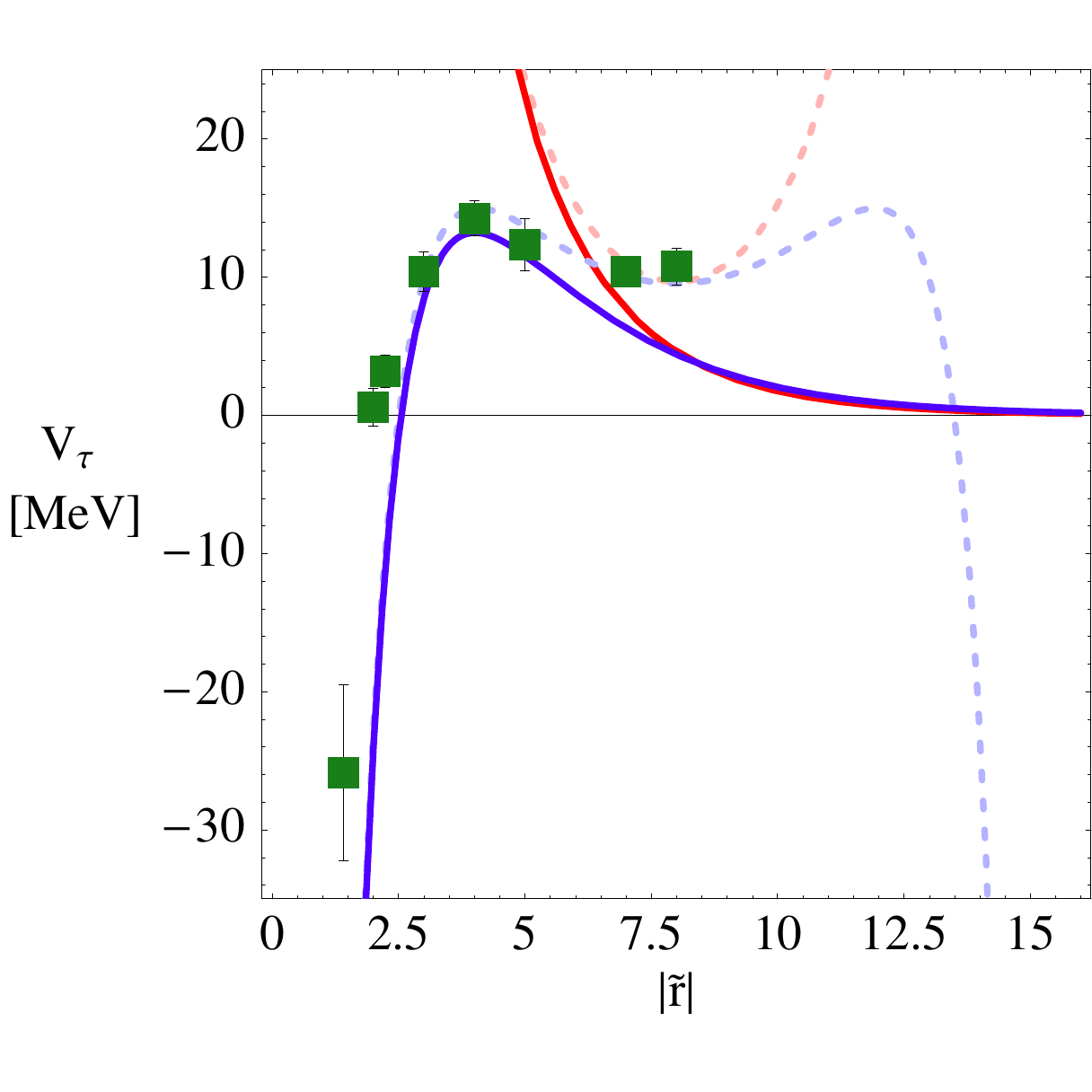,width=6.0cm}}
\vspace*{8pt}
\caption{Fits to the finite-volume isovector t-channel potentials between two
  B-mesons. The dashed lines correspond to the finite-volume fits
    to the lattice data, and the solid curves are the infinite-volume
    extrapolations.
}
\label{fig:BBpotsextrap}
\end{figure}
Since only the longest range contribution to the potential in each channel
can be identified, we fit our results at large separations, $|{\bf
  r}|>\Lambda_\chi^{-1}$, using the finite-volume versions of the simplified infinite-volume
potentials,
\begin{eqnarray}
  \label{eq:Vtausigmasimp}
  V_{\sigma\tau}^{(\infty)}({\bf r}) &\stackrel{|{\bf
      r}|\to\infty}{\longrightarrow} & \ \frac{g^2\ 
    m_\pi^2}{24\pi f_\pi^2} \ \frac{e^{-m_\pi
      |{\bf r}|}}{|{\bf r}|} \ +\ 
\alpha^\prime_\chi \ \frac{e^{-\Lambda_\chi |{\bf r}|}}{|{\bf r}|}
\ \ \ ,
\end{eqnarray}
\begin{eqnarray}
  \label{eq:Vtausimp}
  V_{\tau}^{(\infty)}({\bf r}) &\stackrel{|{\bf r}|\to\infty}{\longrightarrow} &
\ \frac{g_\rho^2}{4\pi} \ \frac{e^{-m_\rho |{\bf r}|}}{|{\bf r}|} \ +\  
\alpha_\chi \ \frac{e^{-\Lambda_\chi |{\bf r}|}}{|{\bf r}|}
\ \ \ .
\end{eqnarray}
The short distance forms in the above equations are entirely model
dependent and are the simplest forms that we could find that provide a
reasonable description of the data. Using the measured values and
uncertainties of $m_\pi$ and $m_\rho$ and the physical value of
$f_\pi$ we first determine the couplings $g$ and $g_\rho$ by setting
$\alpha_\chi=\alpha_\chi^\prime=0$ and fitting the finite-volume
potentials at the two largest separations.\footnote{Simple fits using
  the infinite-volume long range behavior were considered in
  Ref.~\cite{Michael:1999nq}.}  These fits are shown by the dashed red
curves in Fig.~\ref{fig:BBpotsextrap} and the resulting couplings are found
to be
\begin{eqnarray}
  \label{eq:8}
  g_\rho= 2.17\pm 0.08\,, &\quad\quad &
  g = 0.57\pm 0.06\,.
\end{eqnarray}
Having determined these parameters, we reconstruct
the infinite-volume potentials that are shown in the figure as the
solid red lines.  

Clearly the lattice calculations that exist of the potentials between
B-mesons must be viewed as nothing more than exploratory.  Further,
the analysis of the output of the calculations should be
model-independent in order to impact our understanding of the NN
interaction.

\subsubsection{$J/\psi$-Hadron Scattering}
\noindent
The interactions between quarkonia and the light hadrons is
interesting from both the theoretical and experimental standpoints.
Theoretically, near the heavy-quark limit, the quarkonia are compact
objects for which a multipole expansion of the chromo-electric and
magnetic fields can be performed, with the leading interactions
scaling as $r_{\overline{Q}Q}^3$, where $r_{\overline{Q}Q}$ is the
radius of the $\overline{Q}Q$ state.  This has lead to predictions for
the binding of such states to nuclear matter~\cite{Luke:1992tm} and
interactions with light nuclei~\cite{Brodsky:1997gh}.  On the
experimental side, the number and distribution of quarkonia observed
in heavy ion collisions as a function of nuclear size provides
important information on both the production and attenuation of such
states in nuclear matter under extreme conditions.  The first quenched
lattice studies of the low-energy interactions of charmonium with a
$\pi$, $\rho$ or nucleon have been recently
performed~\cite{Yokokawa:2006td}.  This work is quite encouraging, and
we look forward to fully-dynamical calculations at smaller pion
masses.

\section{n-body Interactions}

\begin{figure}[!ht]
\vspace*{4pt}
\centerline{\psfig{file=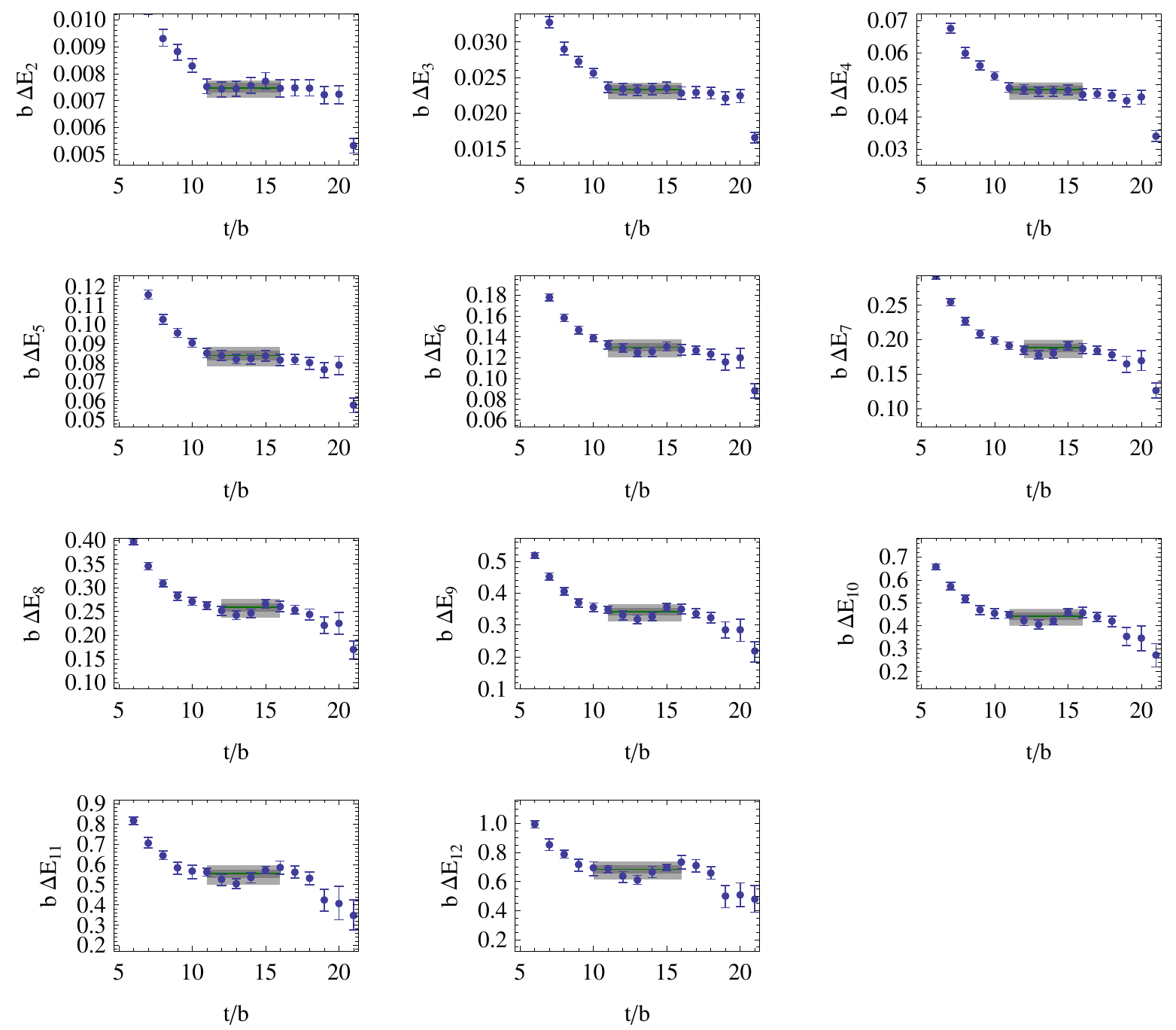,width=15.0cm}}
\vspace*{8pt}
\caption{
The effective energy plots associated with the correlation
functions of $n=2, 3,.., 12$ $\pi^+$'s with the pion rest mass removed for a
pion mass of $m_\pi\sim 290~{\rm MeV}$.
}
\label{fig:npis}
\end{figure}
\noindent 
The simplest multi-hadron systems consist of $n$ pseudoscalar mesons
of maximal isospin. Although such systems are of limited
phenomenological interest for $n>3$ (three pion interferometry is
currently a topic of interest in heavy-ion
collisions~\cite{Aggarwal:2000ex,Bearden:2001ea,Adams:2003vd}), they
serve as a testing ground for more complicated many-body systems.  We
performed the first Lattice QCD calculation of multi-hadron
($n>2$) interactions~\cite{Beane:2007es,Detmold:2008fn} by calculating
the ground-state energies of $\pi^+\pi^+$, $\pi^+\pi^+\pi^+$,
$\pi^+\pi^+\pi^+\pi^+$ up to $(\pi^+)^{12}$ in a spatial volume of
$V\sim (2.5~{\rm fm})^3$ with periodic boundary conditions.  These
systems serve as an ideal laboratory for investigating multi-particle
interactions as chiral symmetry guarantees relatively weak
interactions among pions, and multiple pion correlation functions
computed with Lattice QCD do not suffer from signal to noise issues
(as can be seen in Fig.~\ref{fig:npis}) that are expected to plague
analogous calculations in multi-baryon systems, as discussed in
section~\ref{sec:staterrors}.  The $\pi^+\pi^+$ scattering length is
extracted from the n-pion ($n>2$) systems with precision that is
comparable to (and in some cases better than) the $n=2$
determination~\cite{Beane:2007xs}.

\begin{figure}[!ht]
\vspace*{4pt}
\centerline{
\qquad\psfig{file=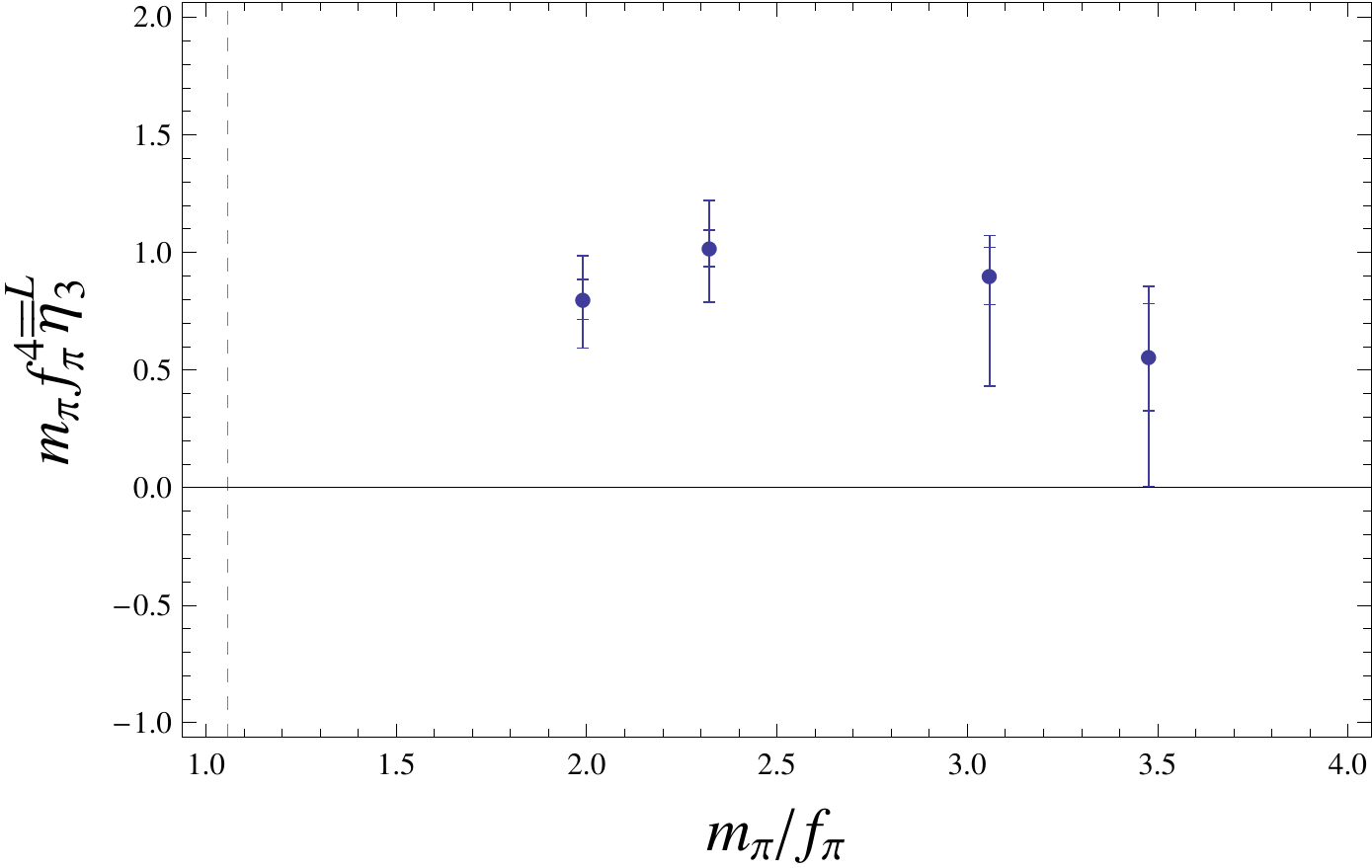,width=10.0cm}}
\vspace*{8pt}
\caption{
The $\pi^+\pi^+\pi^+$ interaction, $\overline{\overline{\eta}}_3^L$,
in units of the NDA
estimate of $1/(m_\pi\ f_\pi^4)$ as a function of $m_\pi/f_\pi$.
}
\label{fig:npisthree}
\end{figure}
Using the expression for the energy shift of $n$-$\pi^+$'s in a finite volume
due to their two- and three-body interactions, 
as given in Eq.~(\ref{eq:Lm7}), the three-$\pi^+$
interaction 
\begin{eqnarray}
\overline{\overline{\eta}}_3^L & = &  \overline{\eta}_3^L\ 
\left[ 1\ -\ 6 \ \left({a\over \pi L}\right) \ {\cal I} \ \right]
\ \ \ ,
\end{eqnarray}
is determined at four different quark masses.
The extracted values of 
$\overline{\overline{\eta}}_3^L$
in units of the NDA estimate of $1/(m_\pi\ f_\pi^4)$ are shown in
Fig.~\ref{fig:npisthree} as a function of $m_\pi/f_\pi$.
There is clear evidence for a three-$\pi^+$ interaction, and further, it is of
a magnitude that is consistent with NDA.

\begin{figure}[!ht]
\vspace*{4pt}
\centerline{
\qquad\psfig{file=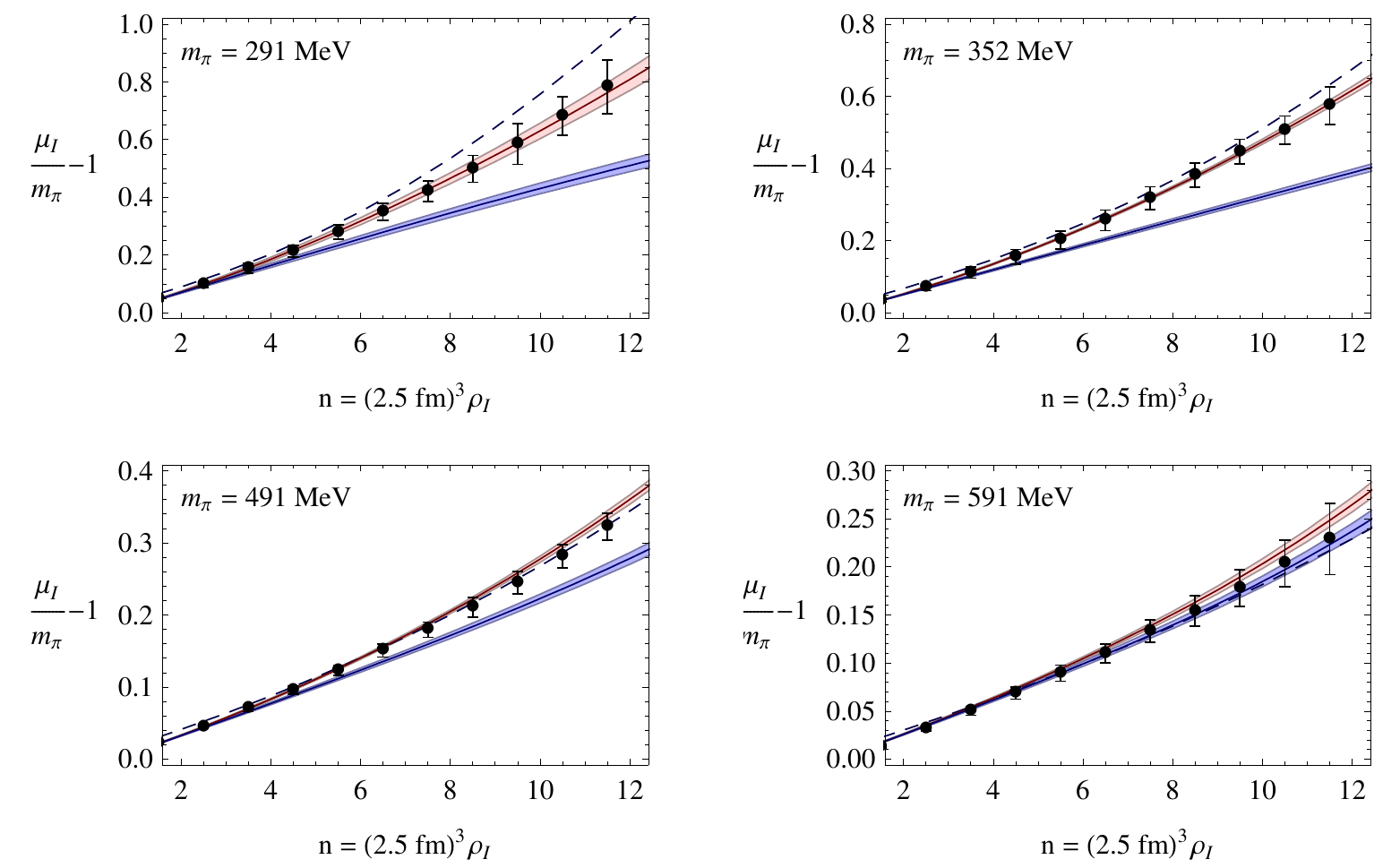,width=15.0cm}}
\vspace*{8pt}
\caption{
The isospin chemical potential as a function of the number of
    pions (equivalent to the isospin density $\rho_I$) at a fixed
    volume with and without the contribution from the
    $\pi^+\pi^+\pi^+$-interaction, \sheep.  The solid red (lighter)
    curves and bands result from the analytic expression for the
    energy of the ground state in the large volume expansion,
    Eq.~(\ref{eq:energyshift}), using the fit values for $\apipi$ and
    \sheep and their correlated uncertainties. The solid blue (darker)
    curves and bands are similarly the results for the fitted value of
    $\apipi$ and \sheep=0. The dashed curve corresponds to the leading
    order prediction of $\chi$-PT.  
}
\label{fig:chem}
\end{figure}
The study of systems with arbitrary numbers of $\pi^+$'s  in a  fixed volume enables
the study of the isospin chemical potential as a function of isospin density.
For the systems containing less than twelve $\pi^+$'s with the coarse MILC
lattices with $L\sim 2.5~{\rm fm}$, the results are shown in
Fig.~\ref{fig:chem}~\cite{Detmold:2008fn}.
This is a first step toward a  general study of multi-hadron systems.

\section{Resource Requirements for Further Progress}
\begin{figure*}
\begin{center}
\vskip 0.2in
\includegraphics[width=0.7 \textwidth,angle=0]{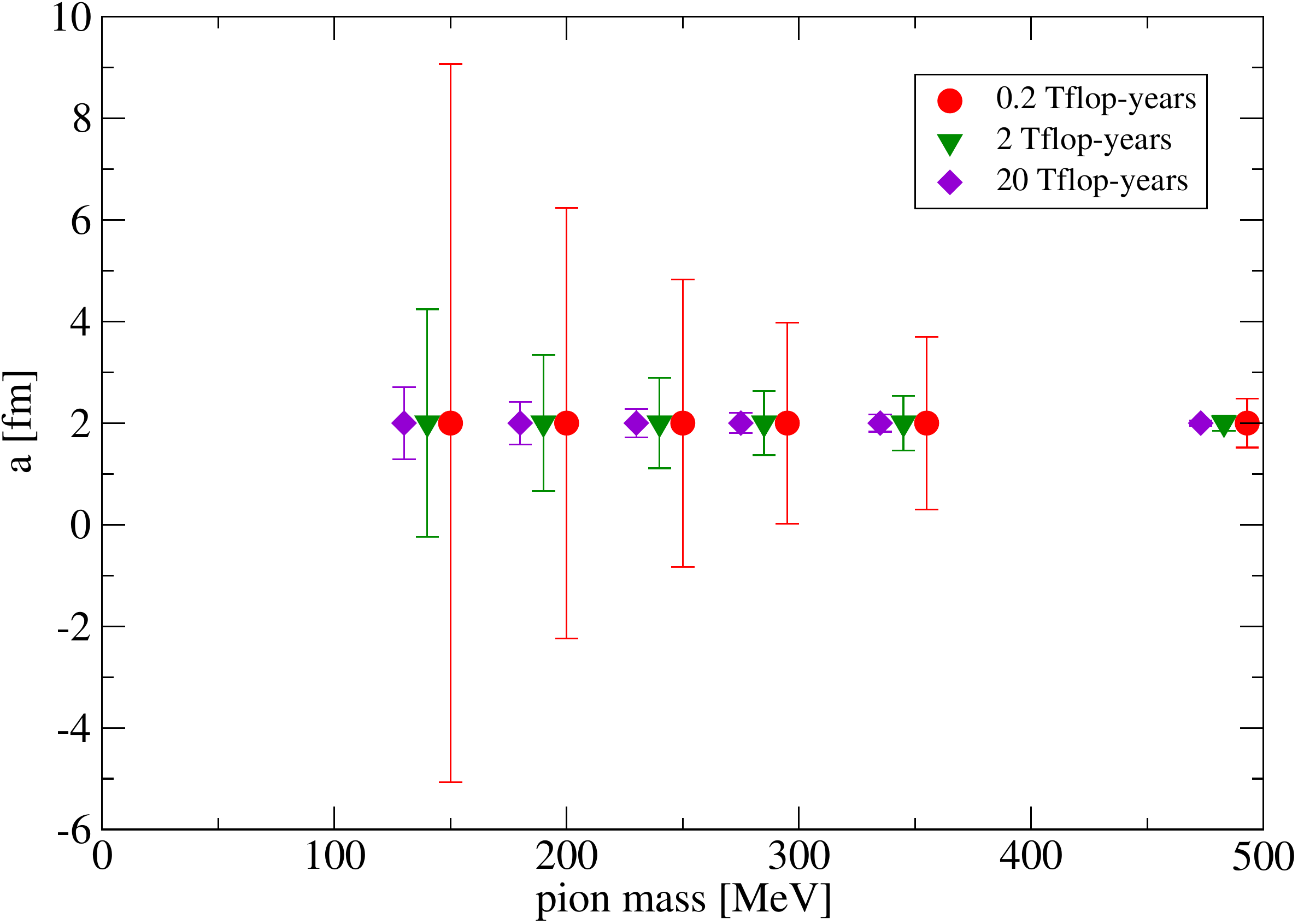}
\end{center}
\caption{ The expected uncertainty in nucleon-nucleon scattering in
the $\siii-\diii$ coupled channels for a $2~{\rm fm}$ scattering
length as a function of the pion mass for domain wall valence quarks
on the coarse MILC lattices ($b\sim 0.125~{\rm fm}$) for given
computational resources (there is a further reduction by an order
of magnitude for clover quarks in the valence sector).  Each data point requires the full
computational resource, i.e. to generate the complete set of six
triangular (green) data requires $12$~Tflop-yrs.  }
\label{fig:resources}
\end{figure*}

\noindent
To understand more concretely the resources that will be required to
compute the interactions between two or more nucleons, we determined
the uncertainty in the nucleon-nucleon scattering length, at a
scattering length of $a=2~{\rm fm}$ (independent of the pion mass) as a
function of the computational resources available to this program.
Our estimates, shown in Fig.~\ref{fig:resources}, are for computations
on lattices with a lattice spacing of $b\sim 0.125~{\rm fm}$ and a
spatial lattice extent of $L\sim 5~{\rm fm}$.  The following procedure
was used to determine the uncertainties in the scattering length:
\begin{enumerate}
\item
The results obtained by NPLQCD indicate that the uncertainty in the scattering
lengths depend (approximately) exponentially upon the pion mass.
The argument and coefficient of the exponential were fit to these results.
\item
The computational requirements for propagator generation scale as
$T_{\rm cpu} = (A + B/m_q) {\rm V}$.
$A$ and $B$ were determined by the timings on the $b\sim 0.125~{\rm fm}$,
$20^3\times 64$ MILC lattices for domain-wall propagator generation.
The lattice volume is in lattice units.
\item
The value of the energy splitting between the two nucleons 
in the volume and two isolated nucleons in the volume was 
tuned to produce a scattering length of $2~{\rm
  fm}$.
\item
The statistical uncertainty  on the projected energy-splitting
was determined by the number of propagators that could be generated
with the given computational resources.
Synthetic data was then processed using the jackknife procedure with the exact 
Luscher formula to produce the
uncertainty in the scattering length.
\item
We have not included the effect of the improved signal in going to larger
volumes, possibly a $1/\sqrt{V}$ factor, as we have no data to extrapolate from.
\end{enumerate}
The quoted resource requirements are for propagator generation ONLY,
and do not include the resource requirements for lattice generation.
Each data point requires the resource indicated in the legend of
Fig.~\ref{fig:resources}.  Therefore, the computational resources
required to generate the data indicated by the purple squares in
Fig.~\ref{fig:resources} with the precision that is shown, is
0.12~Pflop-yrs.  We assume that the lattices will be generated by
others in the USQCD collaboration, and will be made available (in a
timely fashion).  Further, these resources are for the generation of
Domain-Wall fermions.  If, instead, Clover-Wilson fermions are used on
dynamical Clover-Wilson lattices, the resource requirements for
propagator generation are reduced by a factor of $\sim 10$.
Additional factors of $\sim 5 - 8$ (included in these estimates) are 
gained using the newly developed Incremental and EigCG deflation
method~\cite{Stathopoulos:2007zi}. 
Provided that gauge field configurations become available, the
task of computing NN scattering lengths and phase-shifts
from lattice QCD 
could be accomplished with present-day computational resources.

\section{Concluding Remarks}
\label{sec:conclusions}
\noindent
Lattice QCD, when combined with effective field theory, is presently
able to provide quantitative information about the interactions among
hadrons directly from QCD.  During the last five years a number of
two-hadron processes have been investigated in fully-dynamical lattice
QCD, including both meson-meson and baryon-baryon processes.  While,
it is fair to say that the study of hadronic interactions with Lattice
QCD is still in its infancy, the computational resources that will
likely become available during the next five years will allow for a
complete study such systems.  This will amount to nothing short of a
revolution in nuclear physics.

The study of systems containing baryons is much more computationally
intensive than that of systems containing only pions, due to the
exponential degradation of the signal-to-noise in (multi-)baryon
correlators.  This is presently a serious roadblock for the direct
study of nuclei and nuclear interactions with Lattice QCD.
Further, the study of systems such as $\pi^+\pi^-\rightarrow
\pi^+\pi^-$ are presently out of reach due to the need to calculate
disconnected diagrams, which require approximately an order of
magnitude increase in computational time over the connected
contributions.

For the calculations that are of interest to the nuclear physics
community, it is important to step back and consider which lattice
calculations are the most important to perform, and what the criterion
``important'' actually means!  There may be some use in performing a
high precision calculation, say $\sim 0.1\%$-level, of a mesonic
quantity, such as $f_\pi$, which is already known to high precision
experimentally, but there may be equivalent use in performing a low
precision calculation, say at the $\sim 20\%$-level, of the weak
pion-nucleon coupling constant, $h_{\pi NN}^{(1)}$, that is expected
to provide a significant contribution to parity violating observables
in many light nuclei and for which there is still a substantial amount
of uncertainty as regards its value.  This is clearly something for
future discussions and we do not attempt to resolve this issue in this
review.

The near future will see remarkable progress in this field.  One can
expect calculations in large volumes, small lattice spacings and at
the physical quark masses, and also the inclusion of electromagnetism.
With sufficient resources dedicated to this area of Lattice QCD, the
interactions between two and three hadrons will be calculated at high
precision, with complete control over all systematic uncertainties
that arise.  At that time, calculations of processes relevant to areas
of nuclear physics, such as the interactions between hyperons and
nucleons that impact the equation of state of nuclear matter in the
interior of neutron stars, or the interactions between multiple
nucleons, will be calculable directly from QCD.  However, it is
crucial to appreciate the fact that in order for this program in
multi-baryon physics to be realized a significant amount of
computational power will need to be directed exclusively at
multi-baryon calculations.

Finally, in direct analogy with the experimental programs, it is not
sufficient to have just one calculation of any given quantity.
Lattice calculations with different discretizations, lattice volumes,
lattice spacing, performed by different lattice collaborations are
required in order to have confidence in the ``Lattice QCD'' value of
the quantity.  
This concerted effort does not yet exist for any
scattering process, and we look forward to such a coherent
effort taking shape in the near future.

\vskip 0.4in

We would like to thank Paulo Bedaque, William Detmold, Tom Luu,
Elisabetta Pallante, Assumpta Parre\~no, Aaron Torok and Andre
Walker-Loud who have all contributed to the work described in this
review. Our computations were performed at JLab, FNAL, LLNL, NCSA, and
CNdS (Barcelona).  We acknowledge DOE Grants No.~DE-FG03-97ER4014
(MJS), DE-AC05-06OR23177 (KO) and NSF CAREER Grant No. PHY-0645570
(SB).  KO acknowledges the Jeffress Memorial Trust, grant J-813 and a
DOE OJI grant DE-FG02-07ER41527.


\begin{thebibliography}{0}

\bibitem{Page:2006ud}
  D.~Page and S.~Reddy,
  Ann.\ Rev.\ Nucl.\ Part.\ Sci.\  {\bf 56}, 327 (2006)
  [arXiv:astro-ph/0608360].

\bibitem{Pieper:2007ax}
  S.~C.~Pieper,
  ``Quantum Monte Carlo Calculations of Light Nuclei,''
Lecture notes for International School of Physics 'Enrico Fermi': 
Course 169: Nuclear Structure far from Stability: New Physics and New Technology, Varenna, Italy, 17-27 Jul 2007. 
  arXiv:0711.1500 [nucl-th].

\bibitem{Wiringa:1994wb}
  R.~B.~Wiringa, V.~G.~J.~Stoks and R.~Schiavilla,
  Phys.\ Rev.\  C {\bf 51}, 38 (1995)
  [arXiv:nucl-th/9408016].


\bibitem{BarrowTipler}
{\it The Antropic Cosmological Principle}, by J.D. Barrow and F.J. Tipler,
Oxford University Press, 1986.
ISBN 0-19-282147-4.

\bibitem{Symanzik:1983gh}
  K.~Symanzik,
  Nucl.\ Phys.\  B {\bf 226}, 205 (1983).

\bibitem{Edwards:2004sx}
  R.~G.~Edwards and B.~Joo  [SciDAC Collaboration],
  ``The Chroma software system for Lattice QCD,''
  [arXiv:hep-lat/0409003].

\bibitem{McClendon}
C.~McClendon, ``Optimized Lattice QCD Kernels for a Pentium 4 Cluster'', 
JLAB-THY-01-29, 
http://www.jlab.org/~edwards/qcdapi/reports/dslash\_p4.pdf


\bibitem{CreutzBOOK}
{\it Quarks, gluons and lattices}, by M. Creutz, Cambridge Monographs on
Mathematical Physics, Cambridge University Press 1983.
ISBN 0 521 24405 6.

\bibitem{RotheBOOK}
{\it Lattice Gauge Theories An Introduction}, by H.J. Rothe, World Scientific Notes
in Physics - Vol59, World Scientific 1997.
ISBN 981-02-3032-X.

\bibitem{DeGrandDeTarBOOK}
{\it Lattice Methods for Quantum Chromodynamics}, by T DeGrand and C. DeTar,
World Scientific 2006.
ISBN 981-256-727-5.




\bibitem{Luscher:1985dn}
  M.~L\"uscher,
  Commun.\ Math.\ Phys.\  {\bf 104}, 177 (1986).

\bibitem{Luscher:1990ck}
  M.~L\"uscher and U.~Wolff,
  Nucl.\ Phys.\  B {\bf 339}, 222 (1990).


\bibitem{Maiani:1990ca}
  L.~Maiani and M.~Testa,
  Phys.\ Lett.\  B {\bf 245}, 585 (1990).


\bibitem{Huang:1957im}
  K.~Huang and C.~N.~Yang,
  Phys.\ Rev.\  {\bf 105}, 767 (1957).

\bibitem{Luscher:1986pf}
  M.~L\"uscher,
  Commun.\ Math.\ Phys.\  {\bf 105}, 153 (1986).

\bibitem{Luscher:1990ux}
  M.~L\"uscher,
  Nucl.\ Phys.\  B {\bf 354}, 531 (1991).


\bibitem{Mandula:ut}
J.E.~Mandula, G.~Zweig and J.~Govaerts,
{\it Nucl. Phys.} {\bf B228}, 91 (1983).

\bibitem{Hamber:1983vu}
  H.~W.~Hamber, E.~Marinari, G.~Parisi and C.~Rebbi,
  Nucl.\ Phys.\  B {\bf 225}, 475 (1983).

\bibitem{Beane:2003da}
  S.~R.~Beane, P.~F.~Bedaque, A.~Parre\~no and M.~J.~Savage,
  Phys.\ Lett.\  B {\bf 585}, 106 (2004)
  [arXiv:hep-lat/0312004].



\bibitem{Elizalde:1997jv}
E.~Elizalde,
{\it Commun. Math. Phys.}  {\bf 198}, 83 (1998),
{\tt hep-th/9707257}.

\bibitem{Sasaki:2006jn}
  S.~Sasaki and T.~Yamazaki,
  Phys.\ Rev.\  D {\bf 74}, 114507 (2006)
  [arXiv:hep-lat/0610081].




\bibitem{Beane:2007es}
  S.~R.~Beane, W.~Detmold, T.~C.~Luu, K.~Orginos, M.~J.~Savage and A.~Torok,
  Phys.\ Rev.\ Lett.\  {\bf 100}, 082004 (2008)
  [arXiv:0710.1827 [hep-lat]].

\bibitem{Detmold:2008fn}
  W.~Detmold, M.~J.~Savage, A.~Torok, S.~R.~Beane, T.~C.~Luu, K.~Orginos and A.~Parre\~no,
  arXiv:0803.2728 [hep-lat].



\bibitem{Beane:2007qr}
  S.~R.~Beane, W.~Detmold and M.~J.~Savage,
  Phys.\ Rev.\  D {\bf 76}, 074507 (2007)
  [arXiv:0707.1670 [hep-lat]].

\bibitem{Tan:2007bg}
  S.~Tan,
  arXiv:0709.2530 [cond-mat.stat-mech].


\bibitem{Detmold:2008gh}
  W.~Detmold and M.~J.~Savage,
  Phys.\ Rev.\  D {\bf 77}, 057502 (2008)
  [arXiv:0801.0763 [hep-lat]].



\bibitem{Bar:2002nr}
  O.~Bar, G.~Rupak and N.~Shoresh,
  Phys.\ Rev.\  D {\bf 67}, 114505 (2003)
  [arXiv:hep-lat/0210050].

\bibitem{Bar:2003mh}
  O.~Bar, G.~Rupak and N.~Shoresh,
  Phys.\ Rev.\  D {\bf 70}, 034508 (2004)
  [arXiv:hep-lat/0306021].

\bibitem{Bar:2005tu}
  O.~Bar, C.~Bernard, G.~Rupak and N.~Shoresh,
  Phys.\ Rev.\  D {\bf 72}, 054502 (2005)
  [arXiv:hep-lat/0503009].




\bibitem{Chen:2005ab}
  J.~W.~Chen, D.~O'Connell, R.~S.~Van de Water and A.~Walker-Loud,
  Phys.\ Rev.\  D {\bf 73}, 074510 (2006)
  [arXiv:hep-lat/0510024].

\bibitem{Chen:2006wf}
  J.~W.~Chen, D.~O'Connell and A.~Walker-Loud,
  Phys.\ Rev.\  D {\bf 75}, 054501 (2007)
  [arXiv:hep-lat/0611003].



\bibitem{Aubin:2004fs}
  C.~Aubin {\it et al.}  [MILC Collaboration],
  Phys.\ Rev.\ D {\bf 70}, 114501 (2004)
  [arXiv:hep-lat/0407028].

\bibitem{Gasser:1983yg}
  J.~Gasser and H.~Leutwyler,
  Annals Phys.\  {\bf 158}, 142 (1984).

\bibitem{Chen:2007ug}
  J.~W.~Chen, D.~O'Connell and A.~Walker-Loud,
  arXiv:0706.0035 [hep-lat].

\bibitem{Ishii:2006ec}
  N.~Ishii, S.~Aoki and T.~Hatsuda,
  Phys.\ Rev.\ Lett.\  {\bf 99}, 022001 (2007)
  [arXiv:nucl-th/0611096].

\bibitem{Lepage:1989hd}
  G.~P.~Lepage, `The Analysis Of Algorithms For Lattice Field Theory,''
Invited lectures given at TASI'89 Summer School, Boulder, CO, Jun 4-30, 1989.
Published in Boulder ASI 1989:97-120 (QCD161:T45:1989).





 \bibitem{Wilson:1974sk}
   K.~G.~Wilson,
   Phys.\ Rev.\  D {\bf 10}, 2445 (1974).
 

 \bibitem{Kogut:1974ag}
   J.~B.~Kogut and L.~Susskind,
   Phys.\ Rev.\  D {\bf 11}, 395 (1975).

\bibitem{Kaplan:1992bt}
  D.~B.~Kaplan,
  Phys.\ Lett.\ B {\bf 288}, 342 (1992)
  [arXiv:hep-lat/9206013].

\bibitem{Shamir:1993zy}
  Y.~Shamir,
  Nucl.\ Phys.\ B {\bf 406}, 90 (1993)
  [arXiv:hep-lat/9303005].


\bibitem{Furman:1994ky}
  V.~Furman and Y.~Shamir,
  Nucl.\ Phys.\  B {\bf 439}, 54 (1995)
  [arXiv:hep-lat/9405004].


\bibitem{Narayanan:1994gw}
  R.~Narayanan and H.~Neuberger,
  Nucl.\ Phys.\  B {\bf 443}, 305 (1995)
  [arXiv:hep-th/9411108].


 \bibitem{Neuberger:1997fp}
   H.~Neuberger,
   Phys.\ Lett.\  B {\bf 417}, 141 (1998)
   [arXiv:hep-lat/9707022].



\bibitem{Orginos:1999cr}
  K.~Orginos, D.~Toussaint and R.~L.~Sugar  [MILC Collaboration],
  Phys.\ Rev.\ D {\bf 60}, 054503 (1999)
  [arXiv:hep-lat/9903032].

\bibitem{Orginos:1998ue}
  K.~Orginos and D.~Toussaint  [MILC collaboration],
  Phys.\ Rev.\ D {\bf 59}, 014501 (1999)
  [arXiv:hep-lat/9805009].


\bibitem{Shamir:2006nj}
  Y.~Shamir,
  Phys.\ Rev.\  D {\bf 75}, 054503 (2007)
  [arXiv:hep-lat/0607007].

\bibitem{Bernard:2006ee}
  C.~Bernard, M.~Golterman and Y.~Shamir,
  Phys.\ Rev.\  D {\bf 73}, 114511 (2006)
  [arXiv:hep-lat/0604017].

\bibitem{Bernard:2007ma}
  C.~Bernard, M.~Golterman and Y.~Shamir,
  Phys.\ Rev.\  D {\bf 77}, 074505 (2008)
  [arXiv:0712.2560 [hep-lat]].

\bibitem{Bernard:2007eh}
  C.~Bernard, M.~Golterman, Y.~Shamir and S.~R.~Sharpe,
  arXiv:0711.0696 [hep-lat].


\bibitem{Durr:2006ze}
  S.~D\"urr and C.~Hoelbling,
  Phys.\ Rev.\  D {\bf 74}, 014513 (2006)
  [arXiv:hep-lat/0604005].

\bibitem{Durr:2004ta}
  S.~D\"urr and C.~Hoelbling,
  %
  Phys.\ Rev.\ D {\bf 71}, 054501 (2005)
  [arXiv:hep-lat/0411022].


\bibitem{Aubin:2003mg}
C.~Aubin and C.~Bernard,
Phys.\ Rev. D {\bf 68} (2003) 034014
[arXiv:hep-lat/0304014].

\bibitem{Aubin:2003uc}
C.~Aubin and C.~Bernard,
Phys.\ Rev. D {\bf 68} (2003) 074011
[arXiv:hep-lat/0306026].

\bibitem{Bernard:2006zw}
C.~Bernard,
Phys.\ Rev.  D {\bf 73} (2006) 114503
[arXiv:hep-lat/0603011].

\bibitem{Lee:1999zxa}
  W.~J.~Lee and S.~R.~Sharpe,
  Phys.\ Rev.\  D {\bf 60}, 114503 (1999)
  [arXiv:hep-lat/9905023].



\bibitem{Bernard:2006vv}
  C.~Bernard, M.~Golterman, Y.~Shamir and S.~R.~Sharpe,
  Phys.\ Lett.\  B {\bf 649}, 235 (2007)
  [arXiv:hep-lat/0603027].

\bibitem{Creutz:2007rk}
  M.~Creutz,
  PoS {\bf LATTICE2007}, 007 (2006)
  [arXiv:0708.1295 [hep-lat]].


\bibitem{Bernard:2001av}
  C.~W.~Bernard {\it et al.},
  Phys.\ Rev.\ D {\bf 64}, 054506 (2001)
  [arXiv:hep-lat/0104002],
  http://qcd.nersc.gov/.

\bibitem{Edwards:2006zza}
  R.~G.~Edwards {\it et al.}  [LHPC Collaboration],
  PoS {\bf LAT2006} (2006) 195.

\bibitem{Renner:2007pb}
  D.~B.~Renner {\it et al.}  [LHPC Collaboration],
  PoS {\bf LATTICE2007}, 160 (2006)
  [arXiv:0710.1373 [hep-lat]].

\bibitem{Hagler:2007xi}
  Ph.~Hagler {\it et al.}  [LHPC Collaborations],
  arXiv:0705.4295 [hep-lat].


\bibitem{Edwards:2006qx}
  R.~G.~Edwards {\it et al.},
  PoS {\bf LAT2006} (2006) 121
  [arXiv:hep-lat/0610007].


\bibitem{Edwards:2005ym}
  R.~G.~Edwards {\it et al.}  [LHPC Collaboration],
  Phys.\ Rev.\ Lett.\  {\bf 96}, 052001 (2006)
  [arXiv:hep-lat/0510062].


 \bibitem{Alford:1995hw}
   M.~G.~Alford, W.~Dimm, G.~P.~Lepage, G.~Hockney and P.~B.~Mackenzie,
   Phys.\ Lett.\  B {\bf 361}, 87 (1995)
   [arXiv:hep-lat/9507010].


 \bibitem{Toussaint:1998sa}
   D.~Toussaint and K.~Orginos  [MILC Collaboration],
   Nucl.\ Phys.\ Proc.\ Suppl.\  {\bf 73}, 909 (1999)
   [arXiv:hep-lat/9809148].



 \bibitem{Lagae:1998pe}
   J.~F.~Lagae and D.~K.~Sinclair,
   Phys.\ Rev.\  D {\bf 59}, 014511 (1999)
   [arXiv:hep-lat/9806014].


 \bibitem{Lepage:1998vj}
   G.~P.~Lepage,
   Phys.\ Rev.\  D {\bf 59}, 074502 (1999)
   [arXiv:hep-lat/9809157].

 \bibitem{Orginos:1999kg}
   K.~Orginos, R.~Sugar and D.~Toussaint,
   Nucl.\ Phys.\ Proc.\ Suppl.\  {\bf 83}, 878 (2000)
   [arXiv:hep-lat/9909087].


\bibitem{Naik:1986bn}
  S.~Naik,
  Nucl.\ Phys.\  B {\bf 316}, 238 (1989).



\bibitem{Hasenfratz:2001hp}
  A.~Hasenfratz and F.~Knechtli,
  %
  Phys.\ Rev.\ D {\bf 64}, 034504 (2001).
\bibitem{DeGrand:2002vu}
  T.~A.~DeGrand, A.~Hasenfratz and T.~G.~Kovacs,
  %
  Phys.\ Rev.\ D {\bf 67}, 054501 (2003).
\bibitem{DeGrand:2003in}
  T.~A.~DeGrand,
  %
  Phys.\ Rev.\ D {\bf 69}, 014504 (2004).

\bibitem{Durr:2004as}
  S.~D\"urr, C.~Hoelbling and U.~Wenger,
  %
  Phys.\ Rev.\ D {\bf 70}, 094502 (2004).


\bibitem{Renner:2004ck}
  D.~B.~Renner {\it et al.},
  %
  Nucl.\ Phys.\ Proc.\ Suppl.\  {\bf 140}, 255 (2005).

\bibitem{Edwards:2005kw}
  R.~G.~Edwards {\it et al.},
  %
  PoS {\bf LAT2005}, 056 (2005).

\bibitem{Foley:2005ac}
  J.~Foley, K.~Jimmy Juge, A.~O'Cais, M.~Peardon, S.~M.~Ryan and J.~I.~Skullerud,
  Comput.\ Phys.\ Commun.\  {\bf 172}, 145 (2005)
  [arXiv:hep-lat/0505023].

\bibitem{DeGrand:1990ss}
  T.~A.~DeGrand and D.~Toussaint,
{\it  Singapore, Singapore: World Scientific (1990) 750 p}

\bibitem{Beane:2005rj}
  S.~R.~Beane, P.~F.~Bedaque, K.~Orginos and M.~J.~Savage,
  Phys.\ Rev.\ D {\bf 73}, 054503 (2006).







\bibitem{Weinberg:1966kf}
  S.~Weinberg,
  Phys.\ Rev.\ Lett.\  {\bf 17}, 616 (1966).


\bibitem{Bijnens:1995yn}
  J.~Bijnens, G.~Colangelo, G.~Ecker, J.~Gasser and M.~E.~Sainio,
  Phys.\ Lett.\ B {\bf 374}, 210 (1996)
  [arXiv:hep-ph/9511397];

\bibitem{Bijnens:1997vq}
  J.~Bijnens, G.~Colangelo, G.~Ecker, J.~Gasser and M.~E.~Sainio,
  Nucl.\ Phys.\ B {\bf 508}, 263 (1997)
  [Erratum-ibid.\ B {\bf 517}, 639 (1998)]
  [arXiv:hep-ph/9707291].


\bibitem{Pislak:2001bf}
  S.~Pislak {\it et al.}  [BNL-E865 Collaboration],
  Phys.\ Rev.\ Lett.\  {\bf 87}, 221801 (2001)
  [arXiv:hep-ex/0106071].

\bibitem{Pislak:2003sv}
  S.~Pislak {\it et al.},
  Phys.\ Rev.\ D {\bf 67}, 072004 (2003)
  [arXiv:hep-ex/0301040].

\bibitem{Adeva:2005pg}
  B.~Adeva {\it et al.}  [DIRAC Collaboration],
  Phys.\ Lett.\  B {\bf 619}, 50 (2005)
  [arXiv:hep-ex/0504044].


\bibitem{Batley:2005ax}
  J.~R.~Batley {\it et al.}  [NA48/2 Collaboration],
  Phys.\ Lett.\  B {\bf 633}, 173 (2006)
  [arXiv:hep-ex/0511056].


\bibitem{Colangelo:2001df}
  G.~Colangelo, J.~Gasser and H.~Leutwyler,
  Nucl.\ Phys.\ B {\bf 603}, 125 (2001)
  [arXiv:hep-ph/0103088].


\bibitem{Leutwyler:2006qq}
  H.~Leutwyler,
  arXiv:hep-ph/0612112.


\bibitem{Roy:1971tc}
  S.~M.~Roy,
  Phys.\ Lett.\  B {\bf 36}, 353 (1971).

\bibitem{Basdevant:1973ru}
  J.~L.~Basdevant, C.~D.~Froggatt and J.~L.~Petersen,
  Nucl.\ Phys.\  B {\bf 72}, 413 (1974).

\bibitem{Ananthanarayan:2000ht}
  B.~Ananthanarayan, G.~Colangelo, J.~Gasser and H.~Leutwyler,
  Phys.\ Rept.\  {\bf 353}, 207 (2001)
  [arXiv:hep-ph/0005297].


\bibitem{Caprini:2005zr}
  I.~Caprini, G.~Colangelo and H.~Leutwyler,
  Phys.\ Rev.\ Lett.\  {\bf 96}, 132001 (2006)
  [arXiv:hep-ph/0512364].



\bibitem{Bernard:2006wx}
  C.~Bernard {\it et al.}  [MILC Collaboration],
  PoS {\bf LAT2006}, 163 (2006)
  [arXiv:hep-lat/0609053].





 
\bibitem{Sharpe:1992pp}
  S.~R.~Sharpe, R.~Gupta and G.~W.~Kilcup,
  Nucl.\ Phys.\ B {\bf 383}, 309 (1992).

\bibitem{Gupta:1993rn}
R.~Gupta, A.~Patel and S.~R.~Sharpe,
Phys.\ Rev.\ D {\bf 48}, 388 (1993)
[arXiv:hep-lat/9301016].

\bibitem{Kuramashi:1993ka}
  Y.~Kuramashi, M.~Fukugita, H.~Mino, M.~Okawa and A.~Ukawa,
  Phys.\ Rev.\ Lett.\  {\bf 71}, 2387 (1993).

\bibitem{Kuramashi:1993yu}
  Y.~Kuramashi, M.~Fukugita, H.~Mino, M.~Okawa and A.~Ukawa,
  [arXiv:hep-lat/9312016].

\bibitem{Fukugita:1994na}
  M.~Fukugita, Y.~Kuramashi, H.~Mino, M.~Okawa and A.~Ukawa,
  Phys.\ Rev.\ Lett.\  {\bf 73}, 2176 (1994)
  [arXiv:hep-lat/9407012].

\bibitem{Gattringer:2004wr}
  C.~Gattringer, D.~Hierl and R.~Pullirsch  [Bern-Graz-Regensburg
                  Collaboration],
  Nucl.\ Phys.\ Proc.\ Suppl.\  {\bf 140}, 308 (2005)
  [arXiv:hep-lat/0409064].

\bibitem{Fukugita:1994ve}
  M.~Fukugita, Y.~Kuramashi, M.~Okawa, H.~Mino and A.~Ukawa,
  Phys.\ Rev.\ D {\bf 52}, 3003 (1995)
  [arXiv:hep-lat/9501024].

\bibitem{Fiebig:1999hs}
  H.~R.~Fiebig, K.~Rabitsch, H.~Markum and A.~Mihaly,
  Few Body Syst.\  {\bf 29}, 95 (2000)
  [arXiv:hep-lat/9906002].

\bibitem{Aoki:1999pt}
  S.~Aoki {\it et al.}  [JLQCD Collaboration],
  Nucl.\ Phys.\ Proc.\ Suppl.\  {\bf 83}, 241 (2000)
  [arXiv:hep-lat/9911025].

\bibitem{Liu:2001zp}
  C.~Liu, J.~h.~Zhang, Y.~Chen and J.~P.~Ma,
  [arXiv:hep-lat/0109010].

\bibitem{Liu:2001ss}
  C.~Liu, J.~h.~Zhang, Y.~Chen and J.~P.~Ma,
  Nucl.\ Phys.\ B {\bf 624}, 360 (2002)
  [arXiv:hep-lat/0109020].

\bibitem{Aoki:2001hc}
  S.~Aoki {\it et al.}  [CP-PACS Collaboration],
  Nucl.\ Phys.\ Proc.\ Suppl.\  {\bf 106}, 230 (2002)
  [arXiv:hep-lat/0110151].

\bibitem{Aoki:2002in}
  S.~Aoki {\it et al.}  [JLQCD Collaboration],
  Phys.\ Rev.\ D {\bf 66}, 077501 (2002)
  [arXiv:hep-lat/0206011].

\bibitem{Aoki:2002sg}
  S.~Aoki {\it et al.}  [CP-PACS Collaboration],
  Nucl.\ Phys.\ Proc.\ Suppl.\  {\bf 119}, 311 (2003)
  [arXiv:hep-lat/0209056].


\bibitem{Aoki:2002ny}
  S.~Aoki {\it et al.}  [CP-PACS Collaboration],
  Phys.\ Rev.\ D {\bf 67}, 014502 (2003)
  [arXiv:hep-lat/0209124].

\bibitem{Juge:2003mr}
  K.~J.~Juge  [BGR Collaboration],
  Nucl.\ Phys.\ Proc.\ Suppl.\  {\bf 129}, 194 (2004)
  [arXiv:hep-lat/0309075].


\bibitem{Ishizuka:2003nb}
  N.~Ishizuka and T.~Yamazaki,
  Nucl.\ Phys.\ Proc.\ Suppl.\  {\bf 129}, 233 (2004)
  [arXiv:hep-lat/0309168].


\bibitem{Aoki:2005uf}
  S.~Aoki {\it et al.}  [CP-PACS Collaboration],
  Phys.\ Rev.\  D {\bf 71}, 094504 (2005)
  [arXiv:hep-lat/0503025].




\bibitem{Aoki:2004wq}
  S.~Aoki {\it et al.}  [CP-PACS Collaboration],
  Nucl.\ Phys.\ Proc.\ Suppl.\  {\bf 140}, 305 (2005)
  [arXiv:hep-lat/0409063].


\bibitem{Li:2007ey}
  X.~Li {\it et al.}  [CLQCD Collaboration],
  JHEP {\bf 0706}, 053 (2007)
  [arXiv:hep-lat/0703015].


\bibitem{Sasaki:2008sv}
  K.~Sasaki and N.~Ishizuka,
  arXiv:0804.2941 [hep-lat].



\bibitem{Yamazaki:2004qb}
  T.~Yamazaki {\it et al.}  [CP-PACS Collaboration],
  Phys.\ Rev.\ D {\bf 70}, 074513 (2004)
  [arXiv:hep-lat/0402025].


\bibitem{Beane:2007xs}
  S.~R.~Beane, T.~C.~Luu, K.~Orginos, A.~Parre\~no, M.~J.~Savage, A.~Torok and A.~Walker-Loud,
  Phys.\ Rev.\  D {\bf 77}, 014505 (2008)
  [arXiv:0706.3026 [hep-lat]].



\bibitem{Maltman:1996nw}
  K.~Maltman and C.~E.~Wolfe,
  Phys.\ Lett.\  B {\bf 393}, 19 (1997)
  [Erratum-ibid.\  B {\bf 424}, 413 (1998)]
  [arXiv:nucl-th/9610051].

\bibitem{Meissner:1997fa}
  U.~G.~Meissner, G.~Muller and S.~Steininger,
  Phys.\ Lett.\  B {\bf 406}, 154 (1997)
  [Erratum-ibid.\  B {\bf 407}, 454 (1997)]
  [arXiv:hep-ph/9704377].

\bibitem{Knecht:1997jw}
  M.~Knecht and R.~Urech,
  Nucl.\ Phys.\  B {\bf 519}, 329 (1998)
  [arXiv:hep-ph/9709348].


\bibitem{Knecht:2002gz}
  M.~Knecht and A.~Nehme,
  Phys.\ Lett.\  B {\bf 532}, 55 (2002)
  [arXiv:hep-ph/0201033].

\bibitem{BD_kaon2007} 
B.~Bloch-Devaux, 
{\it Recent results from NA48/2 on Ke4 decays and interpretation in term of
  $\pi\pi$ scattering lengths}, Talk presented at Kaon 2007, May 21-25 (2007).

\bibitem{JG_kaon2007} 
J.~Gasser,
{\it Theoretical progress on cusp effect and Kl4}, 
 Talk presented at Kaon 2007, May 21-25 (2007).




\bibitem{DIRACprops}
http://dirac.web.cern.ch/DIRAC/future.html


\bibitem{Buettiker:2003pp}
  P.~Buettiker, S.~Descotes-Genon and B.~Moussallam,
  %
  Eur.\ Phys.\ J.\ C {\bf 33}, 409 (2004)
  [arXiv:hep-ph/0310283].

\bibitem{Ananthanarayan:2001uy}
  B.~Ananthanarayan, P.~Buettiker and B.~Moussallam,
  Eur.\ Phys.\ J.\  C {\bf 22}, 133 (2001)
  [arXiv:hep-ph/0106230].

\bibitem{Ananthanarayan:2000cp}
  B.~Ananthanarayan and P.~Buettiker,
  Eur.\ Phys.\ J.\  C {\bf 19}, 517 (2001)
  [arXiv:hep-ph/0012023].


\bibitem{Jamin:2000wn}
  M.~Jamin, J.~A.~Oller and A.~Pich,
  Nucl.\ Phys.\ B {\bf 587}, 331 (2000)
  [arXiv:hep-ph/0006045].

\bibitem{Bernard:1990kw}
  V.~Bernard, N.~Kaiser and U.~G.~Meissner,
  Nucl.\ Phys.\ B {\bf 357}, 129 (1991).


\bibitem{Bernard:1990kx}
  V.~Bernard, N.~Kaiser and U.~G.~Meissner,
  Phys.\ Rev.\ D {\bf 43}, 2757 (1991).

\bibitem{Kubis:2001bx}
  B.~Kubis and U.~G.~Meissner,
  %
  Phys.\ Lett.\ B {\bf 529}, 69 (2002)
  [arXiv:hep-ph/0112154].

\bibitem{Bijnens:2004bu}
  J.~Bijnens, P.~Dhonte and P.~Talavera,
  JHEP {\bf 0405}, 036 (2004)
  [arXiv:hep-ph/0404150].

\bibitem{Miao:2004gy}
  C.~Miao, X.~i.~Du, G.~w.~Meng and C.~Liu,
  Phys.\ Lett.\ B {\bf 595}, 400 (2004)
  [arXiv:hep-lat/0403028].


\bibitem{DescotesGenon:2006uk}
  S.~Descotes-Genon and B.~Moussallam,
  Eur.\ Phys.\ J.\  C {\bf 48}, 553 (2006)
  [arXiv:hep-ph/0607133].



\bibitem{Beane:2006gj}
  S.~R.~Beane, P.~F.~Bedaque, T.~C.~Luu, K.~Orginos, E.~Pallante, A.~Parre\~no and M.~J.~Savage,
  Phys.\ Rev.\  D {\bf 74}, 114503 (2006)
  [arXiv:hep-lat/0607036].

\bibitem{Gasser:1984gg}
  J.~Gasser and H.~Leutwyler,
  %
  Nucl.\ Phys.\ B {\bf 250}, 465 (1985).

\bibitem{Gasser:1983ky}
  J.~Gasser and H.~Leutwyler,
  %
  Phys.\ Lett.\ B {\bf 125}, 321 (1983).

\bibitem{Beane:2006kx}
  S.~R.~Beane, P.~F.~Bedaque, K.~Orginos and M.~J.~Savage,
  Phys.\ Rev.\  D {\bf 75}, 094501 (2007)
  [arXiv:hep-lat/0606023].

\bibitem{Roessl:1999iu}
  A.~Roessl,
  Nucl.\ Phys.\ B {\bf 555}, 507 (1999)
  [arXiv:hep-ph/9904230].

\bibitem{Schweizer:2005nn}
  J.~Schweizer,
  Phys.\ Lett.\ B {\bf 625}, 217 (2005)
  [arXiv:hep-ph/0507323].

\bibitem{donal}
D.~O'Connell, ``Ginsparg-Wilson Meson Scattering on a Staggered Sea,''
{\it talk at} LATTICE 2006, Tucson, Arizona.

\bibitem{Orginos:2007tw}
  K.~Orginos and A.~Walker-Loud,
  arXiv:0705.0572 [hep-lat].

\bibitem{Nagata:2007zz}
  J.~Nagata, A.~Nakamura and S.~Muroya,
  Nucl.\ Phys.\  A {\bf 790} (2007) 414.

\bibitem{Flynn:2007ki}
  J.~M.~Flynn and J.~Nieves,
  Phys.\ Rev.\  D {\bf 75}, 074024 (2007)
  [arXiv:hep-ph/0703047].


\bibitem{Flynn:2007rs}
  J.~M.~Flynn and J.~Nieves,
  PoS {\bf LAT2007}, 352 (2007)
  [arXiv:0711.3339 [hep-lat]].





\bibitem{Kaplan:1986yq}
  D.~B.~Kaplan and A.~E.~Nelson,
  Phys.\ Lett.\  B {\bf 175}, 57 (1986).



\bibitem{Cramer:2004ih}
  J.~G.~Cramer, G.~A.~Miller, J.~M.~S.~Wu and J.~H.~S.~Yoon,
  Phys.\ Rev.\ Lett.\  {\bf 94}, 102302 (2005)
  [arXiv:nucl-th/0411031].

\bibitem{Miller:2005ji}
  G.~A.~Miller and J.~G.~Cramer,
  J.\ Phys.\ G {\bf 34}, 703 (2007)
  [arXiv:nucl-th/0507004].

\bibitem{Miller:2007gh}
  G.~A.~Miller and J.~G.~Cramer,
  Nucl.\ Phys.\  A {\bf 782} (2007) 251.



\bibitem{Abelev:2006gu}
  B.~I.~Abelev {\it et al.}  [STAR Collaboration],
  Phys.\ Rev.\  C {\bf 74}, 054902 (2006)
  [arXiv:nucl-ex/0608012].

\bibitem{Beane:2007uh}
  S.~R.~Beane, T.~C.~Luu, K.~Orginos, A.~Parre\~no, M.~J.~Savage, A.~Torok and A.~Walker-Loud
                  [NPLQCD Collaboration],
  arXiv:0709.1169 [hep-lat].

\bibitem{Beane:2006mx}
  S.~R.~Beane, P.~F.~Bedaque, K.~Orginos and M.~J.~Savage,
  Phys.\ Rev.\ Lett.\  {\bf 97}, 012001 (2006)
  [arXiv:hep-lat/0602010].


\bibitem{Beane:2002vs}
  S.~R.~Beane and M.~J.~Savage,
  Nucl.\ Phys.\ A {\bf 713}, 148 (2003)
  [arXiv:hep-ph/0206113].

\bibitem{Beane:2002xf}
  S.~R.~Beane and M.~J.~Savage,
  Nucl.\ Phys.\ A {\bf 717}, 91 (2003)
  [arXiv:nucl-th/0208021].

\bibitem{Epelbaum:2002gb}
  E.~Epelbaum, U.~G.~Meissner and W.~Gloeckle,
  Nucl.\ Phys.\ A {\bf 714}, 535 (2003)
  [arXiv:nucl-th/0207089].

\bibitem{Lee:2008fa}
  D.~Lee,
  arXiv:0804.3501 [nucl-th].

\bibitem{DKprivate}
D.B.  Kaplan, 
{\it The OFT}, talk presented at the 
workshop on {\bf Domain Wall Fermions at 10 Years},
Brookhaven National Laboratory
March 15-18, 2007.
{\tt http://www.phys.washington.edu/users/dbkaplan/kaplan\_talks.html}\ .

\bibitem{Beane:2002nu}
  S.~R.~Beane and M.~J.~Savage,
  Phys.\ Lett.\ B {\bf 535}, 177 (2002)
  [arXiv:hep-lat/0202013].

\bibitem{Weinberg:1990rz}
  S.~Weinberg,
  Phys.\ Lett.\ B {\bf 251}, 288 (1990).

\bibitem{Weinberg:1991um}
  S.~Weinberg,
  Nucl.\ Phys.\ B {\bf 363}, 3 (1991).

\bibitem{Ordonez:1995rz}
  C.~Ordo\~nez, L.~Ray and U.~van Kolck,
  Phys.\ Rev.\ C {\bf 53}, 2086 (1996)
  [arXiv:hep-ph/9511380].

\bibitem{Kaplan:1998we}
  D.~B.~Kaplan, M.~J.~Savage and M.~B.~Wise,
  Nucl.\ Phys.\ B {\bf 534}, 329 (1998)
  [arXiv:nucl-th/9802075].

\bibitem{Kaplan:1998tg}
  D.~B.~Kaplan, M.~J.~Savage and M.~B.~Wise,
  Phys.\ Lett.\ B {\bf 424}, 390 (1998)
  [arXiv:nucl-th/9801034].

\bibitem{Beane:2001bc}
  S.~R.~Beane, P.~F.~Bedaque, M.~J.~Savage and U.~van Kolck,
  Nucl.\ Phys.\ A {\bf 700}, 377 (2002)
  [arXiv:nucl-th/0104030].

\bibitem{Tiburzi:2005is}
B.~C.~Tiburzi,
Phys.\ Rev.\ D {\bf 72}, 094501 (2005)
[arXiv:hep-lat/0508019].


\bibitem{hypernuclei-review}
A. Gal, E. Hungerford, Nucl. Phys. {\bf A 754} 1-489 (2005).

\bibitem{Hashimoto:2006aw}
  O.~Hashimoto and H.~Tamura,
  Prog.\ Part.\ Nucl.\ Phys.\  {\bf 57}, 564 (2006).

\bibitem{Ba98}
J. Balewski et al., Phys. Lett. B {\bf 420}, 211 (1998).

\bibitem{Bi98}
R. Bilger et al., Phys. Lett. B {\bf 420}, 217 (1998).

\bibitem{Se99}
S. Sewerin et al., Phys. Rev. Lett. {\bf 83}, 682 (1999).

\bibitem{Ko04}
P. Kowina et al., Eur. Phys. j. {\bf A22}, 293 (2004).


\bibitem{AB04}
M. Abdel-Bary et al., Phys. Lett. B {\bf 595}, 127 (2004).

\bibitem{GHHS04}
A. Gasparyan, J. Haidenbauer, C. Hanhart, J. Speth, Phys. Rev. C {\bf 69}, 
034006 (2004).


\bibitem{Batty:1997zp}
  C.~J.~Batty, E.~Friedman and A.~Gal,
  Phys.\ Rept.\  {\bf 287} (1997) 385.

\bibitem{NNonline}
{\tt http://nn-online.sci.kun.nl/index.html}.

\bibitem{nij99}
V.G.J. Stoks and Th.A. Rijken, Phys. Rev. C {\bf 59}, 3009 (1999);
Th.A. Rijken, V.G.J. Stoks and Y. Yamamoto, Phys. Rev. C {\bf 59}, 21-40 (1999).

\bibitem{nij06}
Th.A. Rijken, Y. Yamamoto, Phys. Rev. C {\bf 73} 044008 (2006).

\bibitem{HHS89}
B. Holzenkamp, K. Holinde, and J. Speth, Nucl. Phys. A {\bf 500} (1989) 485.
\bibitem{RHKS96}
A. Reuber, K. Holinde, H.-C. Kim, and J. Speth, Nucl. Phys. A {\bf 608} 243 (1996).
\bibitem{HM05}
J. Haidenbauer and Ulf G. Mei{\ss}ner, Phys. Rev. C {\bf 72} 044005 (2005).
\bibitem{savage-wise}
  M.~J.~Savage and M.~B.~Wise,
  Phys.\ Rev.\ D {\bf 53}, 349 (1996).
  
\bibitem{KDT01}
C.L. Korpa, A.E.L. Dieperink, and R.G.E. Timmermans, Phys. Rev. C {\bf 65}, 015208 (2001).

\bibitem{Hammer02}
H.W. Hammer, Nucl. Phys. A {\bf 705}, 173 (2002). 


\bibitem{Beane:2003yx}
  S.~R.~Beane, P.~F.~Bedaque, A.~Parre\~no and M.~J.~Savage,
  Nucl.\ Phys.\  A {\bf 747}, 55 (2005)
  [arXiv:nucl-th/0311027].


\bibitem{PHM06}
  H.~Polinder, J.~Haidenbauer and Ulf-G.~Mei\ss ner,
  Nucl.\ Phys.\ A {\bf 779}, 244 (2006)
  [arXiv:nucl-th/0605050].




\bibitem{Beane:2006gf}
  S.~R.~Beane, P.~F.~Bedaque, T.~C.~Luu, K.~Orginos, E.~Pallante, A.~Parre\~no and M.~J.~Savage
                  [NPLQCD Collaboration],
  Nucl.\ Phys.\  A {\bf 794}, 62 (2007)
  [arXiv:hep-lat/0612026].





\bibitem{Michael:1999nq}
  C.~Michael and P.~Pennanen  [UKQCD Collaboration],
  Phys.\ Rev.\  D {\bf 60}, 054012 (1999)
  [arXiv:hep-lat/9901007].

\bibitem{Pennanen:1999xi}
  P.~Pennanen, C.~Michael and A.~M.~Green  [UKQCD Collaboration],
  Nucl.\ Phys.\ Proc.\ Suppl.\  {\bf 83}, 200 (2000)
  [arXiv:hep-lat/9908032].

\bibitem{Green:1999mf}
  A.~M.~Green, J.~Koponen and P.~Pennanen,
  Phys.\ Rev.\  D {\bf 61}, 014014 (2000)
  [arXiv:hep-ph/9902249].

\bibitem{Fiebig:2001mr}
  H.~R.~Fiebig  [LHP collaboration],
  Nucl.\ Phys.\ Proc.\ Suppl.\  {\bf 106}, 344 (2002)
  [arXiv:hep-lat/0110163].

\bibitem{Fiebig:2001nn}
  H.~R.~Fiebig  [LHP Collaboration],
  Nucl.\ Phys.\ Proc.\ Suppl.\  {\bf 109A}, 207 (2002)
  [arXiv:hep-lat/0112010].

\bibitem{Cook:2002am}
  M.~S.~Cook and H.~R.~Fiebig,
  arXiv:hep-lat/0210054.

\bibitem{Takahashi:2006er}
  T.~T.~Takahashi, T.~Doi and H.~Suganuma,
  AIP Conf.\ Proc.\  {\bf 842}, 249 (2006)
  [arXiv:hep-lat/0601006].

\bibitem{Doi:2006kx}
  T.~Doi, T.~T.~Takahashi and H.~Suganuma,
  AIP Conf.\ Proc.\  {\bf 842}, 246 (2006)
  [arXiv:hep-lat/0601008].

\bibitem{Green:2004ia}
  A.~M.~Green,
  arXiv:nucl-th/0409021.

\bibitem{Detmold:2007wk}
  W.~Detmold, K.~Orginos and M.~J.~Savage,
  Phys.\ Rev.\  D {\bf 76}, 114503 (2007)
  [arXiv:hep-lat/0703009].



\bibitem{Luke:1992tm}
  M.~E.~Luke, A.~V.~Manohar and M.~J.~Savage,
  Phys.\ Lett.\  B {\bf 288}, 355 (1992)
  [arXiv:hep-ph/9204219].



\bibitem{Brodsky:1997gh}
  S.~J.~Brodsky and G.~A.~Miller,
  Phys.\ Lett.\  B {\bf 412}, 125 (1997)
  [arXiv:hep-ph/9707382].

\bibitem{Yokokawa:2006td}
  K.~Yokokawa, S.~Sasaki, T.~Hatsuda and A.~Hayashigaki,
  Phys.\ Rev.\  D {\bf 74}, 034504 (2006)
  [arXiv:hep-lat/0605009].




\bibitem{Aggarwal:2000ex}
  M.~M.~Aggarwal {\it et al.}  [WA98 Collaboration],
  Phys.\ Rev.\ Lett.\  {\bf 85}, 2895 (2000)
  [arXiv:hep-ex/0008018].

\bibitem{Bearden:2001ea}
  I.~G.~Bearden {\it et al.}  [NA44 Collaboration],
  Phys.\ Lett.\  B {\bf 517}, 25 (2001)
  [arXiv:nucl-ex/0102013].

\bibitem{Adams:2003vd}
  J.~Adams {\it et al.}  [STAR Collaboration],
  Phys.\ Rev.\ Lett.\  {\bf 91}, 262301 (2003)
  [arXiv:nucl-ex/0306028].

\bibitem{Stathopoulos:2007zi}
  A.~Stathopoulos and K.~Orginos,
  arXiv:0707.0131 [hep-lat].


\end{thebibliography}
\end{document}